\newcommand{\ba}{\begin{eqnarray}}
\newcommand{\ea}{\end{eqnarray}}
\newcommand{\be}{\begin{equation}}
\newcommand{\ee}{\end{equation}}
\newcommand{\nh}{{\bf      h}}

\newcommand{\nk}{{\bf      k}}
\newcommand{\np}{{\bf      p}}
\newcommand{\nq}{{\bf      q}}

\documentclass[10pt]{article}
\usepackage{authblk}
\usepackage{graphics}
\usepackage{xcolor}
\usepackage{amsmath}
\usepackage{graphicx}
\usepackage{anysize}

\usepackage{ulem} 
\usepackage{soul} 

\date{ } 
\title{Neutrino-Nucleus scattering in the SuSA model}
\author[1]{J.E. Amaro}
\author[2,3]{M.B. Barbaro}
\author[4,5]{J.A. Caballero}
\author[6]{T.W. Donnelly}
\author[7]{R. Gonz\'alez-Jim\'enez}
\author[4,8]{G.D. Megias}
\author[1]{I. Ruiz Simo
}
\affil[1]{
Departamento de F\'isica At\'omica, Molecular y Nuclear, and Instituto de F\'isica Te\'orica y Computacional Carlos I, Universidad de Granada, Granada 18071, Spain}
\affil[2]{
Dipartimento di Fisica, Universit\`{a} di Torino and INFN, Sezione di Torino, Via P. Giuria 1, 10125 Torino, Italy}
\affil[3]{
Universit\'e Paris-Saclay, CNRS/IN2P3, IJCLab, 91405 Orsay, France}
\affil[4]{
Departamento de F\'{i}sica At\'omica, Molecular y Nuclear, Universidad de Sevilla, 41080 Sevilla, Spain}
\affil[5]{
Instituto de F\'isica Te\'orica y Computacional Carlos I, Granada 18071, Spain}
\affil[6]{
Center for Theoretical Physics, Laboratory for Nuclear Science and Department of Physics, Massachusetts Institute of Technology, Cambridge, Massachusetts 02139, USA}
\affil[7]{
Grupo de F\'isica Nuclear, Departamento de Estructura de la Materia, F\'isica T\'ermica y Electr\'onica, Facultad de Ciencias F\'isicas, Universidad Complutense de Madrid and IPARCOS, Madrid 28040, Spain}
\affil[8]{
University of Tokyo, Institute for Cosmic Ray Research, Research Center for Cosmic Neutrinos, Kashiwa, Japan
}
\begin{document}
\maketitle
\abstract{
The Super-Scaling Approach (SuSA) model, based on the analogies between electron and neutrino interactions with nuclei, is reviewed and its application to the description of neutrino-nucleus scattering is presented. The contribution of both one- and two-body relativistic currents is considered. A selection of results is presented where theoretical predictions are compared with cross section measurements from the main ongoing neutrino oscillation experiments.
} 

\section{Introduction}
\label{sec:intro}

A huge experimental programme on accelerator-based neutrino oscillation experiments has been developed in recent years with the scope of improving our knowledge of neutrino properties and of potentially answering one of the fundamental open questions of modern physics, that is the origin of the matter/antimatter asymmetry in the universe, by measuring the weak CP-violating phase~\cite{Abe:2019vii}.
The success of this program partly relies on the control of systematic errors, largely due to uncertainties in the description of the neutrino interactions  with the  detector, typically made of medium-weight nuclei like carbon, oxygen or argon. 
With this motivation, intense theoretical activity has been carried out in parallel in order to provide an accurate description of neutrino-nucleus reactions
in the GeV region, relevant for the experiments, for different processes ranging from  quasielastic  up to deep inelastic scattering, encompassing the excitation of nucleon resonances and the emission of two or more nucleons  (see \cite{Alvarez-Ruso:2017oui} for a comprehensive review of recent progresses and open challenges in the field). 

In this article we review our work on the SuSA (Super-Scaling Approach) model, originally developed in Ref.~\cite{Amaro:2004bs} and subsequently refined and improved to the updated version SuSAv2~\cite{Gonzalez-Jimenez:2014eqa,Megias:2016fjk} (see also Refs.~\cite{Amaro20,Barbaro:2021psv} for other recent reviews). 
The original SuSA, as will be explained in more detail in Sect.~\ref{sec:scal}, is a phenomenological model which, while retaining the relativistic aspect of the Relativistic Fermi Gas (RFG), provides  by construction - unlike the RFG - a good description of inclusive electron scattering data $(e,e')$ in the quasielastic region. In the model initial and final state interaction effects, absent in the RFG, are directly extracted from $(e,e')$ data. 
The updated version, SuSAv2, implements inputs from the relativistic mean field model (RMF), which provides a microscopic interpretation of  the basic features of SuSA and also includes a more detailed description of the different spin and isospin channels. 
Moreover, the model has been extended to include two-body currents, able to excite two-particle-two-hole (2p2h) states; these, as will be shown in Sect.~\ref{sec:results}, play an important role in the analysis of neutrino-nucleus experimental data.

The scheme of this review is the following. In Sect.~\ref{sec:incl} we focus on the inclusive reaction, presenting the formalism, describing the model and testing it with electron scattering data.
In Sect.~\ref{sec:semi_electrons} we discuss the semi-inclusive reaction an its relation to the inclusive one. In Sect.~\ref{sec:results} we show  some selected results and discuss the comparison with experimental data. Finally, in Sect.~\ref{sec:concl} we draw our conclusion and outline the future developments of our work.

\section{Inclusive scattering}
\label{sec:incl}

\subsection{General formalism}
\label{sec:form}
In this review we mainly focus on the charged current (CC) inclusive process
\begin{equation}
\nu_l (\overline\nu_l)+A \longrightarrow l^-(l^+) + X \,,
\end{equation}
where a neutrino (antineutrino) of a given flavour $l$ having four-momentum  $K^\mu=(\varepsilon,\nk)$ hits a nucleus $A$ and a charged lepton $l^-$ ($l^+$) is detected in the final state with four-momentum $K^{\prime\mu}=(\varepsilon^\prime,\nk^\prime)$ and scattering angle $\theta$, while the residual system $X$ is unobserved. The four-momentum transferred from the probe to the nucleus is $Q^\mu=(\omega,{\bf q})$, with $\omega=\varepsilon-\varepsilon^\prime$ and ${\bf q}={\bf k}-{\bf k^\prime}$. We shall assume the neutrino to be massless, while the outgoing lepton has finite mass $m_l$.
The corresponding cross section is obtained by contracting the  leptonic and hadronic tensors
\begin{eqnarray}
l_{\mu\nu}(\nq,\omega) = 2\left(K_\mu K^\prime_\nu+K^\prime_\mu K_\nu- K\cdot K^\prime g_{\mu\nu}+ i \chi \epsilon_{\mu\nu\rho\sigma}K^\rho K^{\prime\sigma}\right)
\label{eq:lmunu}
\\
W^{\mu\nu}(\nq,\omega) = \sum_n <A|J^{\mu\dagger}(\nq,\omega)|n><n|J^\nu(\nq,\omega)|A> \,\delta(\omega+E_A-E_n) ,
\label{eq:Wmunu}
\end{eqnarray}
where $\chi=+1 (-1)$ in the neutrino (antineutrino) case, $|A>$ is the initial nuclear ground state and $|n>$ are all the intermediate hadronic states that can be reached through the weak current operator $J^\mu$. 
The double differential cross section with respect to the momentum and scattering angle of the outgoing lepton can be  expressed
in terms of five response functions~\cite{Amaro:2004bs} (for the semi-inclusive reaction $(\nu_l,l^-N)$, which will be briefly discussed in Sect.~\ref{sec:semi_electrons}, the response functions are ten)     
\begin{equation}
\frac{d^2\sigma}{d\varepsilon^\prime d\cos\theta}
=
\frac{(G_F\cos\theta_c)^2}{4\pi}
v_0 \frac{k^\prime}{\varepsilon^\prime}
\left(
V_{CC} R_{CC}+
2{V}_{CL} R_{CL}
+{V}_{LL} R_{LL}+
{V}_{T} R_{T}
+
2\chi{V}_{T'} R_{T'}
\right), 
\label{eq:cs}
\end{equation}
where $G_F$ the Fermi weak constant,
$\theta_c$ the Cabibbo angle and $v_0= (\varepsilon+\varepsilon^\prime)^2-{\bf q}^2$.  The indices $C$, $L$, $T$ refer to the Coulomb, longitudinal and transverse components of the leptonic and hadronic currents
with respect to  $\nq$.

The leptonic coefficients $V_K$ are related to the components of the tensor \eqref{eq:lmunu} and are given by
\begin{eqnarray}
&&V_{CC } = 1-\delta^2 \frac{|Q^2|}{v_0},\ 
V_{CL }  = \frac{\omega}{q}+\frac{\delta^2}{\rho^\prime} \frac{|Q^2|}{v_0},\ 
V_{LL }  =  \frac{\omega^2}{q^2}+ \delta^2 \frac{|Q^2|}{v_0}
\left(1+\frac{2\omega}{q\rho^\prime}
+\rho\delta^2\right),\nonumber\\
&&V_{T } = \frac{|Q^2|}{v_0}
+\frac{\rho}{2}-\delta^2\frac{|Q^2|}{v_0}
\left(\frac{\omega}{q\rho^\prime}+\frac{\rho\delta^2}{2}\right) ,\ 
V_{T' }  = \frac{|Q^2|}{v_0}
\left(\frac{1}{\rho^\prime}-\frac{\omega\delta^2}{q}\right) ,
\end{eqnarray}
where the variables $\delta^2=\frac{m_l^2}{|Q^2|}$, $\rho=\frac{|Q^2|}{q^2}$ and $\rho^\prime=\frac{q}{\varepsilon+\varepsilon^\prime}$  have been introduced.

The response functions 
\begin{equation}
R_{CC } =  W^{00 },\ 
R_{CL }  =  -\frac{1}{2}\left(W^{03 }+W^{30}\right),\ 
R_{LL }  =  W^{33 },\ 
R_{T } =  W^{11 } + W^{22 },\ 
R_{T' }  =  -\frac{i}{2}\left(W^{12 }-W^{21}\right)
\end{equation}
embody the nuclear dynamics and are specific components of the hadronic tensor $W^{\mu\nu}$, depending only upon $\omega$ and $\bf q$ .

The above decomposition into response functions is valid for all reaction channels  - elastic, quasielastic, inelastic - each of which characterized by a 
different current  operator $J^\mu$ in the hadronic tensor \eqref{eq:Wmunu}. In the case of quasielastic scattering, corresponding to the interaction of the  probe with a single nucleon, the weak current operator is
\begin{equation}
J^\mu({\bf q},\omega)=F_1(Q^2)\gamma^\mu+\frac{iF_2(Q^2)}{2m_N}\sigma^{\mu\nu}Q_\nu-G_A(Q^2)\gamma^\mu\gamma_5-\frac{G_P(Q^2)}{m_N}Q^\mu\gamma_5,
\end{equation}
where $F_1$, $F_2$, $G_A$ and $G_P$ are the Pauli, Dirac, axial and pseudoscalar weak form factors, respectively.

In neutrino oscillation experiments the incident beam is not monochromatic. As a consequence, before comparing with experimental data, the cross
section \eqref{eq:cs} must be folded with the normalized neutrino flux $\phi(\varepsilon)$ $(\int d\varepsilon \,\phi(\varepsilon)=1)$:
\begin{equation}
\langle \frac{d^2\sigma}{d\varepsilon^\prime d\cos\theta} \rangle  =
\int d\varepsilon \,\phi(\varepsilon) \,\frac{d^2\sigma}{d\varepsilon^\prime d\cos\theta}.
\end{equation}

\subsection{Scaling and SuSAv2}
\label{sec:scal}

The phenomenon of scaling is sometimes found to occur in different fields, including solid state, atomic, molecular, nuclear and hadronic physics, when an interacting probe scatters from composite many-body systems. Under some circumstances it is found that the response of the complex system no longer depends on two independent variables, but only on a particular combination of those, called the scaling variable. This phenomenon is very well known in high energy physics where the inelastic nucleon response functions are shown to depend only on the Bjorken variable $x$~\cite{bjorkxy}. A similar phenomenon is observed in lepton-nucleus scattering, where the nuclear response functions are found to scale with a single scaling variable, denoted as $y$. This indicates that the probe (lepton) interacts with the nucleus' constituents, in this case with the nucleons in the nucleus. 

The phenomenon of $y$-scaling emerges from the analysis of quasielastic (QE)
$(e,e')$ reactions and has been studied in detail in \cite{Day90,DS199,DS299}. The scaling function is defined as the QE $(e,e')$ differential cross section divided by a single-nucleon cross section averaged  over the Fermi gas. 
For high enough values of the momentum transfer, $q$, the scaling function does only depend
on a single variable, $y$, given as a particular combination of the energy ($\omega$) and momentum ($q$) transferred in the process.
The scaling variable $y$ is (up to a sign) the minimum value of the missing momentum allowed by 
kinematics~\cite{Day90,DS199,DS299}.
In the QE domain the basic mechanism in $(e,e')$ reactions on nuclei corresponds to elastic scattering from
individual nucleons in the nuclear medium. This implies that the inclusive $(e,e')$ cross section is mainly constructed from the
exclusive $(e,e'N)$ process, including the contribution of all nucleons in the target and integrating over all (unobserved) ejected
nucleon variables (see also the following section). This approach constitutes the basis of the Impulse Approximation (IA). Thus, the double differential ($e,e'$) inclusive cross section is given as the sum of two response functions corresponding 
to the longitudinal, $L$, and transverse, $T$, channels,
\begin{equation}\label{ecsec}
 \displaystyle \frac{d^2\sigma}{d\Omega_e d\omega}=\sigma_{Mott}\left[v_LR_L(q,\omega)+v_TR_T(q,\omega)\right],
\end{equation}
where $\sigma_{Mott}$ is the Mott cross section and the $v_{L,T}$
are kinematical factors that involve leptonic variables (see~\cite{Amaro20,Kel96} for explicit expressions). 
In terms of the scaling functions the nuclear responses are
\begin{equation}\label{eresp}
\displaystyle R_{L,T} (q,\omega)=\frac{1}{k_F}\Bigl[f_{L,T}(q,\omega)G_{L,T}(q,\omega)\Bigr] ,
\end{equation}
where  $k_F$ is the Fermi momentum and $G_{L,T}$ are defined as the
responses of a moving nucleon and include relativistic corrections arising from the presence of the medium. 
Their explicit expressions can be found in \cite{Gonzalez-Jimenez:2014eqa,Amaro20,Amaro07}. The terms $f_{L,T}$
are the scaling functions that show a mild dependence 
upon the momentum transfer $q$ (first-kind scaling) and very weak dependence on the nuclear
system considered (second-kind scaling). The occurrence of both types of scaling is denoted as superscaling.
Hence to the extent  that at some kinematics the above $f$ functions are the same for all nuclei and do only depend on a
single variable, denoted as the scaling variable $\psi(q,\omega)$, but not separately on $q$, one says that superscaling occurs.
The superscaling variable $\psi$ is given in terms of $q$ and $\omega$ as
\begin{equation}
\psi = \frac{1}{\sqrt{\xi_F} } \frac{\lambda-\tau}{\sqrt{(1+\lambda)\tau+\kappa\sqrt{\tau(1+\tau)}}} \, ,
\label{eq:psi}
\end{equation}
where the dimensionless Fermi kinetic energy $\xi_F=\frac{T_F}{m_N}$, energy transfer $\lambda=\frac{\omega}{2m_N}$, momentum transfer
$\kappa=\frac{q}{2m_N}$ and squared four-momentum transfer $\tau=\kappa^2-\lambda^2$ have been introduced.
Here the $y$-variable introduced above is given approximately by
\begin{equation}
y \cong k_F \psi
\label{ytopsi}
\end{equation}
as discussed in \cite{DS299}.
A phenomenological energy shift $E_{shift}$,  fitted for each nucleus to the experimental  position of the quasielastic peak~\cite{MDS02}, is included in the definition of the scaling variable $\psi'\equiv \psi(q,\omega-E_{shift})$. 

\begin{figure}[htbp]
\centering  
\includegraphics[width=.325\textwidth,angle=270]{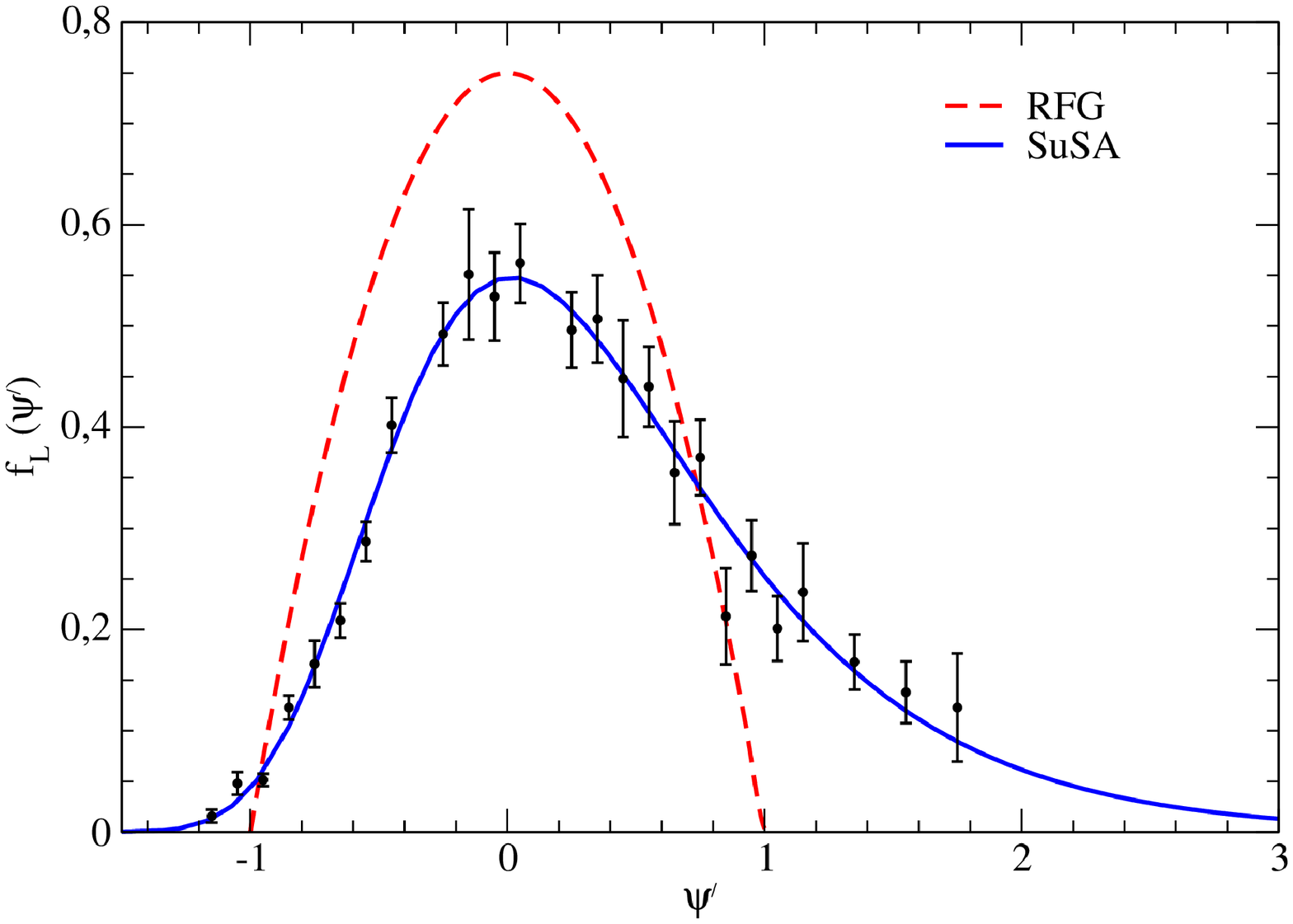}\hspace*{0.25cm}
\includegraphics[width=.325\textwidth,angle=270]{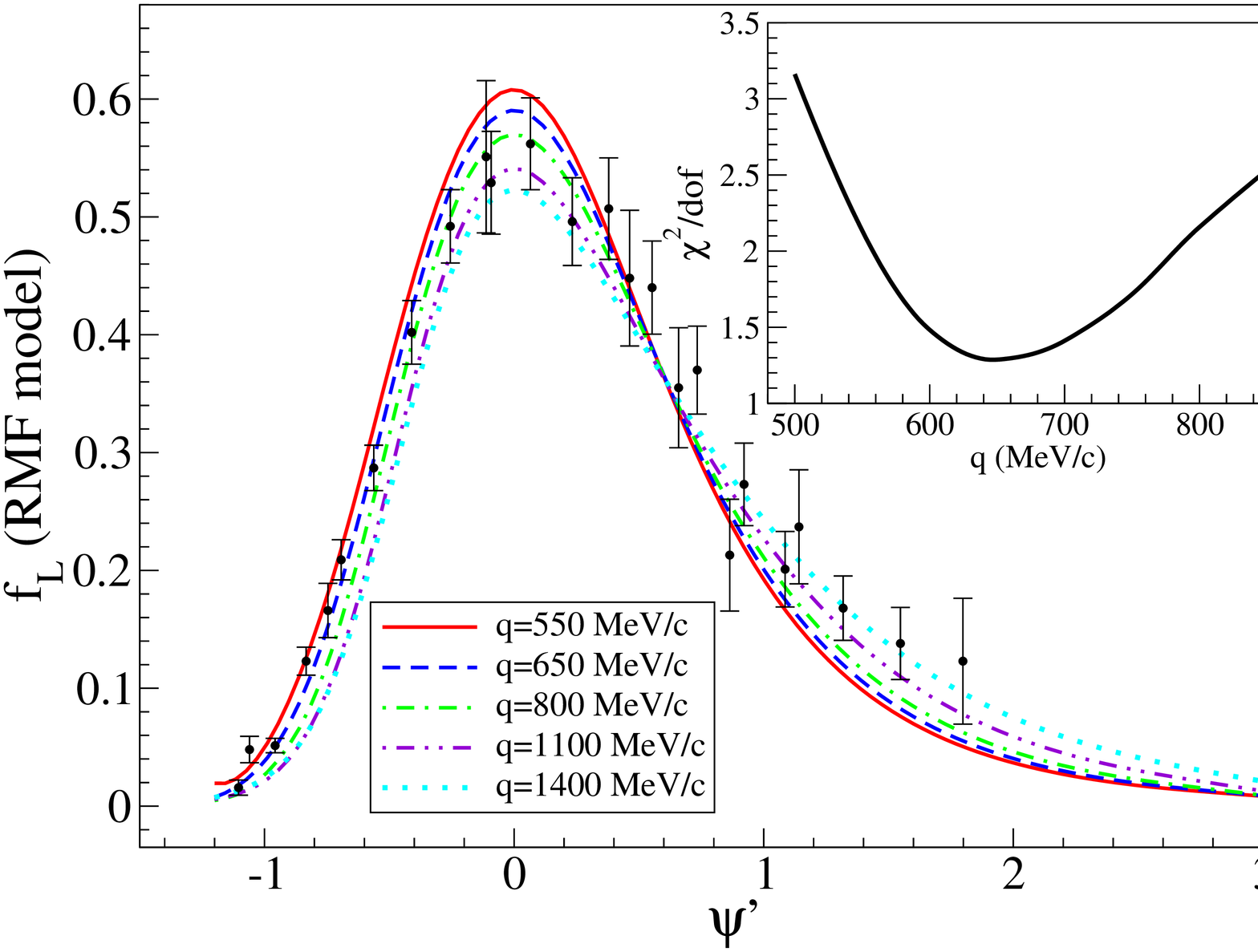}
\caption{{\it Left panel:} Phenomenological SuSA scaling function versus $\psi'$ in the QE region together with ($e,e'$) longitudinal scaling data from~\cite{Jourdan}. The RFG scaling function is also shown as reference. 
 {\it Right panel:} RMF longitudinal scaling functions for ($e,e'$) at different $q$ values compared with ($e,e'$) longitudinal scaling data from~\cite{Jourdan}. In the inner smaller plot a reduced-$\chi^2$ analyses shows a minimum at $q=650$ MeV/c. Figure adapted from~\cite{Gonzalez-Jimenez:2014eqa}.}\label{fig:scaling}
\end{figure}

Scaling and superscaling properties of electron-nucleus interactions have been analyzed in detail in a series of previous 
works~\cite{Day90,DS199,DS299,MDS02,Barbaro:1998gu,PRL05,jac06} . The importance of this phenomenon to test the validity of any nuclear model aiming to describe electron scattering reactions has been clearly proven. The model, denoted as the Superscaling Approach (SuSA), is entirely based on the phenomenology, making use of a unique, universal, scaling function extracted from the analysis of the longitudinal electron scattering data (see Fig.~\ref{fig:scaling}). Notice that the behavior and properties of the experimental superscaling function constitute a 
strong constraint for any theoretical model describing QE electron
scattering. Not only should the superscaling behavior be fulfilled, but also the specific shape of the longitudinal scaling function, $f^{exp}_L$, must be reproduced. The SuSA model assumes the longitudinal phenomenological scaling function to be valid also in describing the transverse 
channel, {\it i.e.,} $f_L=f_T$.
\vspace*{-0.15cm}

In recent years we have explored in detail the extension of SuSA to neutrino-nucleus scattering. Due to the complexity of the weak process, with an increased number of nuclear responses compared with the pure electromagnetic one, and the particular role played by the axial term in the weak current, we have developed an improved version of the superscaling model, called SuSAv2~\cite{Gonzalez-Jimenez:2014eqa}.
Contrary to the original SuSA~\cite{Amaro:2004bs,Amaro07,Amaro:2006tf} where a universal scaling function based on electron scattering data is used, the new SuSAv2 model incorporates relativistic mean field (RMF) effects~\cite{PRL05,jac06} in the longitudinal and 
transverse nuclear responses, as well as in the isovector and isoscalar channels that is of great importance in order to describe charged-current (CC) neutrino reactions that are purely isovector~\cite{Caballero:2007tz}.

The origin of the SuSAv2 approach is based on the capability of the RMF to describe properly the scaling behavior of the electron scattering data. As shown in previous works~\cite{PRL05,jac06}, RMF is one of the few microscopic models capable of reproducing the asymmetric shape of the phenomenological scaling function with a long tail extended to high values of the transfer energy (large values of $\psi^\prime$). Moreover, RMF produces an enhancement in the transverse scaling function, a genuine dynamical relativistic effect linked to the lower components in the wave functions, that is supported by the analysis of data. The RMF framework to finite nuclei has proven to successfully reproduce the scaling behavior shown by data at low to intermediate $q$ values (see Fig.~\ref{fig:scaling}). However, the model clearly fails at higher momentum transfers where Final State Interactions (FSI) are expected to be weaker. This is due to the RMF strong energy-independent scalar and vector potentials used in the final state that lead to too much asymmetry in the scaling functions and shift the QE peak to very high transfer energies, in clear disagreement with data. To remedy this shortfall of the RMF model, the SuSAv2 incorporates both the RMF scaling functions at low-to-intermediate $q$ values and the Relativistic Plane Wave Impulse Approximation (RPWIA) ones at higher $q$ by using a $q$-dependent blending function that smoothly connects the two regimes (see~\cite{Megias:2016fjk,Megias:2016lke} for details). A similar solution to this drawback of the RMF model has been taken in the recent Energy-Dependent RMF (ED-RMF) approach~\cite{Gonzalez-Jimenez_edRMF,Gonzalez-Jimenez_edRMF2} where RMF potentials are multiplied by a blending function inspired by the SuSAv2 one that scales them down as the kinetic energy of the scattered nucleon increases, also preventing non-orthogonality issues. This model predicts both lepton and nucleon kinematics, showing a similar agreement on electron and neutrino data with SuSAv2. 

In summary, the SuSAv2 model and, for extension, the ED-RMF one reproduce the experimental longitudinal scaling data, gives rise to an enhancement in the electromagnetic transverse channel, {\it i.e.}, $f_T^{(e,e')} > f_L^{(e,e')}$, takes into account the differences in the isoscalar/isovector scaling functions, of crucial interest for neutrino scattering processes, and finally avoids the problems of the RMF model in the region of high momentum transfer, where FSI effects are negligible. One of the basic merits of SuSAv2 is the translation of sophisticated and demanding microscopic calculations into relatively straightforward parametrizations and, hence, easing its implementation in the MonteCarlo simulations employed in the analysis of neutrino oscillation experiments.

To conclude, it is important to point out that SuSAv2 is not restricted to the QE kinematics domain. On the contrary, it can be extended to the inelastic region by generalizing the superscaling variable \eqref{eq:psi} to the excitation of any inelastic state having invariant mass $W$. One can thus define~\cite{Barbaro:2003ie}
\begin{equation}
\psi_W = \frac{1}{\sqrt{\xi_F} } \frac{\lambda-\tau\rho_W}{\sqrt{(1+\lambda\rho_W)\tau+\kappa\sqrt{\tau(1+\tau\rho_W^2)}}}
\,,\ \ \rho_W=1+\frac{1}{4\tau}\left(W^2/m_N^2-1\right)
\end{equation}
and evaluate the inelastic hadronic tensor as the integral
\begin{equation}
W^{\mu\nu}_{inel} = \frac{1}{m_N^2} \int_{W_{\rm min}}^{W_{\rm max}} dW W f(\psi_W)\, w^{\mu\nu}_{inel} 
\end{equation}
between the limits $W_{\rm min}, W_{\rm max}$ imposed by the kinematics, where  $f(\psi_W)$ is the SuSAv2 superscaling function previously described.
This procedure is based on the assumption that the nuclear effects, encoded in the function $f$, are the same in the QE and inelastic regions and requires the knowledge of the elementary tensor $w^{\mu\nu}_{inel}$ across the full inelastic spectrum. 
\begin{figure}[htbp]
\centering  
\includegraphics[width=.345\textwidth,angle=270]{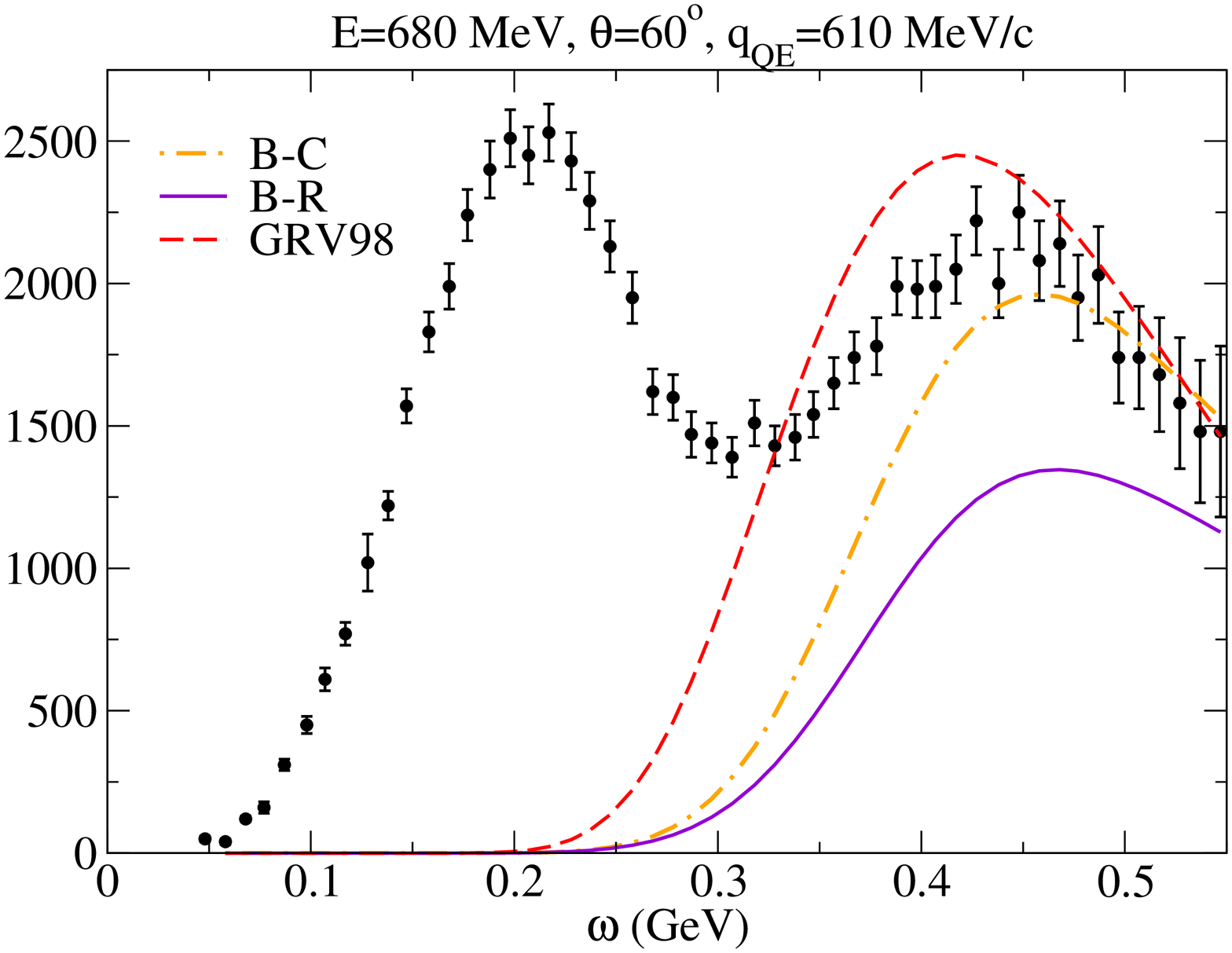}\hspace*{0.25cm}
\includegraphics[width=.345\textwidth,angle=270]{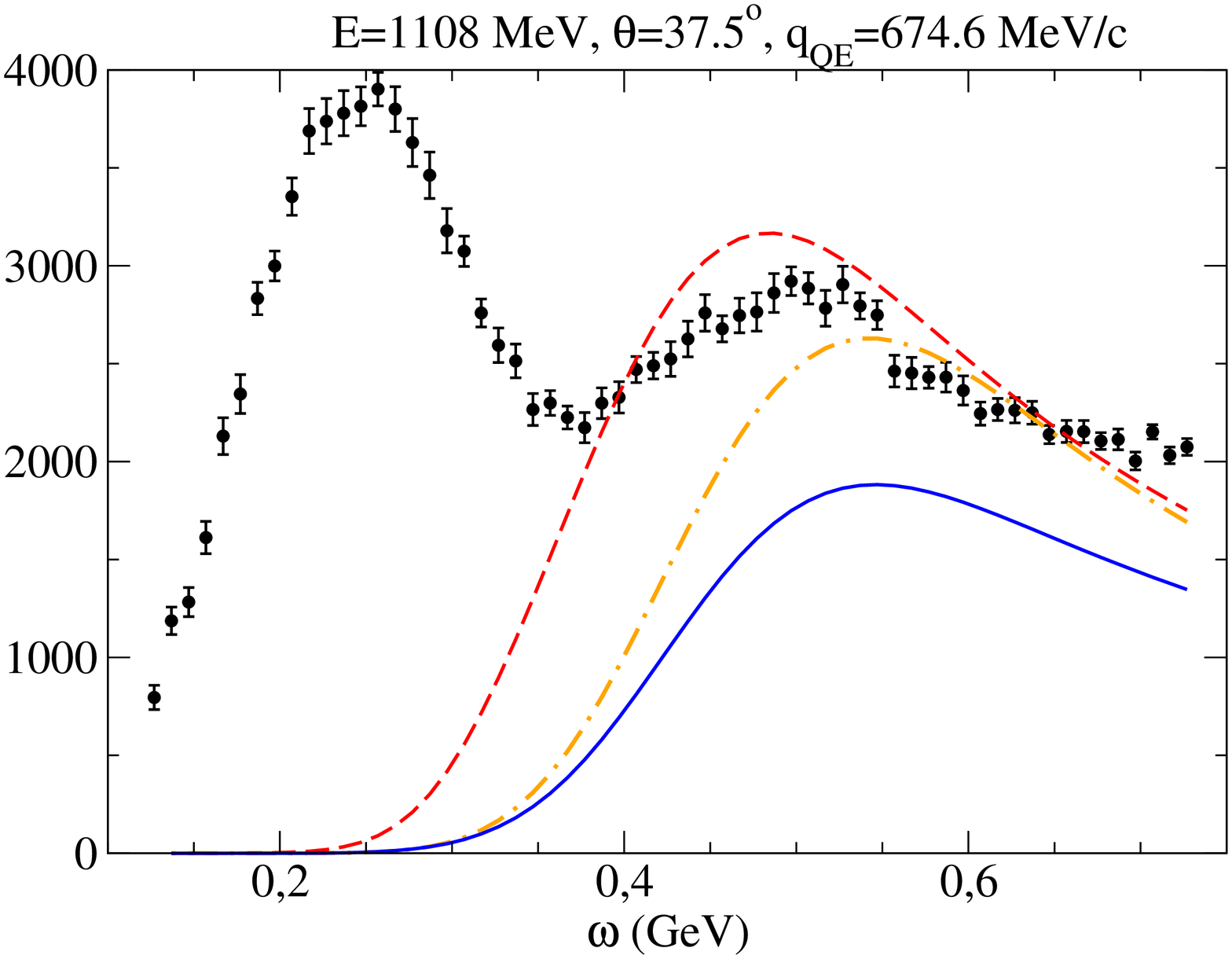}
\caption{ 
Comparison of inclusive $^{12}$C($e,e'$) double differential cross sections and predictions for the inelastic regime of the Bosted-Christy parametrization~\cite{Bosted:2007xd,Christy:2007ve} (dot-dashed lines), Bodek-Ritchie parametrization~\cite{BodekRitchie1,BodekRitchie2} (solid lines) and GRV98 PDFs~\cite{GRV98_1,GRV98_2} (dashed lines) at different kinematics (incident electron beam and scattering angle) in terms of the energy transferred to the nucleus ($\omega$). Experimental data taken from~\cite{QESarchive,QESarxiv}. The y-axis represents $d^2\sigma/d\Omega/d\omega$ in nb/GeV/sr. The value of $q$ at the QE peak ($q_{QE}$) is shown as reference.}\label{fig:susav2inelastic}
\end{figure}

The SuSAv2 inelastic model has been applied to the electron scattering case using phenomenological fits of the single-nucleon inelastic structure functions $w_1$ and $w_2$ extracted from e-p and e-d data~\cite{Bosted:2007xd,Christy:2007ve}. In Fig.~\ref{fig:susav2inelastic}, a comparison of the SuSAv2-inelastic model with $^{12}$C($e,e'$) data using different inelastic structure functions shows a preference for the Bosted-Christy parametrization~\cite{Bosted:2007xd,Christy:2007ve}.
In general, the SuSAv2 model (QE+inelastic) provides a very good description of data for very different kinematics once Meson Exchange Currents (MEC) are also incorporated, denoted as SuSAv2-MEC (see Sect.~\ref{sec:electron} and \cite{Megias:2016fjk,Megias:2017cuh,Megias:2018ujz}). 
The inclusion of the full inelastic spectrum in the SuSAv2 model for weak interactions is still in progress. 

An alternative approach to the study of the resonant pion production was taken in Refs.~\cite{Amaro:2004bs,Maieron:2009an,Ivanov16}, where a scaling function to be used in the $\Delta$ resonance region, different from the quasielastic one, was extracted from electron scattering data and multiplied by the appropriate weak $N\to\Delta$ transition form factors to get the neutrino-nucleus cross section in this region. This method provides a phenomenological description valid at transferred energies below the $\Delta$ peak, while at higher $\omega$ it fails due to the opening of other inelastic channels. Some results corresponding to this method, referred to as SuSA-$\Delta$ approach, will be shown in Sect.~\ref{sec:results}.

\subsection{Relativistic model for CC MEC and 2p2h responses}
\label{sec:2p2h} 

\begin{figure}[htbp]
\centering
\includegraphics[width=8cm,bb=110 310 500 690]{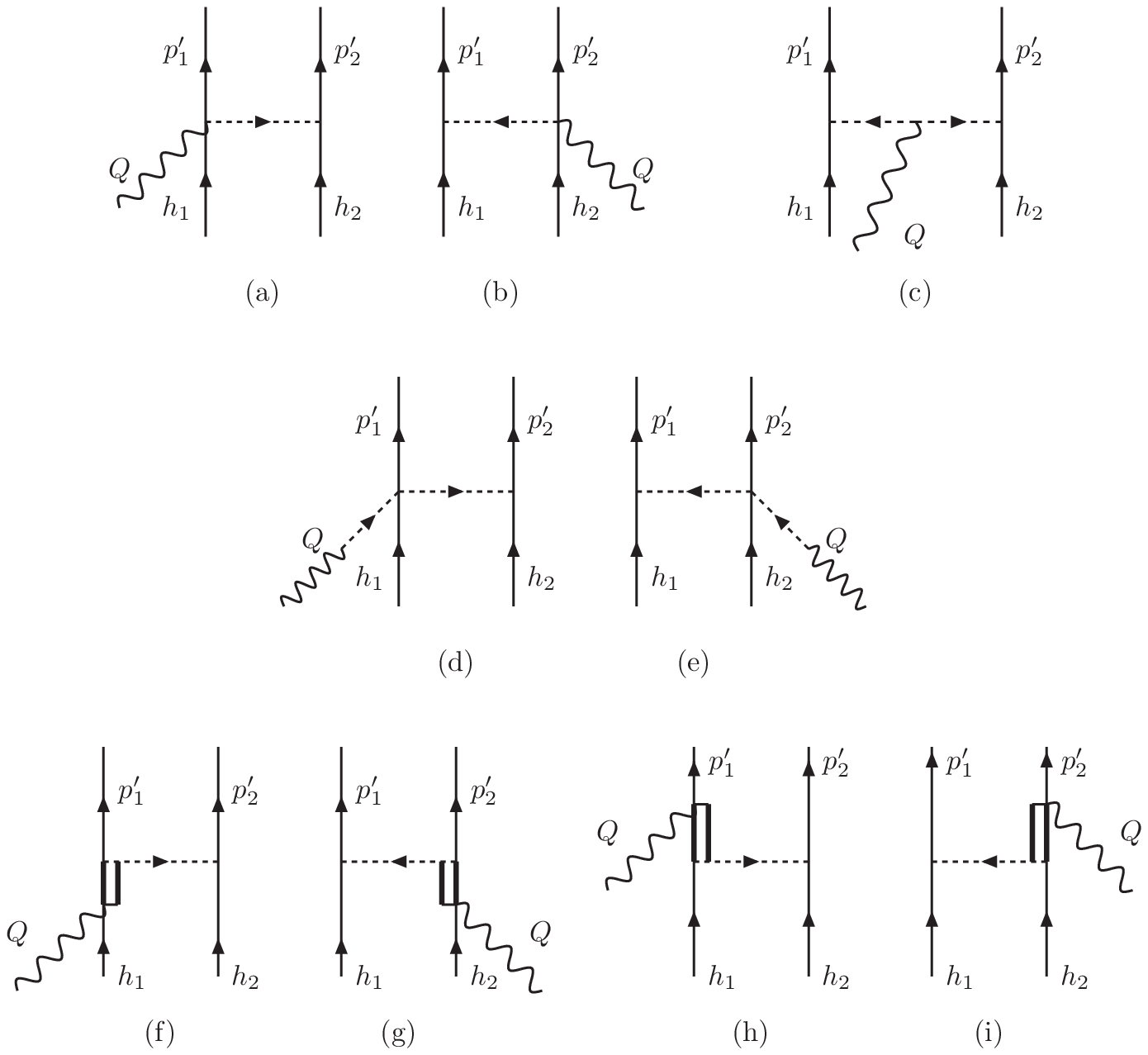}
\caption{Feynman diagrams for the electroweak MEC.  }\label{fig_feynman}
\end{figure}

Multinucleon knockout processes give a non-negligible contribution to
the inclusive neutrino cross section for the intermediate energies
involved in the experiments \cite{Megias:2016fjk,Mar09,Nieves:2011pp,Rocco:2018mwt,Sob20}.
In Ref. \cite{Sim16} we developed a model
 of two-particle two-hole excitations of the RFG induced by weak
meson-exchange currents for inclusive
CC neutrino scattering.  The model is fully relativistic and includes
the diagrams of Fig. \ref{fig_feynman}, involving one-pion exchange
and $\Delta$ excitation, taken from the pion production model of
\cite{Her07}.

The 2p2h matrix element of MEC 
depends on the momenta, spin and isospin coordinates 
$(1',2';1,2) \equiv  (\np'_1s'_1t'_1,\np'_2s'_2t'_2;\nh_1s_1t_1,\nh_2s_2t_2$)
of the two 
holes, $\nh_1$, $\nh_2$, and
the two particles, $\np'_1$, $\np'_2$. It is the
sum of four contributions 
\begin{equation}
j^\mu(1',2';1,2) 
\equiv j^\mu(\np'_1s'_1t'_1,\np'_2s'_2t'_2;\nh_1s_1t_1,\nh_2s_2t_2) =
j_{\rm sea}^\mu +j_{\rm \pi}^\mu + j_{\rm pole}^\mu + j_{\rm \Delta}^\mu,
\end{equation}
corresponding in Fig. \ref{fig_feynman} to the seagull (diagrams
a,b), pion in flight (c), pion-pole (d,e) and $\Delta(1232)$
excitation (f,g,h,i).  Their explicit expressions are given in Ref. 
\cite{Sim16}.

The inclusive hadronic tensor in the 2p-2h channel is computed by integration over all 
the 2p-2h excitations of the RFG.
Momentum conservation enforces $\bf p'_2= h_1+h_2+q-p'_1$.  Hence
\begin{equation}
W^{\mu\nu}_{\rm 2p2h}
=
\frac{V}{(2\pi)^9}\int
d^3p'_1
d^3h_1
d^3h_2
\frac{m_N^4}{E_1E_2E'_1E'_2}
w^{\mu\nu}(\np'_1,\np'_2;\nh_1,\nh_2)
\Theta(p'_1,h_1)\Theta(p'_2,h_2)
\delta(E'_1+E'_2-E_1-E_2-\omega) ,
\label{amaro-hadronic}
\end{equation}
where  $\Theta(p',h) \equiv
\theta(p'-k_F)
\theta(k_F-h)$. The tensor inside the integral is
\begin{equation}
w^{\mu\nu}(\np'_1,\np'_2;\nh_1,\nh_2) \equiv \frac{1}{4}
\sum_{s_1s_2s'_1s'_2}
\sum_{t_1t_2t'_1t'_2}
j^{\mu}(1',2';1,2)^*_A
j^{\nu}(1',2';1,2)_A \, ,
\label{amaro-elementary}
\end{equation}
where $j^\mu(1',2',1,2)_A$ is the antisymetrized MEC matrix element 
\begin{equation} \label{amaro-anti}
j^{\mu}(1',2',1,2)_A
\equiv j^{\mu}(1',2',1,2)-
j^{\mu}(1',2',2,1) \,.
\end{equation}
 The factor $1/4$ in Eq.~(\ref{amaro-elementary}) accounts for the
 antisymmetry of the two-body wave function in isospin formalism, 
to avoid double counting
 in the number of final 2p-2h states.

Due to azimuthal symmetry around the $z$ axis
---in the $\nq$ direction--- we fix the azimuthal angle of particle 1'
$\phi'_1=0$, and multiply by a factor $2\pi$. The
energy delta-function enables integrating over $p'_1$.
Then Eq.~(\ref{amaro-hadronic}) is reduced to a seven
dimensions integral that is computed numerically
\cite{Sim14a,Sim14b}.
The Dirac matrix elements of the currents are also computed numerically.

The 2p-2h inclusive cross section requires one to compute 
the five weak response functions, $R^{CC,CL,LL,T,T'}$, for
$(\nu_\mu,\mu^-)$.  All of these responses were computed and analyzed in
Ref.~\cite{Sim16}.
The five response functions have
been parametrized in the kinematic range $100< q< 2000$ 
\cite{Megias:2016fjk}; 
the parametrization is
convenient because in neutrino scattering there is an additional
integration over the incident neutrino flux.
 This parametrization 
has been implemented in the Monte Carlo event generator GENIE
\cite{Dol20,Megias:2018ujz}.

Using the parametrization of \cite{Megias:2016fjk} does not allow one to modify 
the internal parameters of the MEC model. This is why 
we have developed approximations to the 2p2h responses in order to speed up the calculation. Specifically, in the so-called frozen nucleon approximation we set the momenta of the two holes $h_1=h_2=0$, allowing us to integrate over the initial states analytically
\cite{Rui17}.
In the modified convolution approximation (MCA) \cite{Rui18} we 
write the 2p2h responses as a convolution of two 1p-1h responses 
multiplied by the elementary hadronic tensor 
$w^{\mu\nu}(\np'_1,\np'_2;\nh_1,\nh_2)$ evaluated for convenient 
averaged values of $\nh_1$ and $\nh_2$. The resulting MCA responses are 
a good approximation to the exact result with 4D integration only
over the momentum and energy comunicated to one of the nucleons. 

\subsection{Validation vs electron scattering}
\label{sec:electron}

The model introduced in the previous sections allows one to describe inclusive lepton-nucleus  scattering in the kinematic region including the quasielastic
reaction, the excitation of 2p2h states and the inelastic spectrum. Validation  against the large amount of existing high-precision electron scattering data is a necessary test to be performed before using nuclear models in the analysis of neutrino oscillation experiments.
Such tests have been successfully carried out for the SuSAv2 model in a very wide range of kinematics from low/intermediate energies up to the highly inelastic regime. Only at very low energy and momentum transfers ($q<$300 MeV/c and $\omega<$50 MeV) does the model fail to reproduce the data, as do all models based on the impulse approximation, which is clearly not appropriate in the low-energy regime.
SuSAv2 predictions have been extensively compared with the available $(e,e')$ data on carbon and oxygen in~\cite{Megias:2016lke,Megias:2017cuh} and with the more recent JLab data on argon and titanium in~\cite{SuSAv2JLab}. Here we just show a few representative examples, choosing the kinematics  
 of particular interest for ongoing neutrino experiments.

\begin{figure}[htbp]
\centering  
\includegraphics[width=.29\textwidth,angle=270]{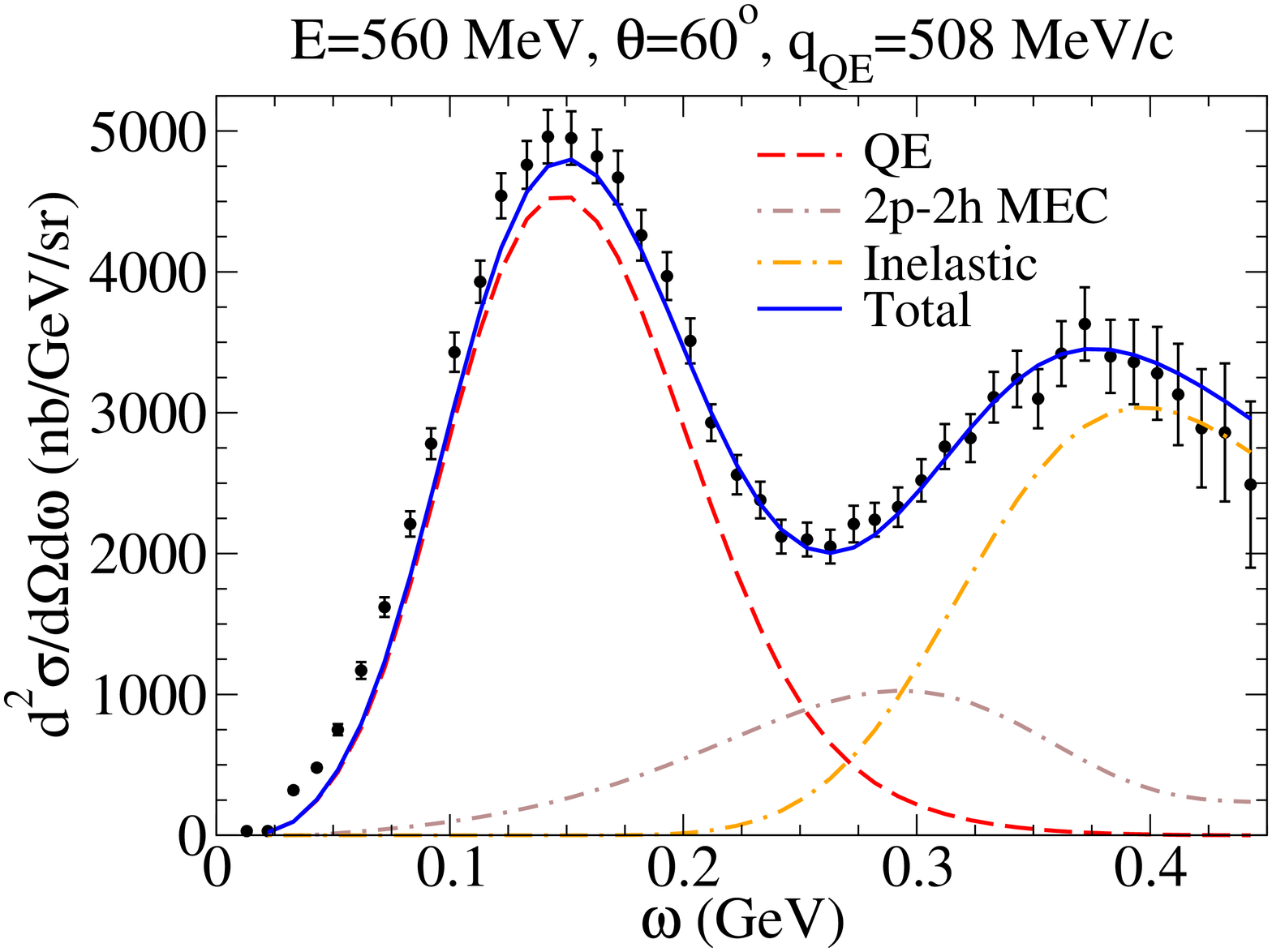}
\includegraphics[width=.29\textwidth,angle=270]{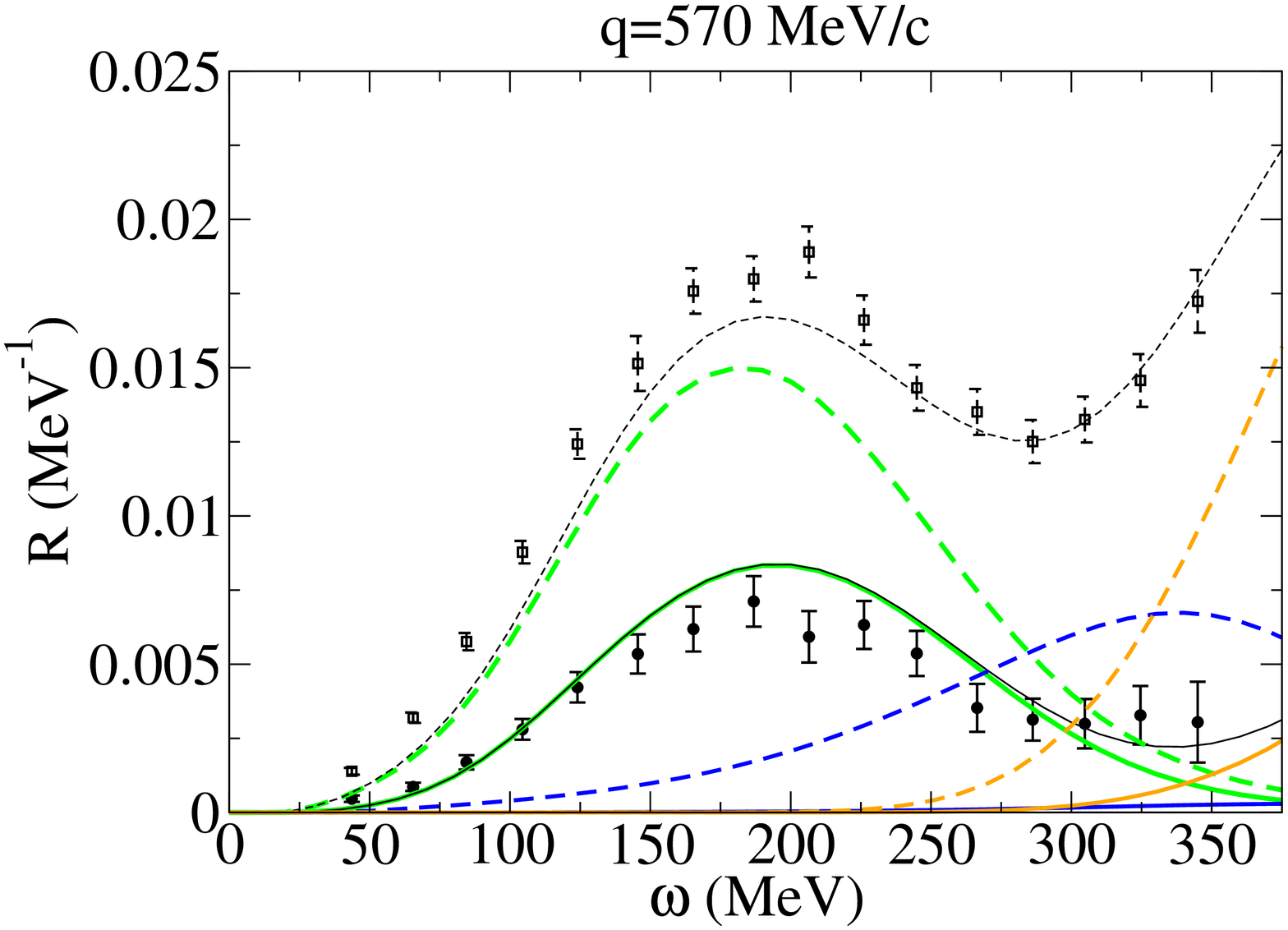}\\\vspace*{-0.295cm}
\includegraphics[width=.29\textwidth,angle=270]{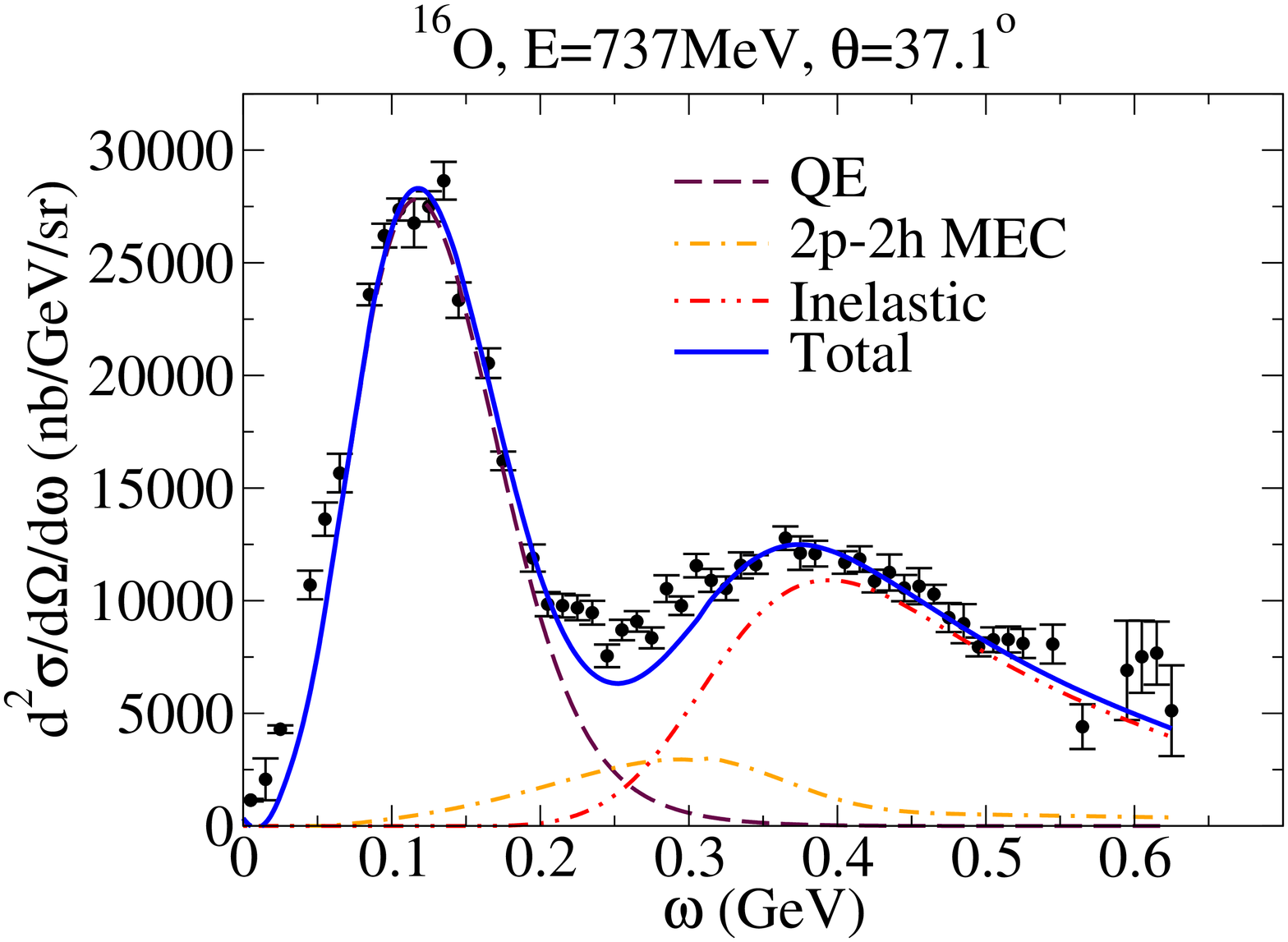}
\includegraphics[width=.29\textwidth,angle=270]{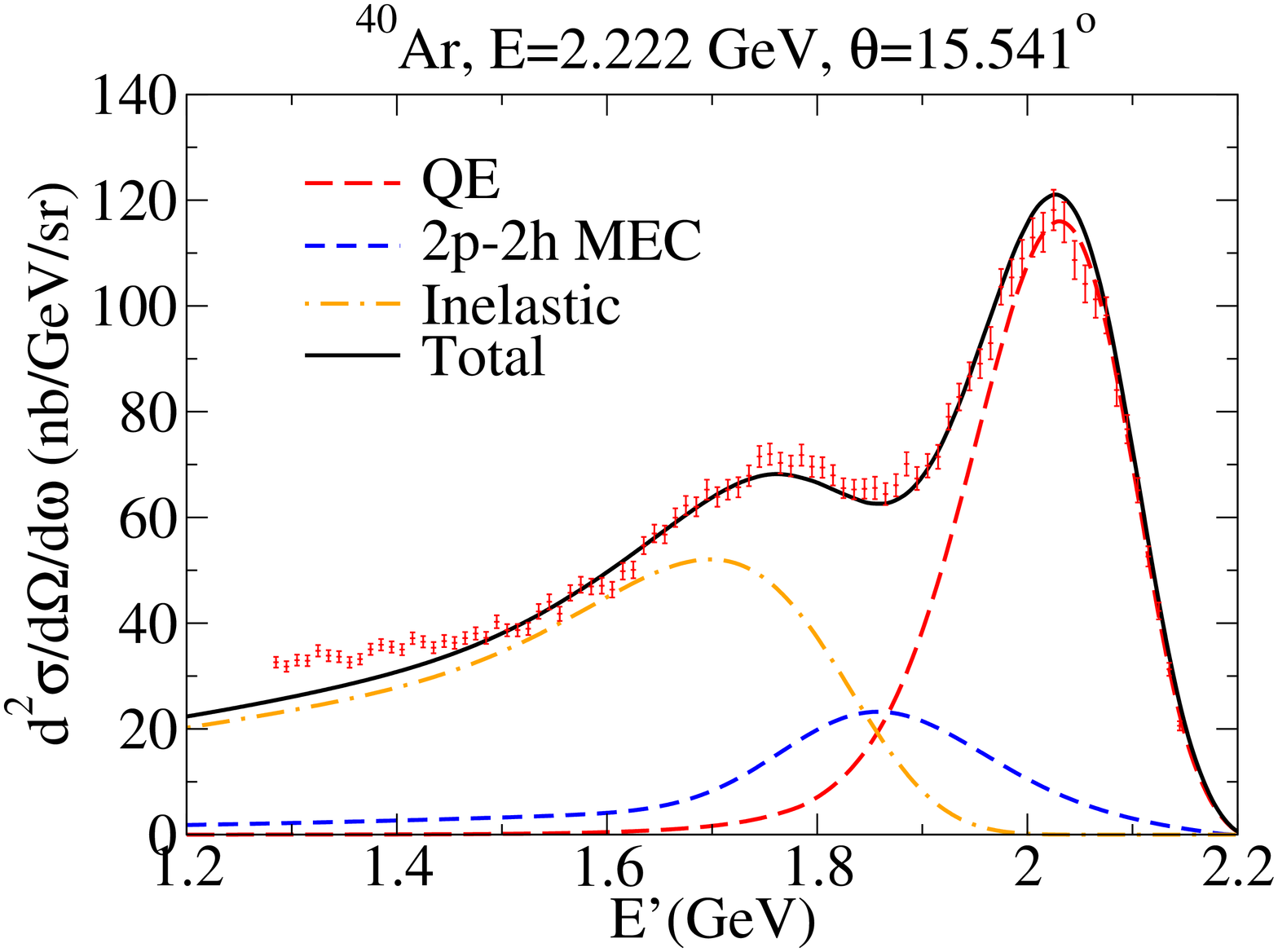}\vspace*{-0.15cm}
\caption{{\it Top left panel:} The $^{12}$C$(e,e')$ cross sections from~\cite{QESarchive,QESarxiv} compared with SuSAv2-MEC predictions. The separate QE, 2p-2h and inelastic contributions are also shown. 
 {\it Top right panel:} $^{12}$C$(e,e')$ longitudinal (solid) and transverse (dashed) responses at $q=$ 570 MeV/c. QE, 2p2h and inelastic contributions are shown, respectively,  as green, blue and orange lines.  The total response is shown by the black lines. Data from~\cite{Jourdan}. {\it Bottom panels:} The $(e,e')$ cross section (left) for $^{16}$O from~~\cite{Anghinolfi:1996vm} and for $^{40}$Ar (right) from~\cite{Dai:2018gch} compared with the SuSAv2-MEC model. The   separate QE,  2p-2h and inelastic contributions are also displayed.}\label{fig:Ceep}
\end{figure}
In Fig.~\ref{fig:Ceep} (top left panel)  the double differential $^{12}$C$(e,e')$ cross section 
is shown as a function of  the energy transfer $\omega$, for an incident electron beam of energy $E$ 
 and scattering angle $\theta$, 
 corresponding to the momentum transfer $q_{_{\rm QE}}$ 
 at the quasielastic peak.
The separate  QE, 2p2h  and inelastic  contributions are shown.  The agreement with the data is excellent across the full spectrum. Note that 2p2h excitations are essential in order to describe the "dip" region between the QE and $\Delta$ peaks.
 
An important feature of the SuSAv2 model, based on the RMF theory, is its capability to reproduce not only cross sections but also
the separated longitudinal and transverse responses, as shown in the top right panel of Fig.~\ref{fig:Ceep}, where $R_L$ (solid) and $R_T$ (dashed) are plotted versus $\omega$ for $q=$570 MeV/c, together with the separate QE, 2p2h and inelastic contributions. Note that while the longitudinal response is almost purely quasielastic, with a small 2p2h contributions arising from relativistic corrections, the transverse response receives contributions from all the three processes, which sizeably overlap at these kinematics.

In Fig.~\ref{fig:Ceep} (bottom panels) the same kind of comparison is shown for the oxygen (left) and argon (right) nuclei, showing that the superscaling approach allows for a consistent description of different nuclei.

These results give us confidence on the reliability of the model for the application to neutrino scattering. Before showing the comparison with neutrino data, in the next section we address the study of semi-inclusive electron scattering.

\section{Semi-inclusive scattering}
\label{sec:semi_electrons}

A pivotal difference between electron and neutrino scattering experiments is that the neutrino energy is known only as a broad distribution, while the energy in electron beams is typically defined with high precision. Hence, in electron scattering experiments by, for instance, detecting one nucleon in coincidence with the scattered electron, the hadronic final state is completely determined if one restricts one's attention to excitation energies of the residual system below the two-nucleon emission threshold. On the contrary, in the neutrino case, even in scenarios in which one or a few hadrons are detected in coincidence with the scattered lepton, there is no way of fully determining the hadronic final state because the beam energy and consequently the excitation energy of the residual system is unknown. This means that one needs to `integrate' for all possible final states compatible with the given kinematics. Formally, this integration is equivalent to the integration over the neutrino flux (or flux folding procedure) introduced in Sect.~\ref{sec:form}.

The semi-inclusive process for CCQE reactions was recently discussed in Refs.~\cite{VanOrden19,Gonzalez-Jimenez21}.
 
To obtain a somewhat deeper understanding of how the inclusive cross sections and scaling emerge, in this section we briefly discuss semi-inclusive (coincidence) electron scattering, focusing on the reaction $(e,e'N)$ in which the scattering electron and a nucleon are assumed to be detected in the final state. Then, in addition to the electron kinematical variables introduced above,  we have an outgoing nucleon with 4-momentum $P^\mu_N = (E_N,{\bf p}_N)$ involving 3-momentum ${\bf p}_N$ and polar and azimuthal angles $\theta_N$ and $\phi_N$, respectively, together with energy $E_N = \sqrt{p_N^2 + m_N^2}$. No other particles are assumed to be detected, although, depending on the specific kinematics, they must be present (see below). The magnitude of the nucleon's 3-momentum is given by $p_N=|\np_N|$. 
Apart from the detected nucleon, the final state contains an undetected hadronic system having missing 4-momentum $(E_B, {\bf p}_B)$, namely, a total energy of $E_B$ and a missing 3-momentum ${\bf p}_B \equiv {\bf p}_m$. In the following we shall assume that the detected nucleon is a proton. One then has 
\ba
{\bf p}_m = {\bf q} - {\bf p}_N\,. 
\ea
The undetected hadronic system has invariant mass $M_B$ ($M_B^0$ at threshold with $M_B \ge M_B^0$) and total energy 
\ba
E_B =T_B+M_B=\sqrt{ (M_B)^2 + {p_m}^2 }\,,\label{Emiss1}
\ea
which defines the kinetic energy of the unobserved final-state system, $T_B$. From (\ref{Emiss1}) one has
\ba
E_B = \varepsilon - \varepsilon^\prime - T_N + (M­_A^0-m_N)\,,\label{Em1}
\ea 
where $M_A^0$ is the target ground-state mass and $T_N=E_N-m_N$ is the kinetic energy of the detected nucleon. This leads to an expression for the so-called missing-energy,
\be
E_m = (M_B-M_B^0)+E_s 
    = \varepsilon - \varepsilon^\prime - T_N - T_B\,,\label{eq:E_m}
\ee
where $E_s=M_B^0+m_N-M_A^0$ is the separation energy and the (typically very small) recoil kinetic energy difference has been neglected. Defining the excitation energy of the residual system ${\cal E}=M_B-M_B^0$, we have
\ba
E_m = {\cal E} +E_s\,.\label{eq:calE}
\ea

The magnitude of the missing-momentum $p_m$ is given by 
\be
p_m=\bigl[k^2+{k^\prime}^2+p_N^2-2kk^\prime\cos\theta_l-2kp_N\cos\theta_N
+2k^\prime p_N(\cos\theta_l \cos\theta_N+\sin\theta_l \sin\theta_N\cos\phi_N)\bigr]^\frac{1}{2}\,.\label{eq:p_missing}
\ee

Depending on the value of the missing-energy, the residual system may be the daughter nucleus in its ground state (this defines the threshold for the reaction to become possible); or it may be in a discrete excited state (lying below the threshold where a second nucleon can be ejected), and, although they de-excite by $\gamma$-decay, that process is slow on the nuclear timescale and thus these states may be treated effectively as stationary states. Then, at a well-defined threshold a second nucleon must be emitted (this is not optional: there are no nuclear states involving one nucleon and a residual bound nucleus above this point). As $E_m$ continues to increase more and more particles enter in the final state in addition to the one special nucleon that is assumed to be detected. At even larger missing energy (roughly 140 MeV) pion production becomes possible (still with the lepton and one nucleon assumed to be detected). 

\begin{figure}[htbp]
\centering  
\includegraphics[width=.34\textwidth,angle=270]{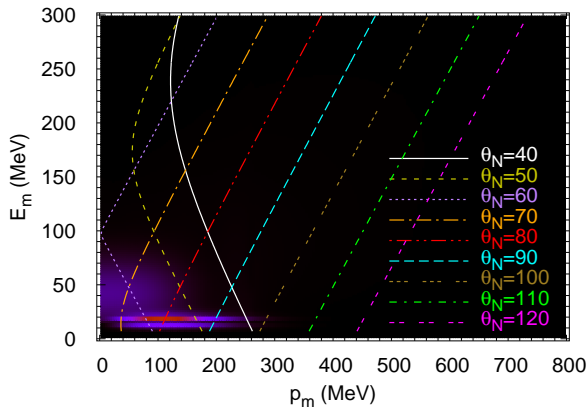}
\caption{The $E_m-p_m$ trajectories are shown for selected ``typical'' kinematics: $E_l=3800$ MeV, $\theta_l=7$ deg, $T_N=140$ MeV, $\phi_N=180$ deg for the reaction $^{16}$O$(e,e'p)$. Each line corresponds to a different value of the proton scattering angle $\theta_N$ (in  degrees). Here, we plot the Rome spectral function as a background to allow one to easily identify the different regions of the spectral function that are crossed by the trajectories. }\label{fig:trajectories}
\end{figure}

From (\ref{eq:E_m} -- \ref{eq:p_missing}) it is clear that for fixed values of the observable parameters the values of $E_m$ and $p_m$ are determined for each value of $E$.  In Fig.~\ref{fig:trajectories} trajectories are shown for selected kinematics, namely, $E_l=3800$ MeV, $\theta_l=7$ deg, $T_N=140$ MeV, $\phi_N=180$ deg, representing a ``typical'' situation. What is varied here is the polar angle for the detected proton, $\theta_N$. As stated above, as one goes along a given trajectory the electron energy $E$ that determines where on the trajectory one finds oneself must vary. The lower boundary defines the threshold for the semi-inclusive reaction to occur. In effect, each event where an electron and a proton are detected in coincidence corresponds to a specific trajectory and point on that trajectory.  

One sees a striking pattern to the behavior one should expect when going along a given trajectory. The strength in the Rome spectral function~\cite{Benhar94,Benhar05}, which should provide a good starting point for the characteristics to be expected in semi-inclusive reactions, is extremely localized. One sees the largest concentration of strength where the $p$-shells are located (at around $E_m = 20$ MeV) with less where the broad $s$-shell is located (at around $E_m = 50$ MeV); at still larger values of $E_m$ (and $p_m$) the spectral function does have some strength, although it is spread over a wide region in the $E_m-p_m$ plane and is too small to be seen in this representation. Furthermore, we note that pion production cannot occur until one reaches $E_m \sim m_\pi$ and that it is not appreciable until $E_m \sim m_\Delta - m_N \sim 300$ MeV. 

We notice that in the case of the CCQE interaction, studied in Ref.~\cite{Gonzalez-Jimenez21}, one gets a $E_m-p_m$ trajectory plot that is almost identical to the one in Fig.~\ref{fig:trajectories}. The reason is that the muon mass does not play a strong role in these trajectories, especially for the kinematics shown here for which $E_l$ is much larger than the muon mass.

In passing we note that, while other choices of two variables to replace $E_m$ and $p_m$ can of course be made, the generic behavior seen here strongly suggests that the present choice is a good one and that other choices may not reflect the highly localized nature of the nuclear response.

\begin{figure}[htbp]
\centering  
(a)\includegraphics[width=.3\textwidth,angle=270]{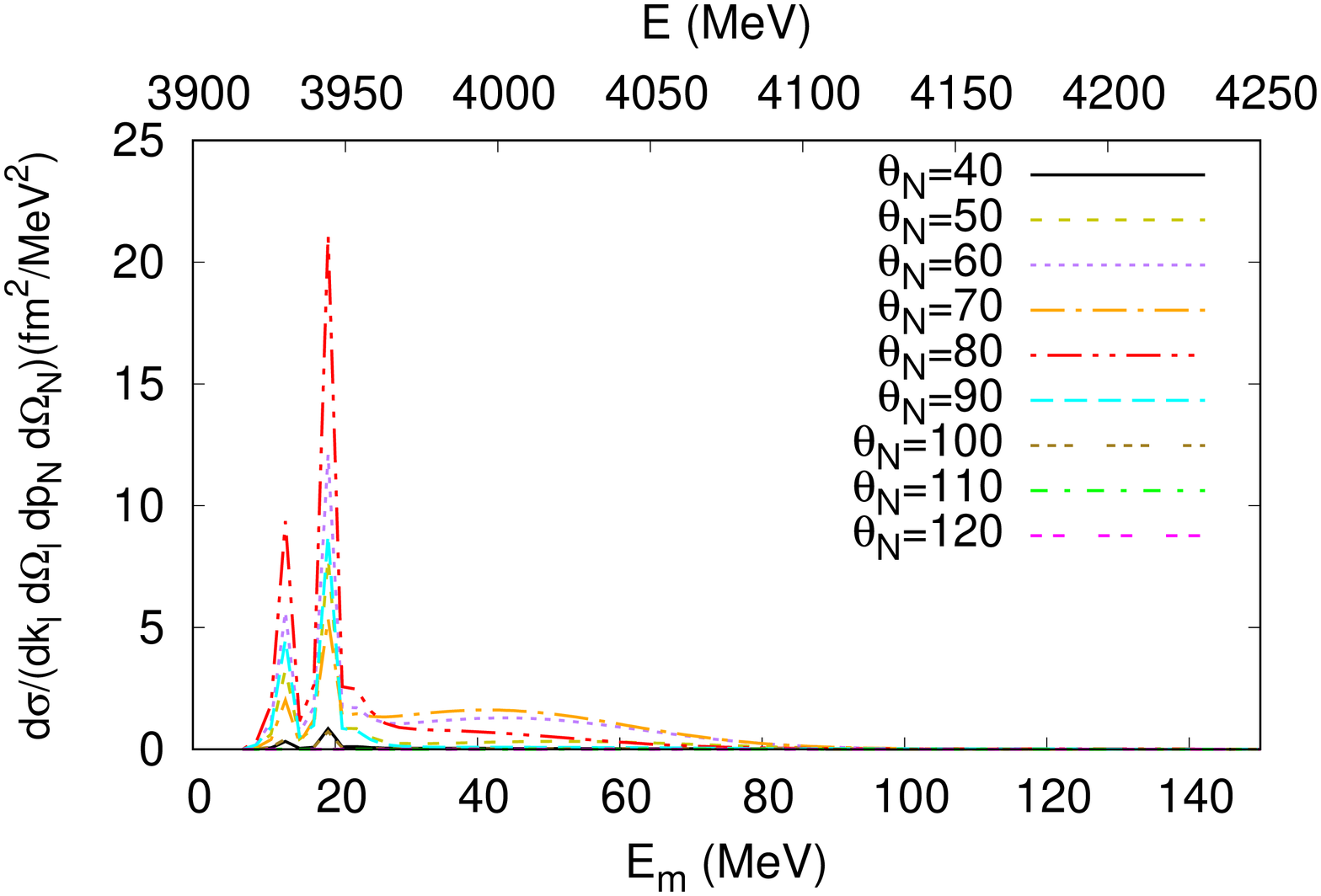}\ 
(b)\includegraphics[width=.3\textwidth,angle=270]{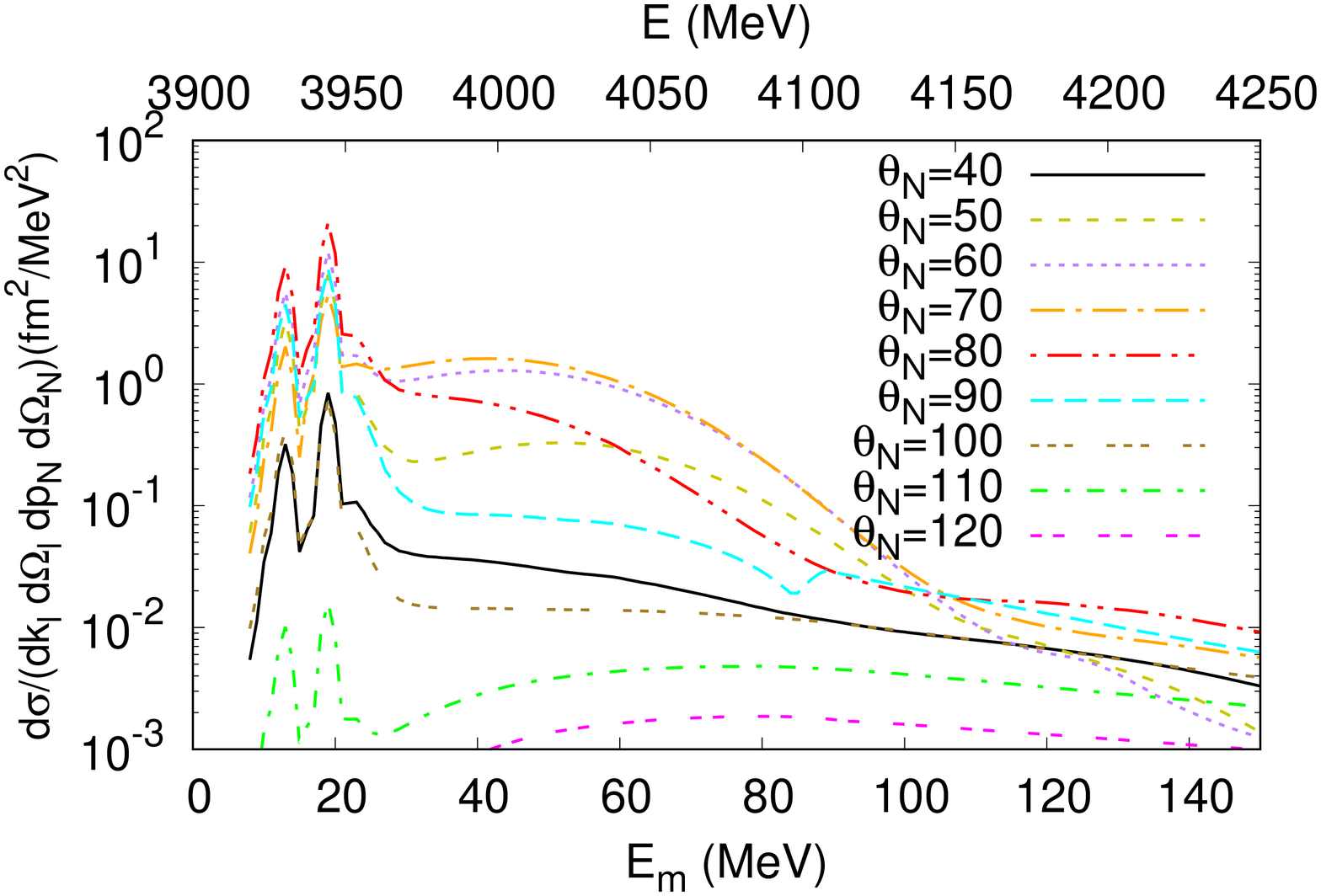}
\caption{The six-fold $^{16}$O$(e,e'p)$ differential cross section as a function of the missing-energy $E_m$ (lower x-axis) and the neutrino energy (upper x-axis) on linear (a) and semi-log (b) scales. The electron and proton variables are fixed to: $E_l=3800$ MeV, $\theta_l=7$ deg, $T_N=140$ MeV, $\phi_N=180$ deg, as in Fig.~\ref{fig:trajectories}. }\label{fig:6fold}
\end{figure}

In Fig.~\ref{fig:6fold} we represent the six-fold differential cross section for fixed electron and proton kinematics as a function of the missing-energy. The model used was PWIA together with the Rome spectral function~\cite{Benhar94,Benhar05}, as described, {\it e.g.,} in Refs.~\cite{VanOrden19,Franco-Patino20}.
Equation~\eqref{eq:E_m} tells us that for fixed nucleon and final lepton energies, if one neglects the kinetic energy of the residual system $T_B$, which under typical QE conditions is always very small, then one finds a one-to-one linear relation between the initial lepton energy and the missing energy. Thus, in Fig.~\ref{fig:6fold} the beam energy is shown as a second x-axis.
By varying the angle $\theta_N$, the cross section changes its magnitude and its shape. One observes two prominent peaks corresponding to the $p$-shells, a wide bump for the $s$-shell and a background that extends up to high missing-energies. Clearly, as expected, the $p$-shell strength is largest, the $s$-shell strength is smaller and the high-$E_m$ strength is completely negligible, being down by several orders of magnitude. By examining these results in the light of the trajectories shown in Fig.~\ref{fig:trajectories} we see that the general behavior we expect to occur is borne out. For example, the trajectories for $\theta_N = 60$ and $80$ degrees both pass though the $p$-shell region near its peak. However, one trajectory intersects the $s$-shell region more than the other one does and this results in relatively different amounts from the $s$-shell compared with the $p$-shell. 
Or, large values of $\theta_N$ correspond to very small cross sections, as they should, since neither the $p$- nor $s$-shell regions are crossed.

\begin{figure}[htbp]
\centering  
\includegraphics[width=.75\textwidth,angle=0]{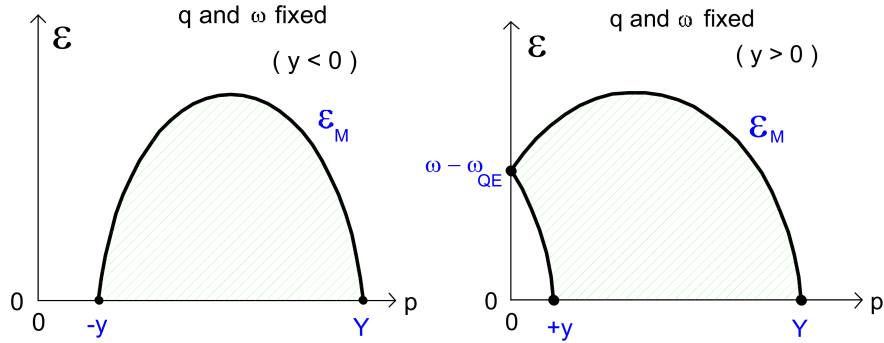}
\caption{Excitation energy ${\cal E} = E_m - E_s$ versus missing momentum $p=p_m$. The shaded area represents the kinematically allowed region for $y<0$ (left panel) and $y>0$ (right panel).}\label{fig:regions}
\end{figure}

We do have some knowledge about this generic behavior of the distribution of strength from inclusive electron scattering, $(e,e')$. Inclusive scattering corresponds to performing integrals over specific regions in the $E_m-p_m$ plane~\cite{Amaro20,Day90,Moreno14}. In Fig.~\ref{fig:regions} we show the typical situation for kinematics at values of $q$ and $\omega$ where one is below the QE peak ($y < 0$, left panel) or above the peak ($y > 0$, right panel).  In each case the inclusive QE cross section is obtained by integrating the semi-inclusive $(e,e'N)$ cross sections over the shaded regions. Accordingly one can see that the two classes of cross section are intrinsically related. Clearly, if one had complete knowledge of the semi-inclusive response for a wide range of kinematics then the integrations could be performed to yield the inclusive response. Unfortunately this is not the case. Note that having a model for the inclusive cross section (the total hadronic cross section) does not mean that such a model will be valid for the semi-inclusive cross section (which constitutes the integrand of the former). An example of this is the RFG model which is not unreasonable for the total (inclusive) QE cross section, but is poor for the semi-inclusive response.

An example is that of the model used for the semi-inclusive response discussed above. If employed for the inclusive cross section one finds for $^{16}$O that somewhat over 50\% of the inclusive cross section stems from the $p$-shells, about 25\% comes from the $s$-shell region and the rest comes from a broad region at higher missing-energy. This strength at higher missing energy is partially responsible for the asymmetry found in the scaling functions. Note that inclusive scattering at high momentum transfers and hence the scaling functions involve broad integrals, whereas semi-inclusive scattering at fixed final-state electron and nucleon momenta involves a trajectory in the $E_m - p_m$ plane as the beam energy is varied. In detailed analyses one finds much more strength in that case coming from the valence knockout region, with much less arising from the high-$E_m$ region. This has consequences for CC$\nu$ reactions as discussed in Ref.~\cite{Gonzalez-Jimenez21}.

\section{Results }
\label{sec:results}

In Sect.~\ref{sec:electron}, the validation of the SuSAv2-MEC model based on the RMF theory with ($e,e'$) data has been proven as a solid benchmark to assess the validity of a nuclear model before its application to neutrino reactions. Here, we show the capability of the SuSAv2-MEC model to describe a wide range of kinematics of interest for neutrino oscillation experiments. In particular, we focus on the comparisons of the SuSAv2-MEC model with charged-current neutrino cross sections from different experiments, in particular T2K and MINERvA. Our analysis is mainly related to charged-current quasielastic-like events, also called CC0$\pi$, which are characterized by having no pions detected in the final state and are dominated by the QE and 2p2h channels. Finally, we extend the analysis to cross section measurements where pion production 
is also included. 

  \begin{figure}[htbp]
	\begin{center}\vspace{0.80cm}
			\hspace*{-0.84cm}\includegraphics[scale=0.167, angle=270]{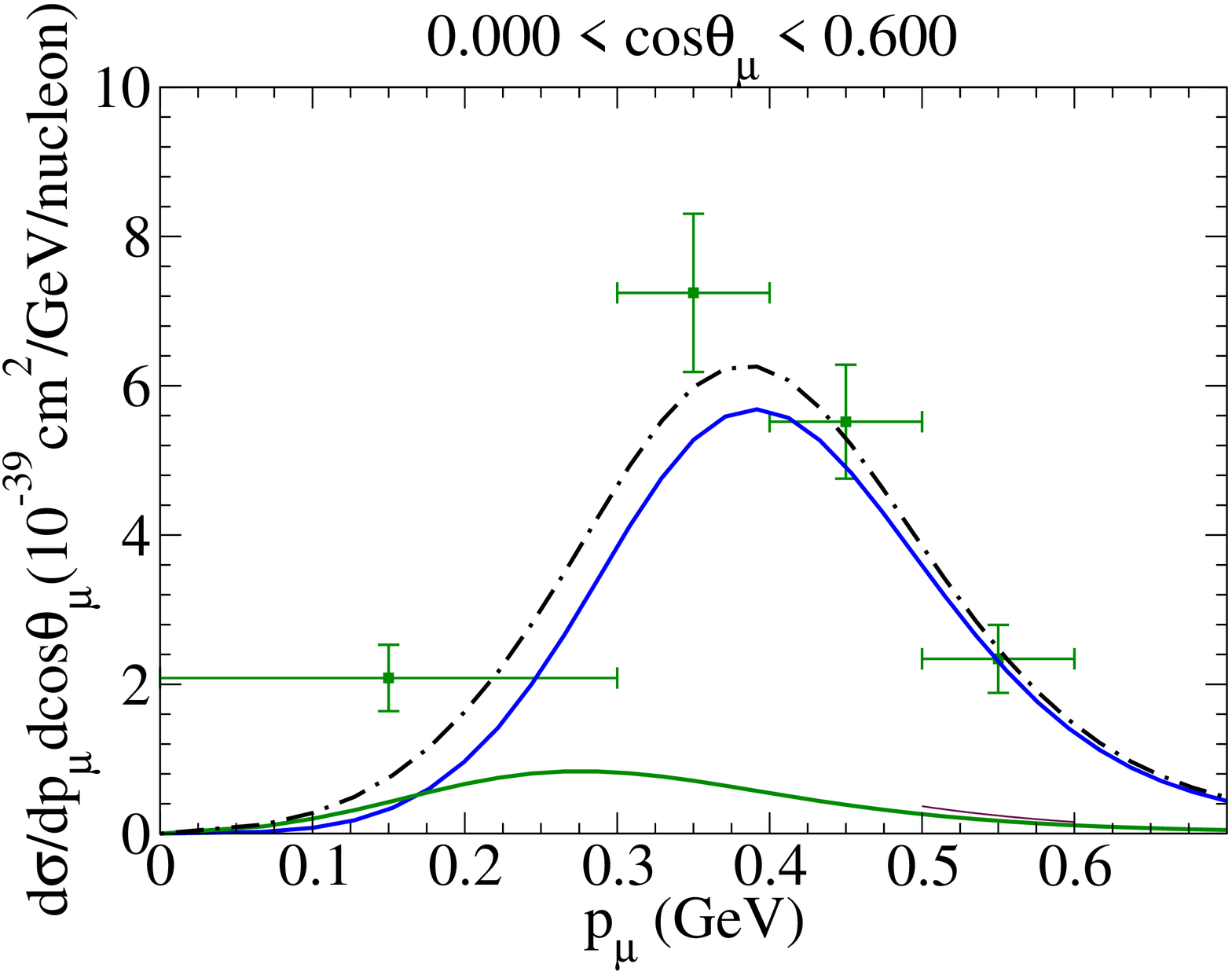}\hspace*{-0.64cm}%
		\includegraphics[scale=0.167, angle=270]{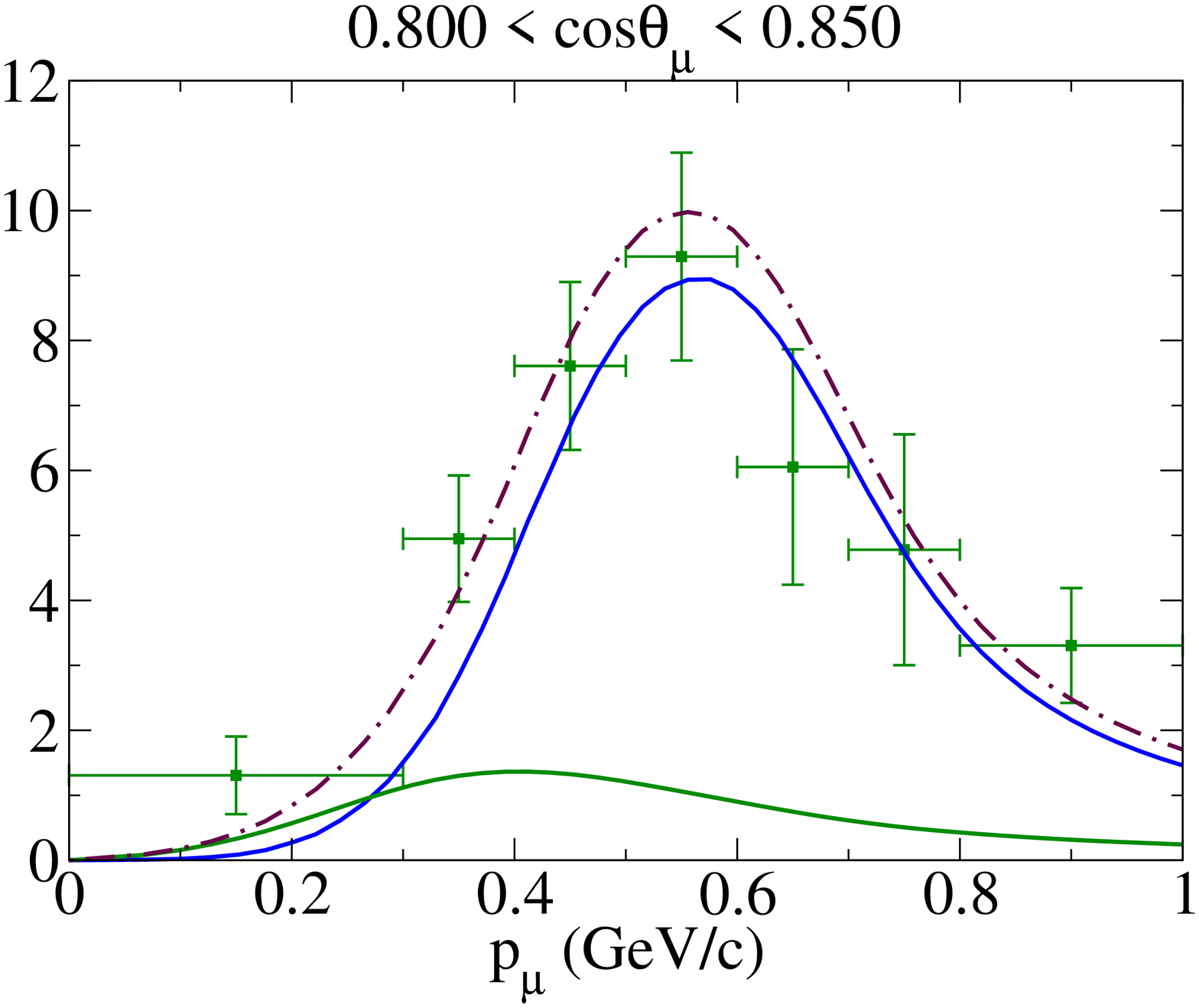}\hspace*{-0.74cm}
		\includegraphics[scale=0.167, angle=270]{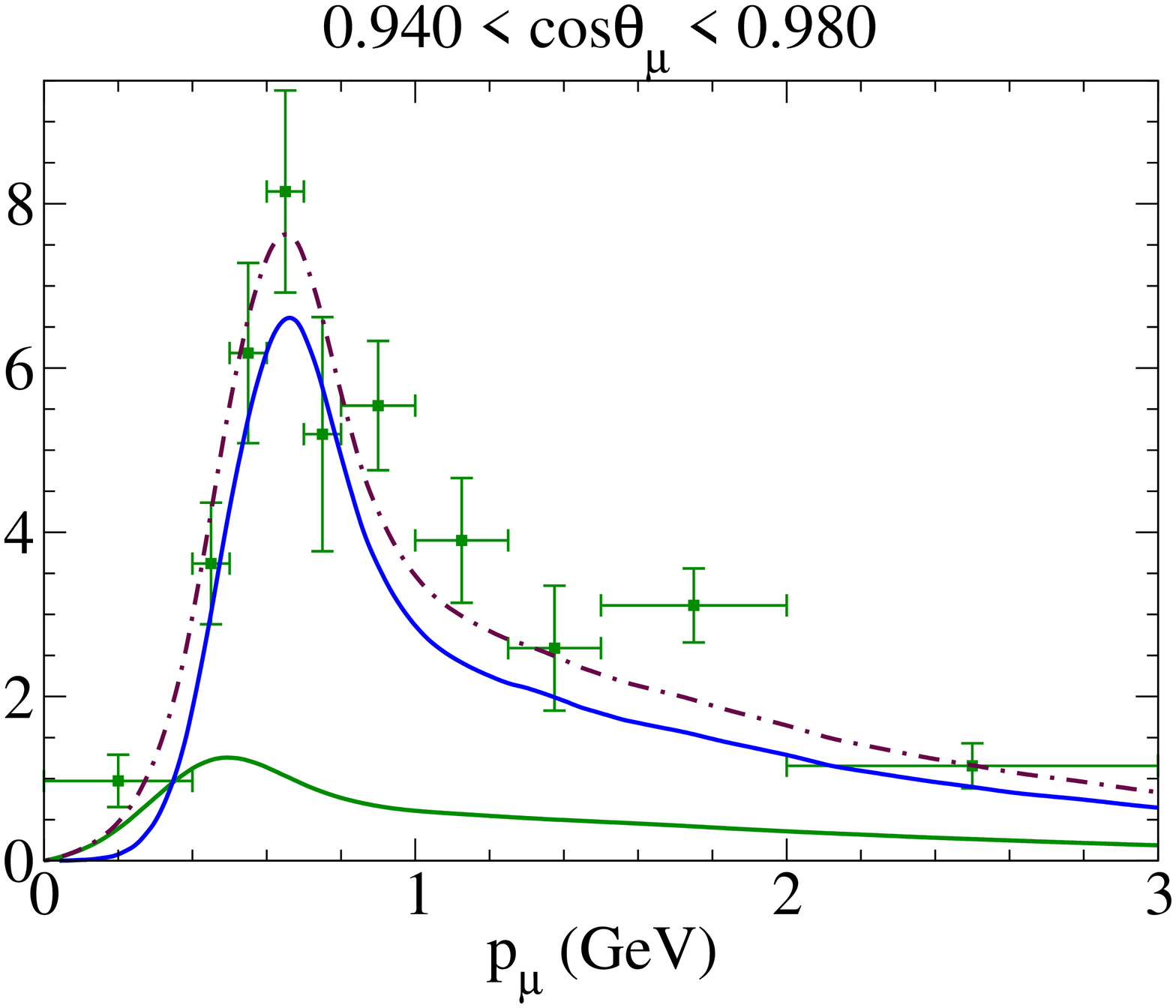}\hspace*{-0.64cm}%
		\includegraphics[scale=0.167, angle=270]{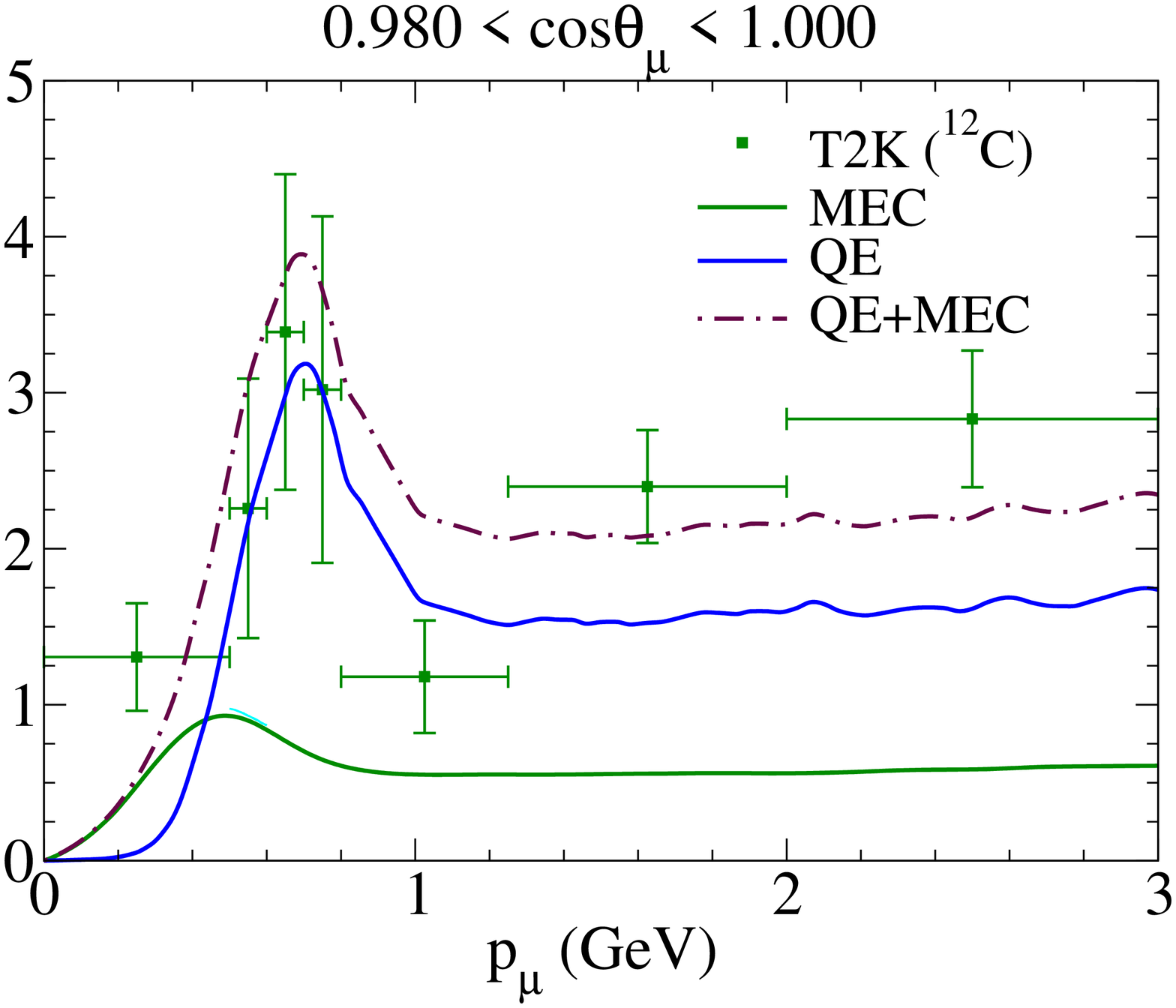}\\
		\hspace{-0.84cm}\includegraphics[scale=0.167, angle=270]{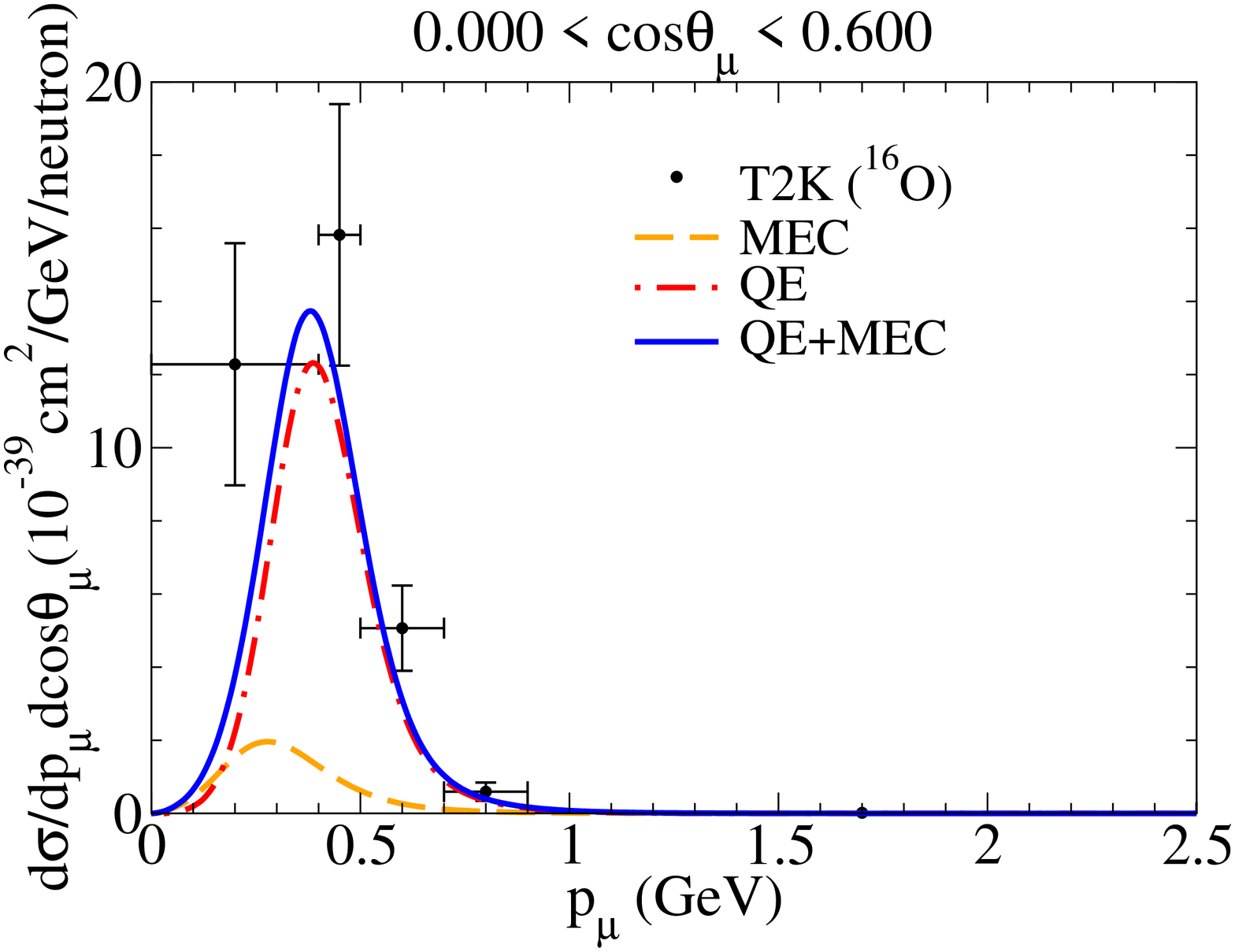}\hspace{-0.64cm}\includegraphics[scale=0.167, angle=270]{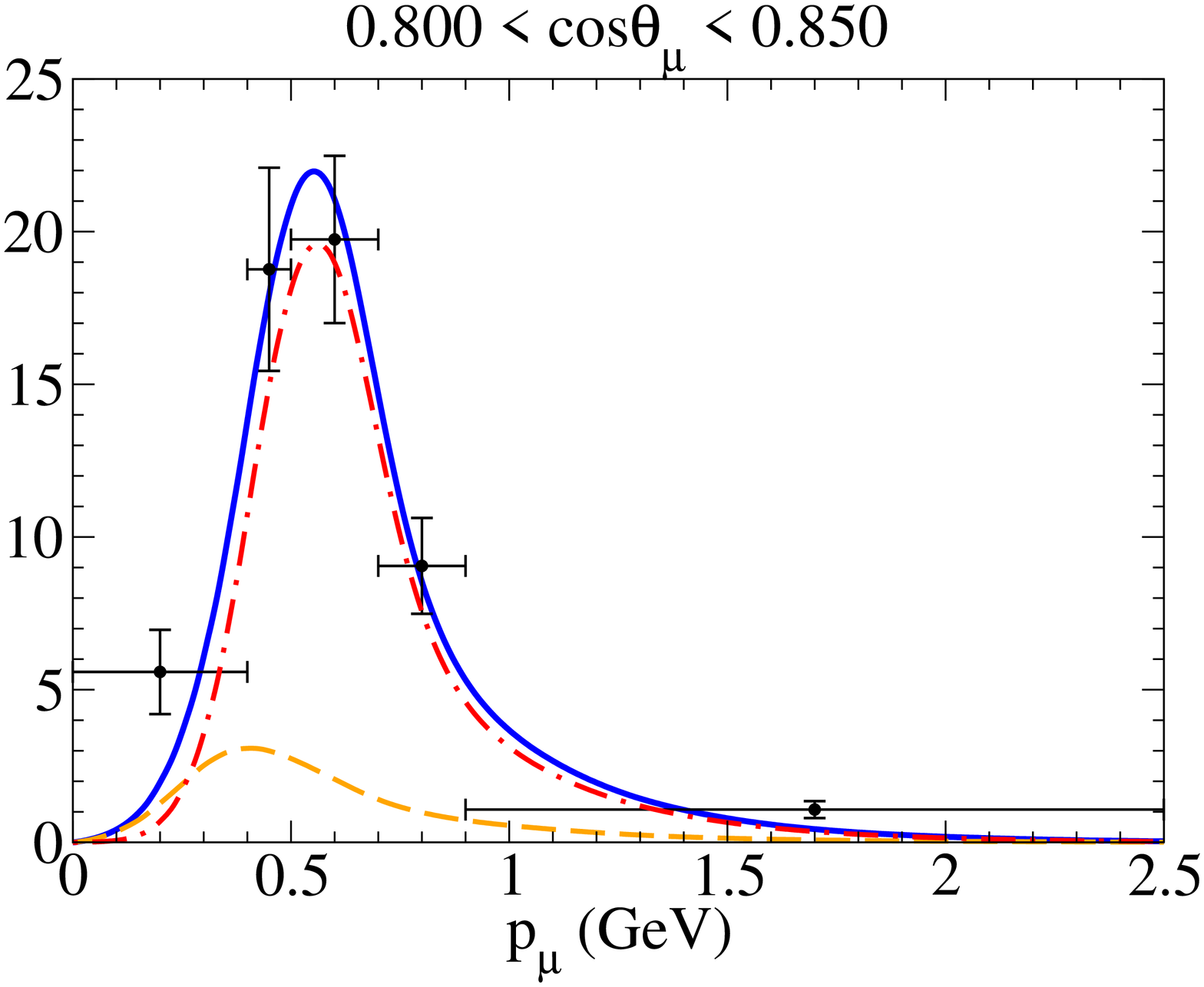}\hspace{-0.64cm}\includegraphics[scale=0.167, angle=270]{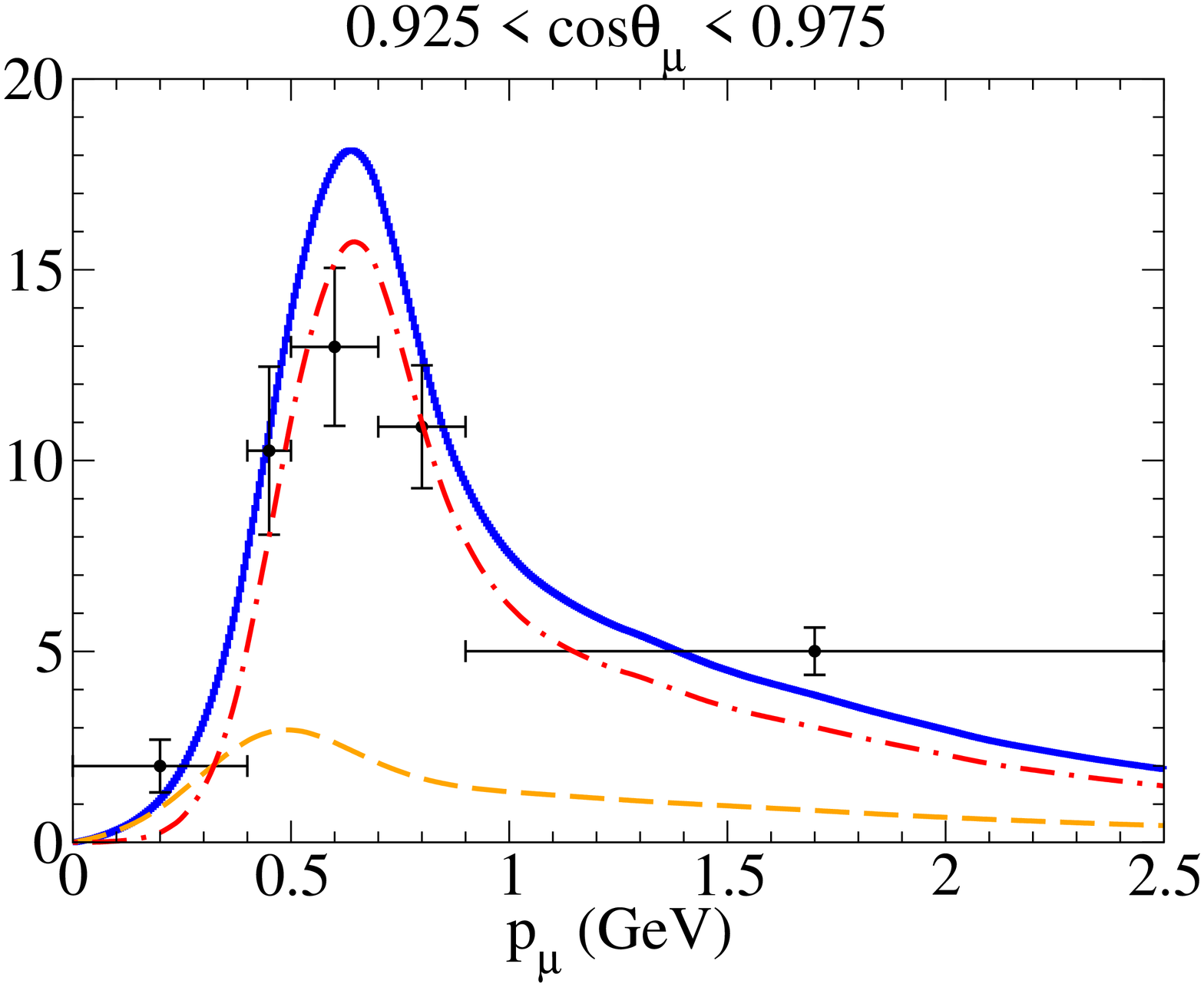}\hspace{-0.64cm}\includegraphics[scale=0.167, angle=270]{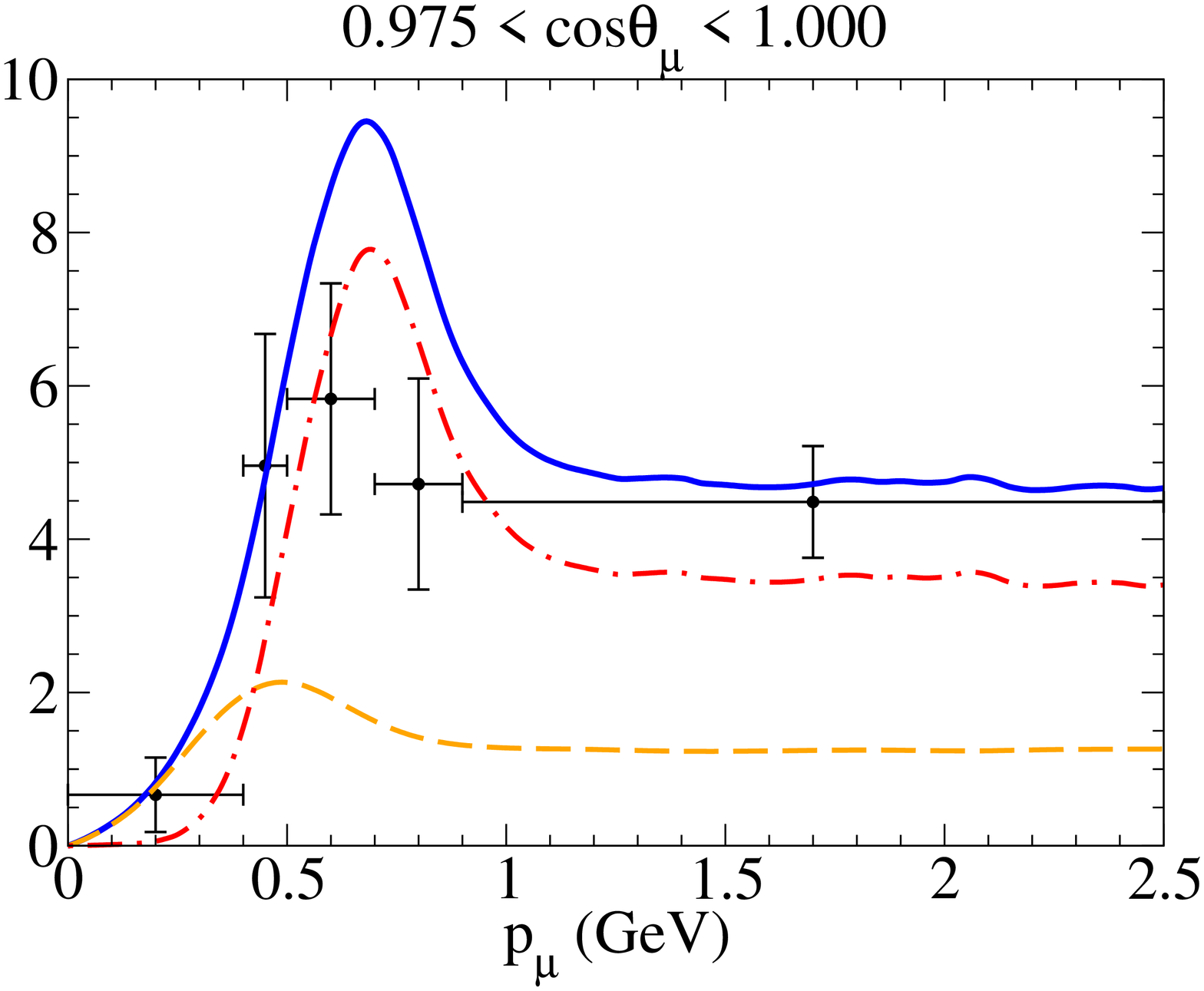}\hspace{-0.4cm}
	\end{center}
	\caption{Comparison of the SuSAv2-MEC model with the T2K flux-integrated CCQE and 2p2h
          double-differential cross section for neutrino for
          scattering on $^{12}$C (top panels) and $^{16}$O (bottom panels) in units of 10$^{-39}$ cm$^2$/GeV per nucleon (for carbon) and per nucleon target (for oxygen).
          The CC$0\pi$ T2K data are from Ref.~\cite{Abe:2016tmq} and~\cite{T2Kwater}.
	\label{fig:T2K_d2snew}}
\end{figure}
In Fig.~\ref{fig:T2K_d2snew} we show the comparison of the SuSAv2-MEC model with T2K CC0$\pi$ double differential cross sections on $^{12}$C (top panels) and $^{16}$O (bottom panels). These  data are compared with the QE and 2p2h MEC contributions showing an overall good agreement. Monte Carlo event generators also include other contributions that can mimic a CCQE-like event in the data analysis such as pion-absorption processes in the nucleus. Nevertheless, as it will be discussed later, these contributions are not particularly relevant at T2K kinematics.
It is also important to note that the QE and 2p2h contributions in the SuSAv2-MEC model can be easily extrapolated from one nucleus to another by means of scaling rules, which are different for each nuclear regime. This is based on the assumption of 2nd-kind scaling, already mentioned in Sect.~\ref{sec:scal}, and that has been proven in the SuSAv2-MEC approach for the QE and 2p2h channels~\cite{SuSAv2JLab,2p2hscaling}. It is worth commenting on the case of the most forward-angle region, associated with low kinematics, {\it i.e.,} low energy and momentum transferred to the nucleus.
\begin{figure}[htbp]
	\begin{center}
		\hspace*{-0.26cm}\includegraphics[scale=0.16, angle=270]{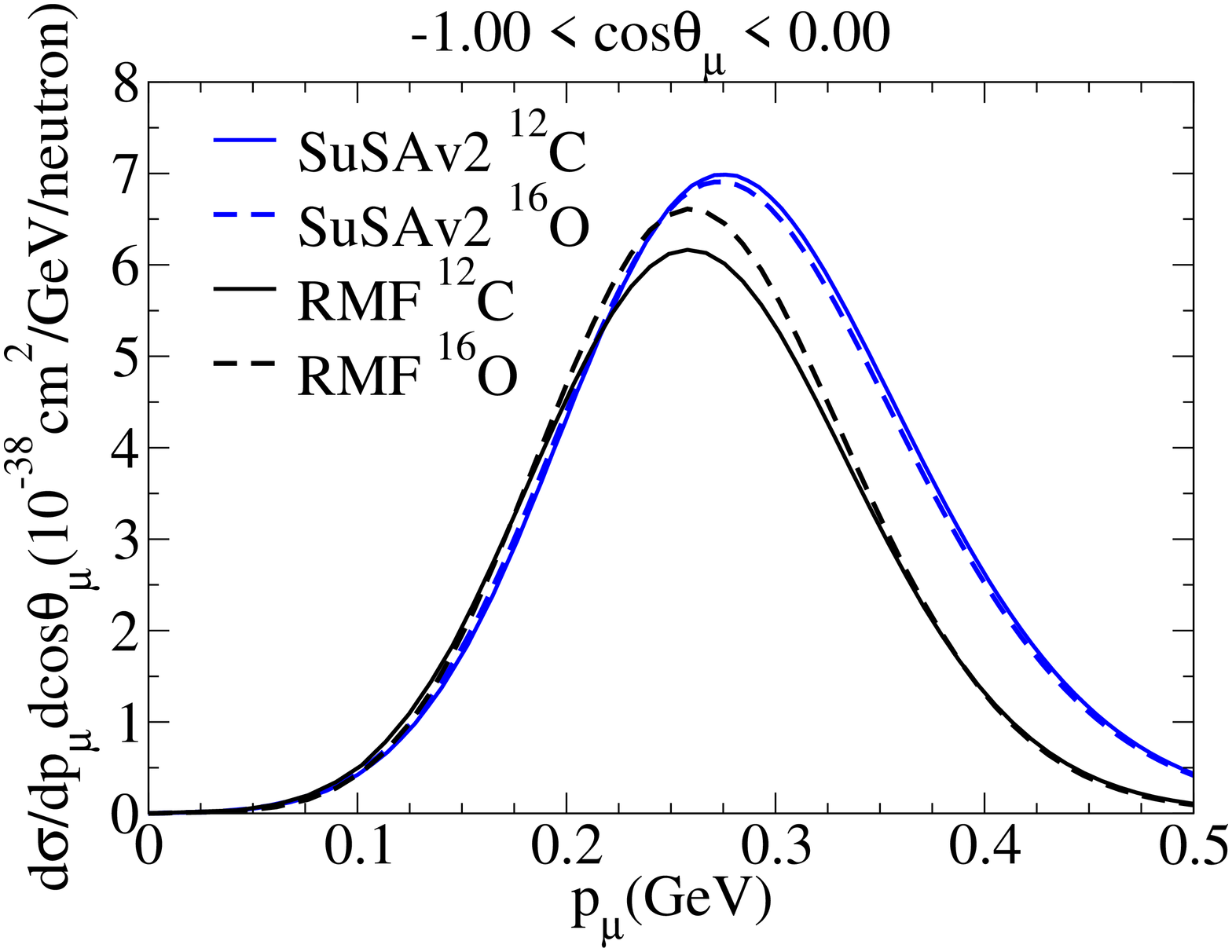}\hspace*{-0.6cm}\includegraphics[scale=0.16, angle=270]{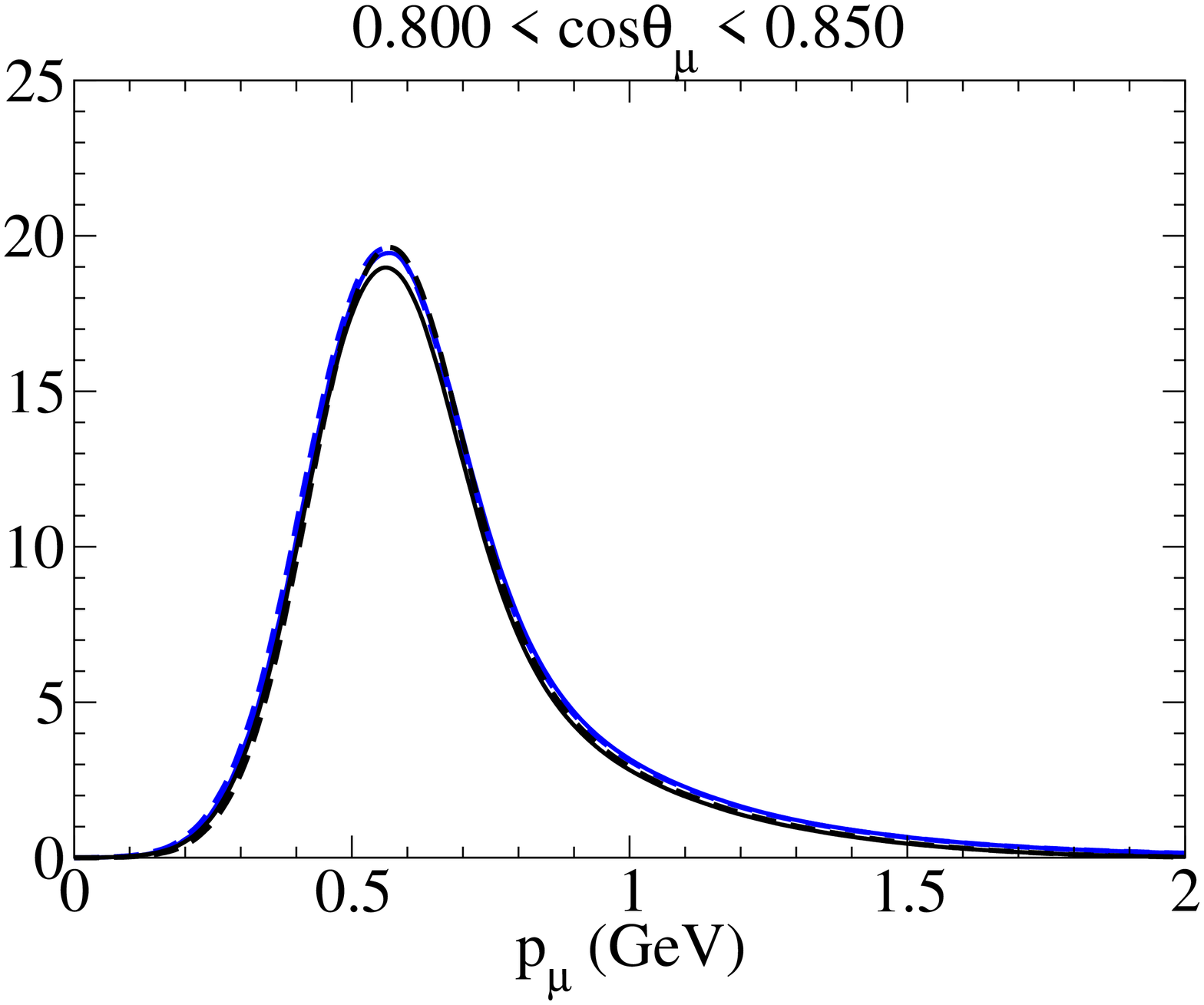}\hspace*{-0.6cm}\includegraphics[scale=0.16, angle=270]{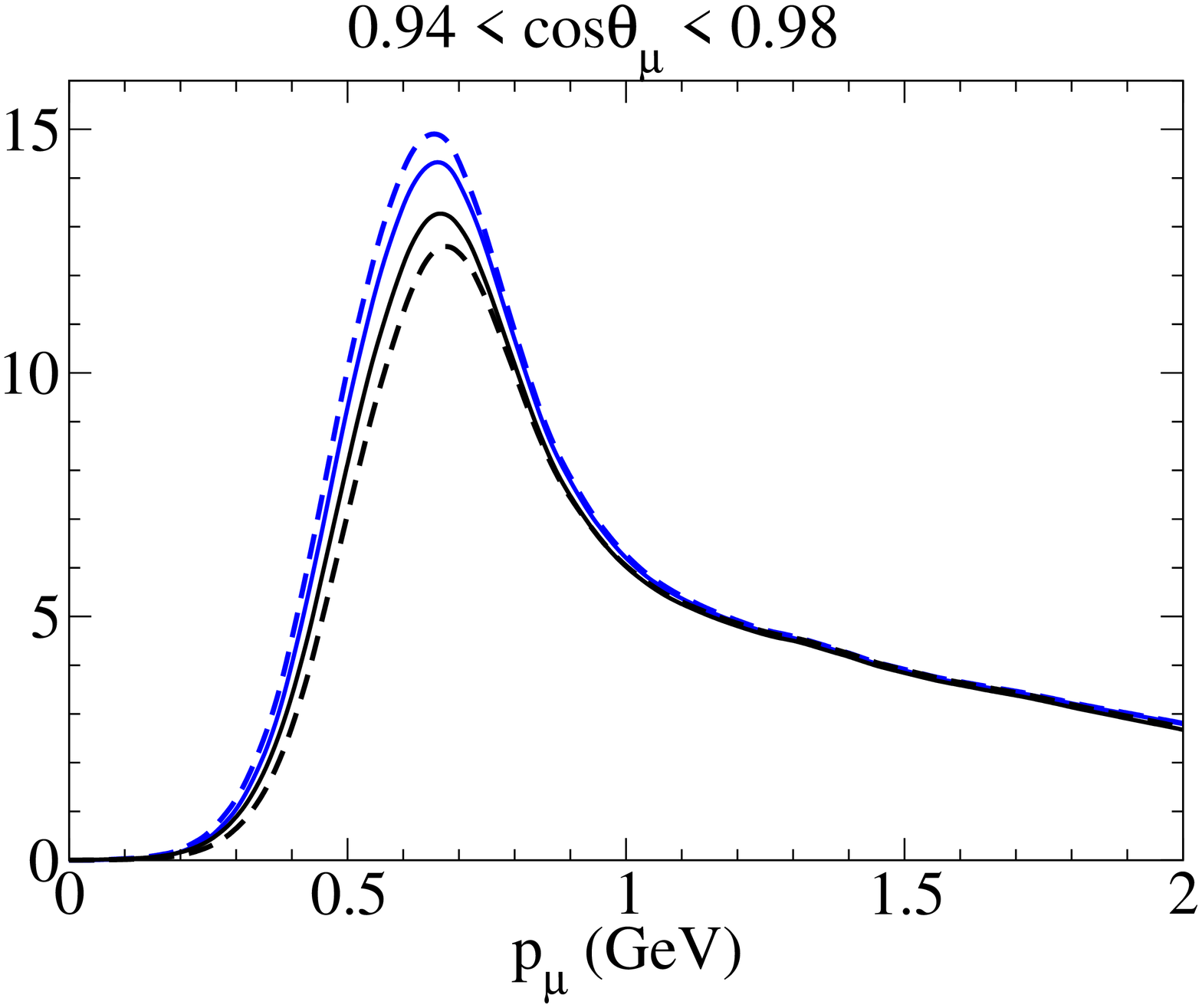}		\hspace*{-0.7cm}\includegraphics[scale=0.16, angle=270]{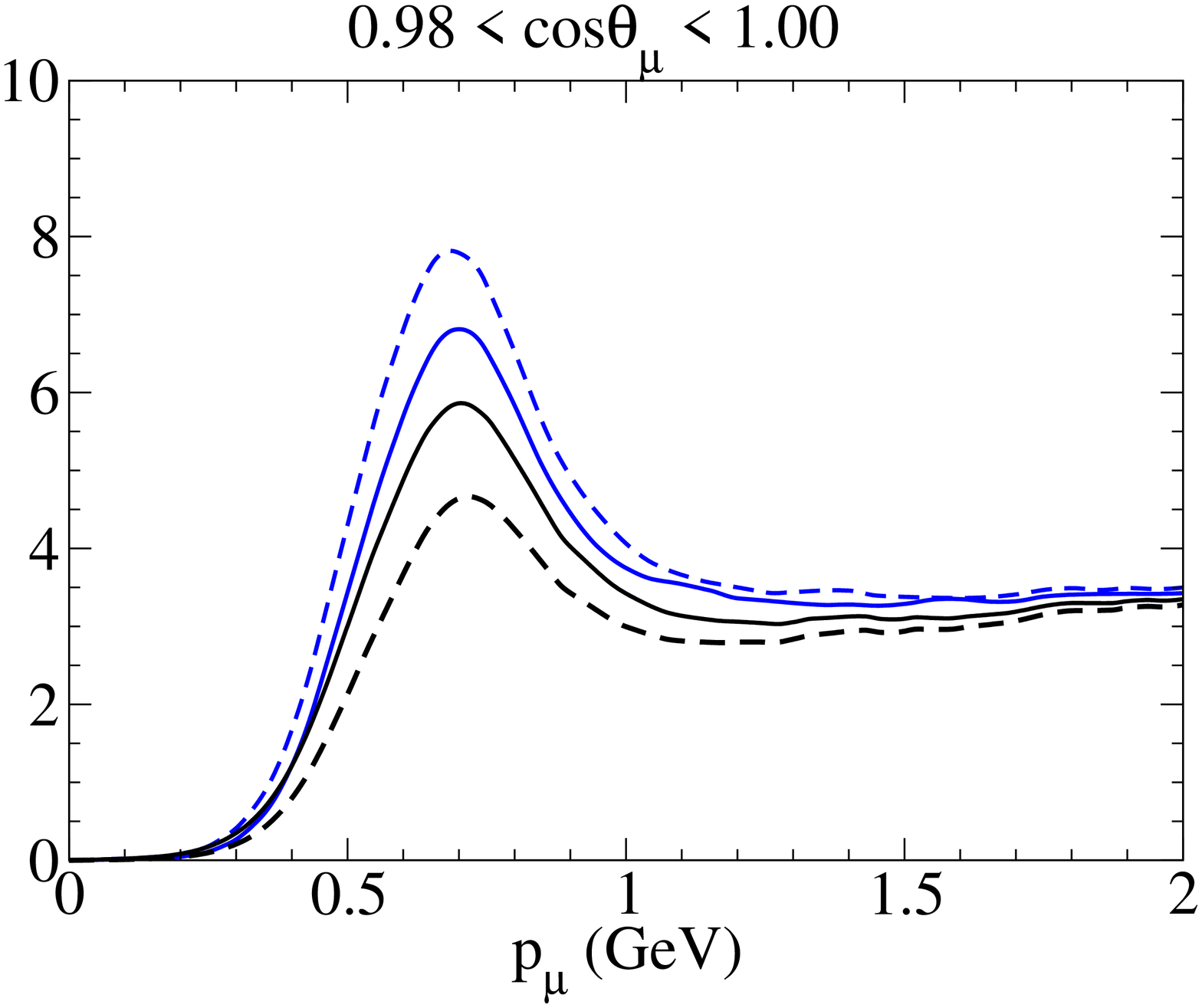}
		\begin{center}
		\end{center}
	\end{center}\vspace*{-0.79cm}
	\caption{Comparison of double differential cross sections on
          $^{12}$C (solid lines) and $^{16}$O (dashed lines) at T2K
          kinematics within the SuSAv2 (1p1h) and RMF models. Results are
          displayed from forward to backward angles.}\label{rmfv2}
\end{figure}
  Within this region, the SuSAv2-MEC model slightly overestimates the T2K data for C and O, being more noticeable in the latter case. This is mainly due to RMF scaling violations, related to low-energy nuclear effects and to different binding energies for each nucleus, which are not properly accounted for in the SuSAv2 approach. This limitation of the model has been solved in the recent ED-RMF approach~\cite{Gonzalez-Jimenez_edRMF,Gonzalez-Jimenez_edRMF2} where the goodness of the RMF strong vector and scalar potentials at low-intermediate kinematics are present while also retaining the benefits of the SuSAv2 model description at larger kinematics, as described in Sect.~\ref{sec:scal}. The differences introduced by this more accurate description of  low-energy nuclear effects in the RMF (and ED-RMF) model can be observed in Fig.~\ref{rmfv2} where large discrepancies between oxygen and carbon predictions are observed between the RMF model and the SuSAv2-MEC one at very forward angles, being smaller as we move to more backward kinematics. At the same time, these differences are more prominent in the case of oxygen. 
This is connected with the assumptions made in the SuSAv2 approach which is based on the RMF analysis on $^{12}$C but on the application of scaling rules to describe other nuclear targets instead of relying on particular RMF predictions for each nucleus. Therefore, at kinematics where scaling violations are present, these differences, while noticeable in the nucleus of reference ($^{12}$C), can be even more important for other nuclear targets. On the contrary, SuSAv2 assumes that scaling works well even at low kinematics, thus implying minor differences between $^{12}$C and $^{16}$O, mainly due to the scaling rule and differences in the energy shift introduced for each nucleus. This analysis is of relevance for T2K~\cite{T2KCO} where the accuracy of the C to O extrapolation plays an important role on the oscillation analysis and whose predictions have shown some discrepancies between C and O for very forward-going muons which may be explained by the RMF predictions.

\begin{figure}[ht]\vspace{-0.128cm}
		\hspace*{-0.295cm}\includegraphics[scale=0.214, angle=270]{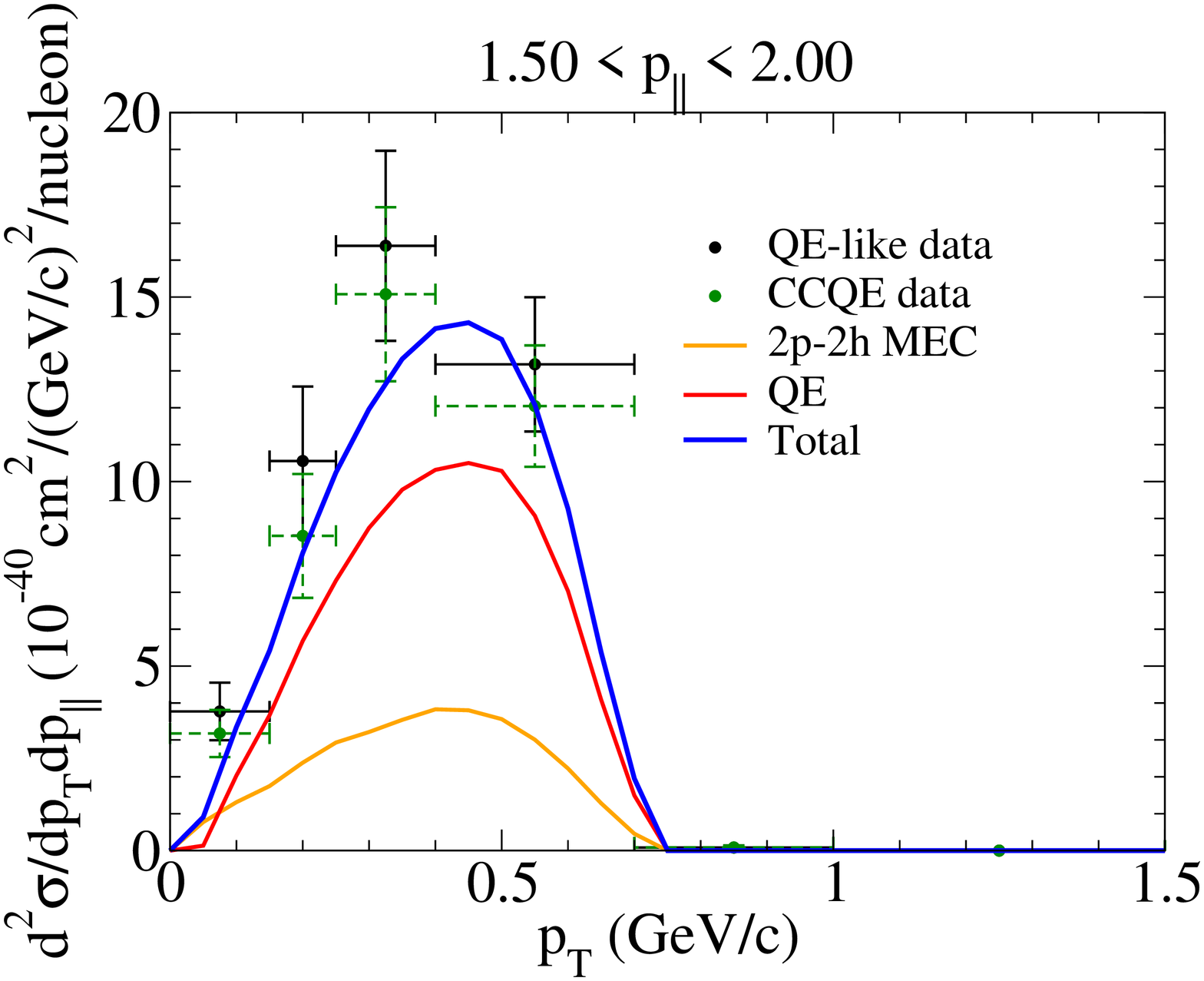}\hspace*{-0.584cm}%
		\hspace*{-0.295cm}\includegraphics[scale=0.214, angle=270]{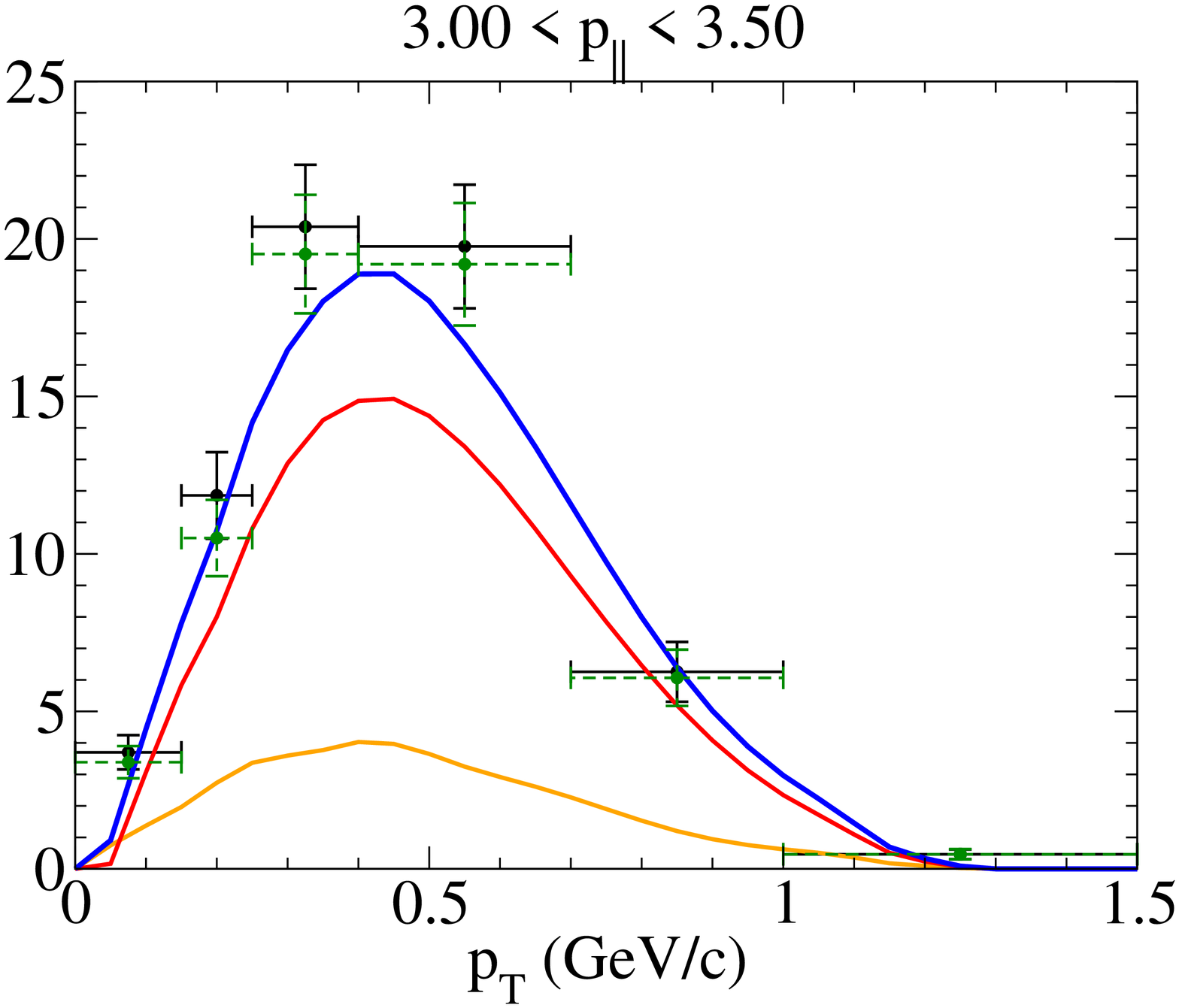}\hspace*{-0.584cm}%
		\hspace*{-0.295cm}\includegraphics[scale=0.214, angle=270]{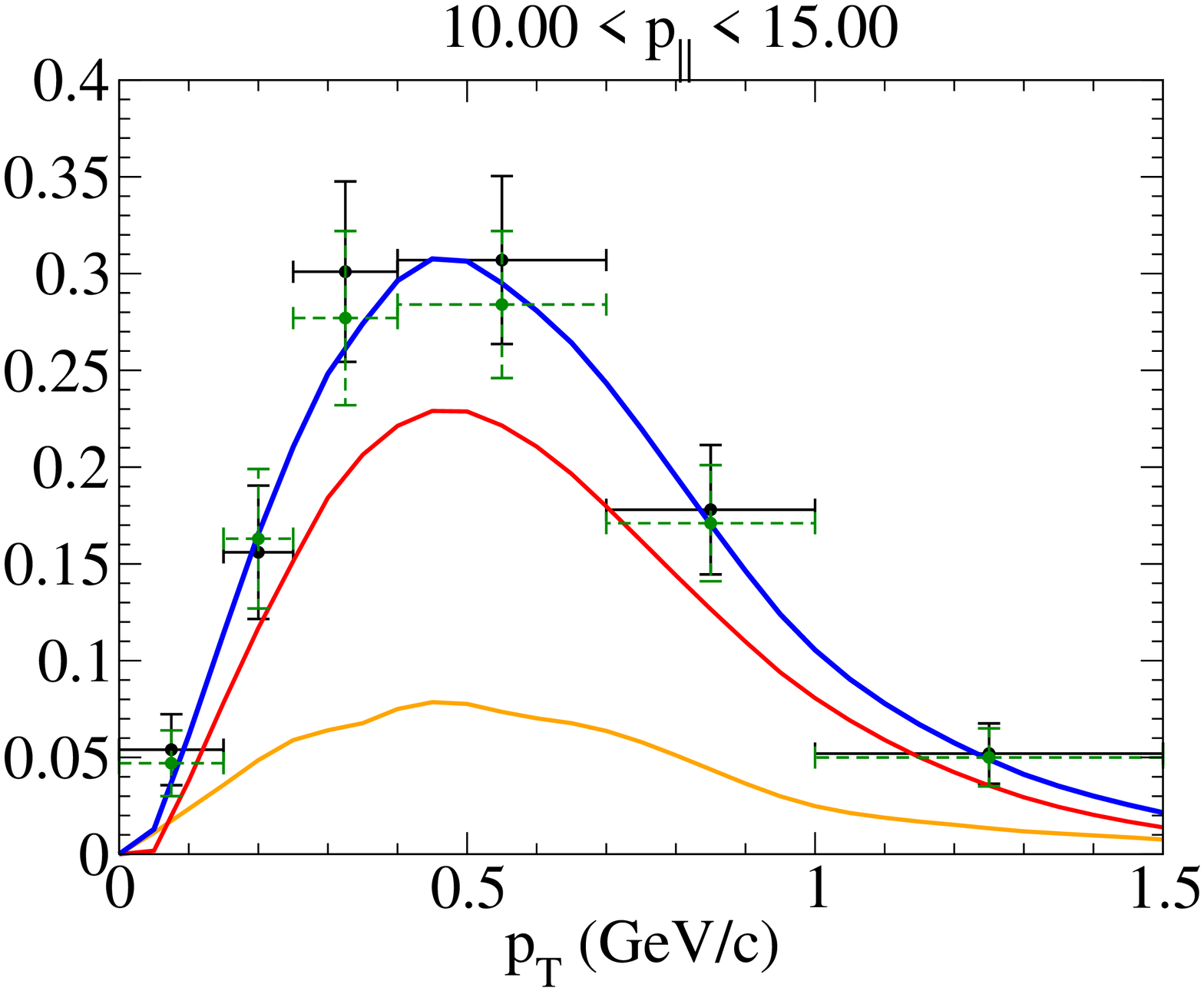}\hspace*{-0.584cm}\\
			\hspace*{-0.295cm}\includegraphics[scale=0.214, angle=270]{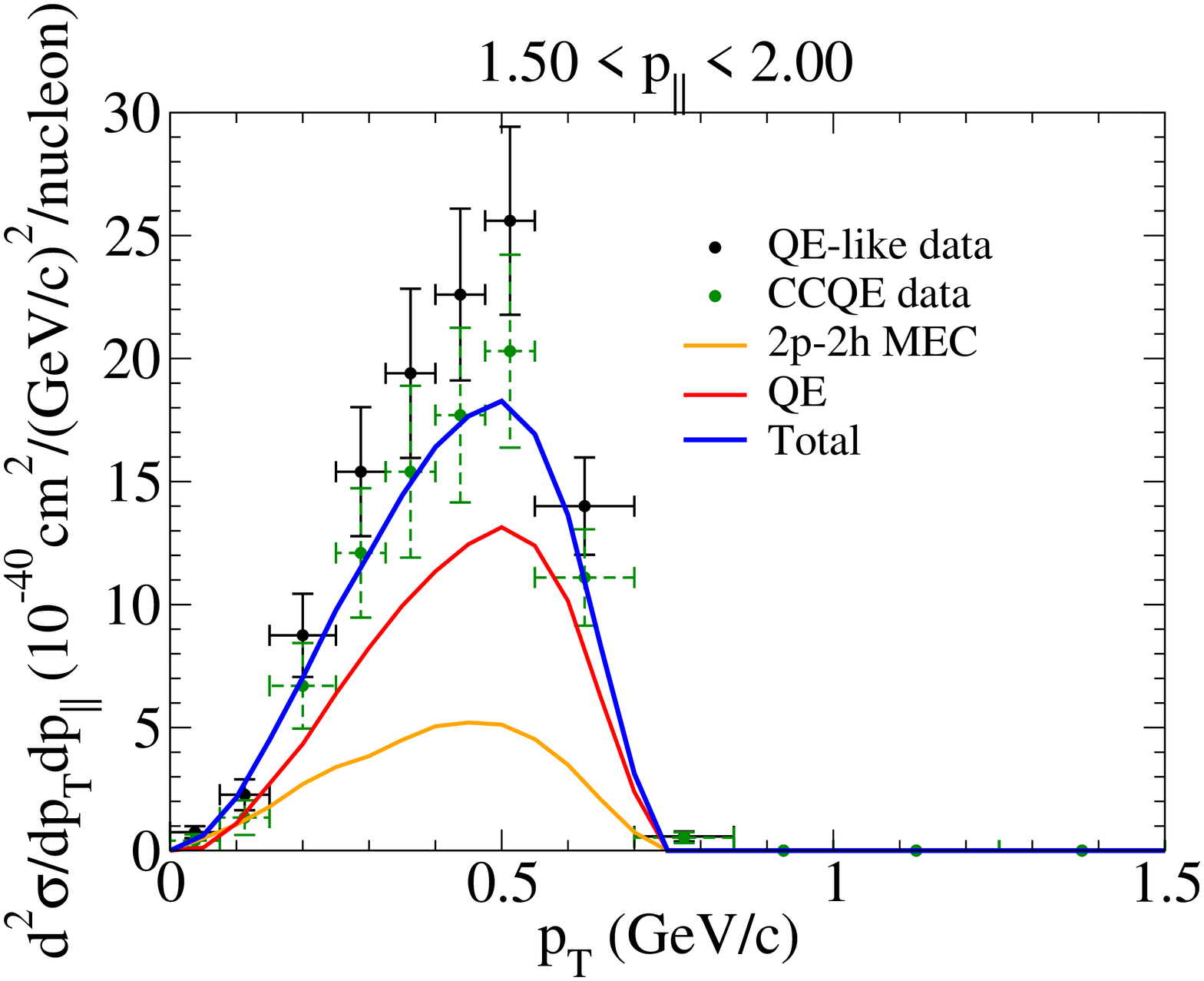}\hspace*{-0.584cm}%
		\hspace*{-0.295cm}\includegraphics[scale=0.214, angle=270]{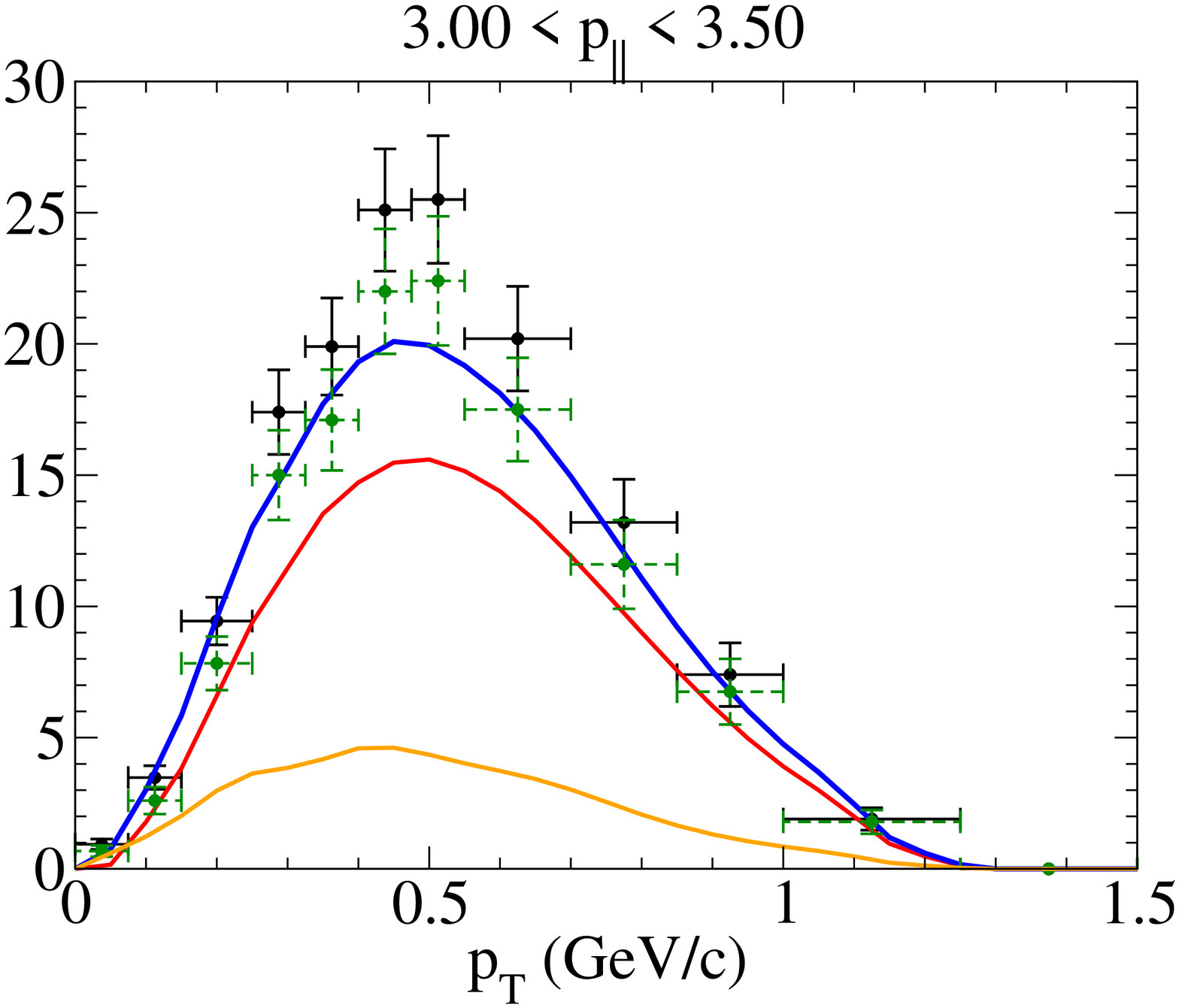}\hspace*{-0.584cm}%
		\hspace*{-0.295cm}\includegraphics[scale=0.214, angle=270]{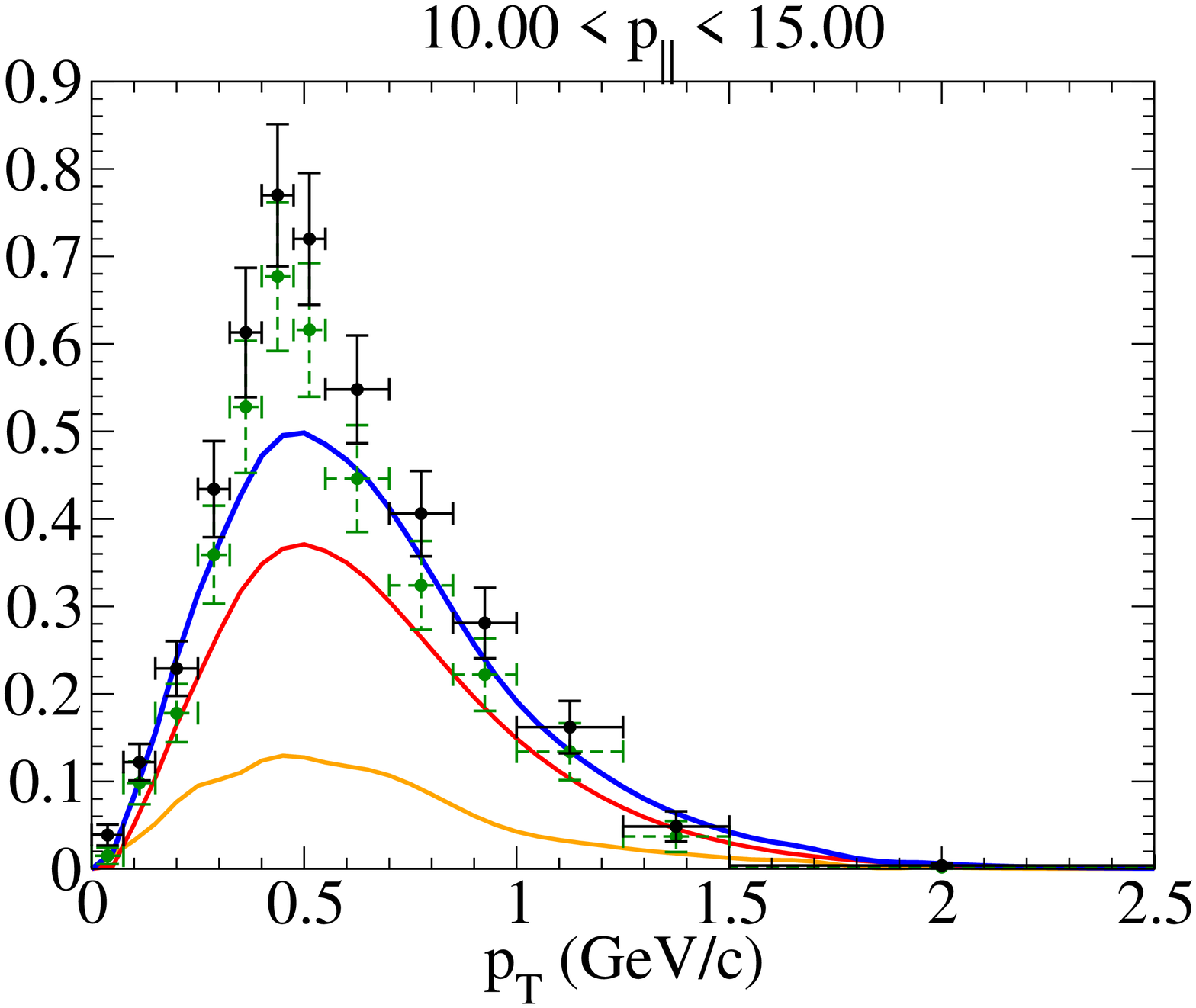}	
	\caption{(Color online) The MINERvA ``QE-like" and ``CCQE"
          double differential cross sections for $\bar\nu_\mu$ (top
          panels) and $\nu_\mu$ (bottom panels) scattering on
          hydrocarbon versus the muon transverse momentum, in bins of
          the muon longitudinal momentum (in GeV/c). The curves
          represent the prediction of the SuSAv2+2p2h-MEC (blue) as
          well as the separate quasielastic (red) and 2p2h-MEC
          (orange) channels.  MINERvA data and experimental
          fluxes are from Refs.~\cite{Patrick:2018gvi}
          and~\cite{PhysRevD.99.012004}. 
         }
	\label{fig:fig1mnv}
\end{figure}
The SuSAv2-MEC model is also compared in Fig.~\ref{fig:fig1mnv} with the MINERvA CCQE-like double differential (anti)neutrino measurements on hydrocarbon (CH) in terms of the transverse momentum of the outgoing muon (with respect to the antineutrino beam), in bins of the muon longitudinal momentum. Due to its relativistic nature, the SuSAv2-MEC model is well suited to describe these data~\cite{Megias:2018ujz,Patrick:2018gvi} where the mean neutrino energy is around 3.5 GeV. An overall good agreement is reached without resorting to any tuning or additional parameters. According to Refs.~\cite{Megias:2018ujz,Patrick:2018gvi}, the MINERvA ``QE-like'' cross sections entail, besides pure quasielastic contributions, events that have post-FSI final states without mesons, prompt photons above nuclear de-excitation energies, heavy baryons, or protons above a kinetic energy threshold of 120 MeV, thus including zero-meson final states arising from resonant pion production followed
by pion absorption in the nucleus and from multi-nucleon interactions. 
Apart from the ``QE-like'' points, the MINERvA ``CCQE'' signal is also shown, and corresponds to events initially generated in the GENIE event generator~\cite{Andreopoulos:2009rq} as quasielastic (that is, no resonant or deep inelastic scatters, but including scatters from nucleons in correlated pairs with zero-meson final states), regardless of the final-state particles produced, thus
including CCQE and 2p2h reactions.  
The difference between the two data sets is mainly related to pion production plus re-absorption and goes from $\sim15\%$ to $\sim5\%$ depending on the kinematics. The present SuSAv2-MEC results do not include processes
corresponding to these pion re-absorption processes inside the
nucleus and the comparison should be done with the
``CCQE'' data rather than with the ``QE-like'' ones. 
 A more detailed analysis of these results together with a $\chi^2$ test can be found in~\cite{Megias:2018ujz}, where the SuSAv2 $\chi^2$ shows its compatibility with data and with the MINERvA/GENIE predictions.
It is also important to note that due to MINERvA's acceptance, the muon scattering angle is limited to $\theta_\mu<$ 20$^{\circ}$ as well as the muon kinematics (1.5 GeV $< p_{||} < $ 15 GeV, $p_T<$ 1.5 GeV) in both experimental and theoretical results. This implies important phase-space restrictions for large energy and momentum transfer to the nuclear target and makes the available phase space not so different from the T2K one, as shown in~\cite{megias2019axial}.

%
\begin{figure}[htbp]
	\begin{center}
		\includegraphics[width=0.349\linewidth, angle=0]{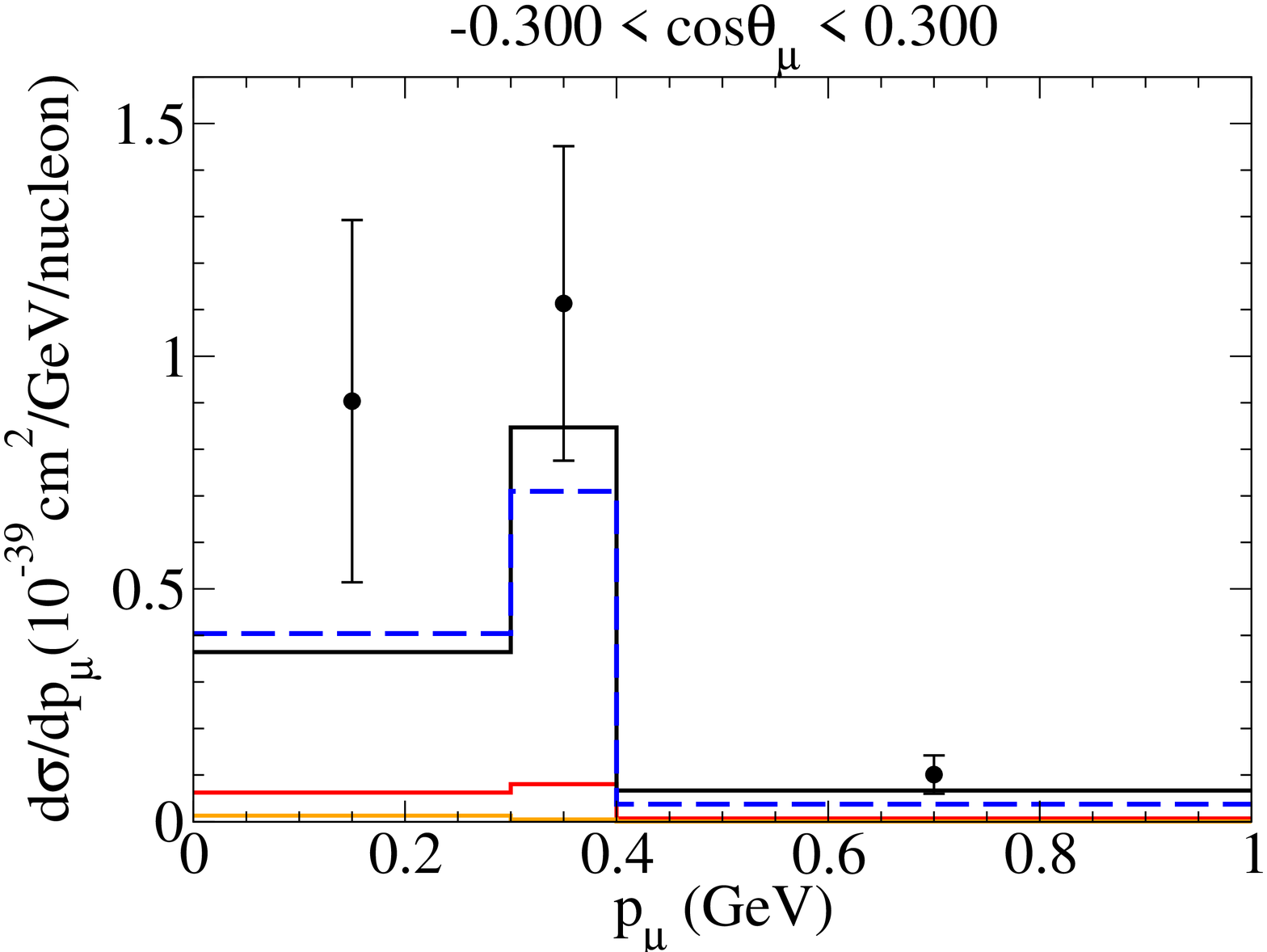}\hspace*{-0.295cm}
		\includegraphics[width=0.349\linewidth, angle=0]{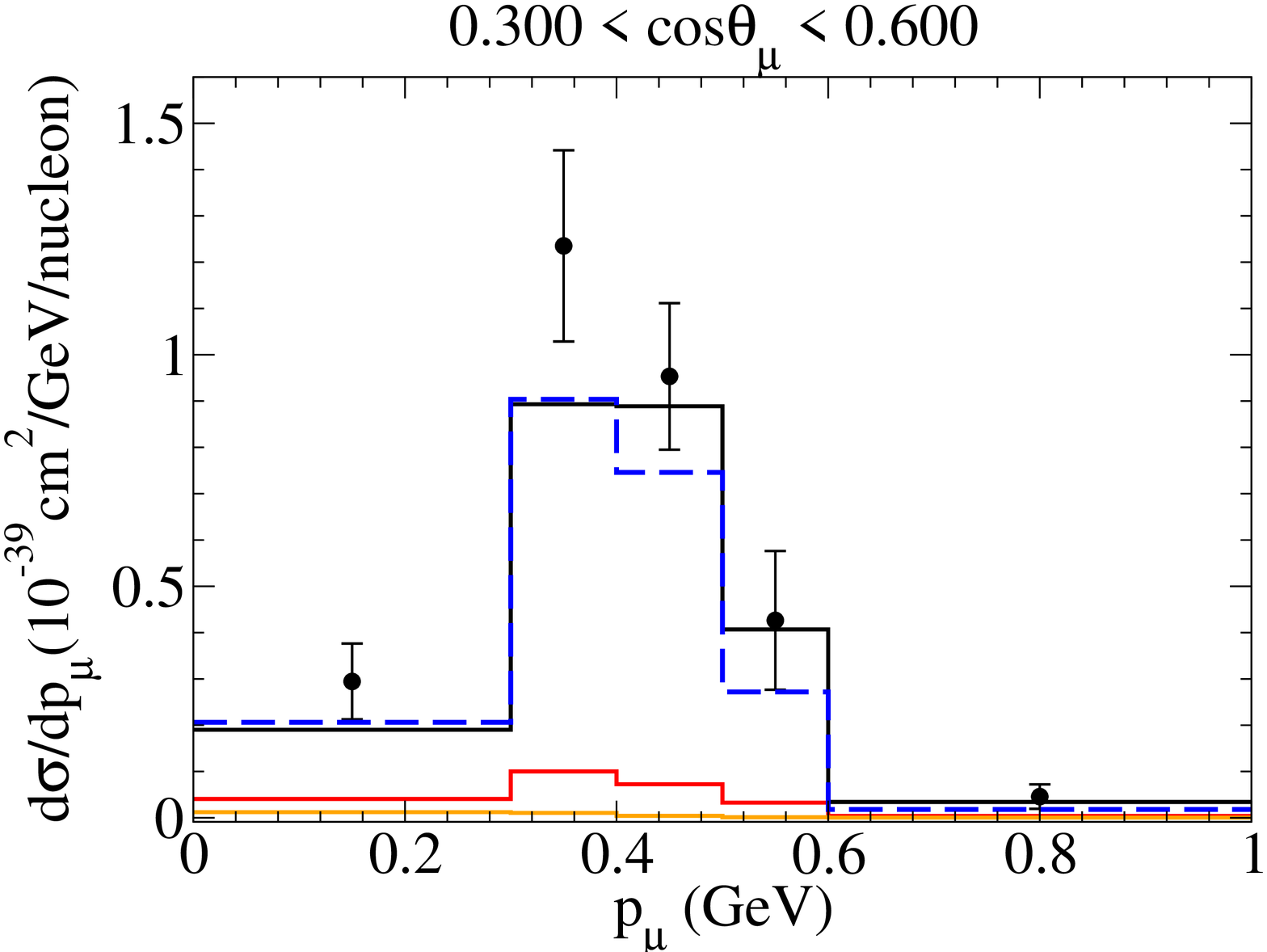}\hspace*{-0.295cm}
		\includegraphics[width=0.349\linewidth, angle=0]{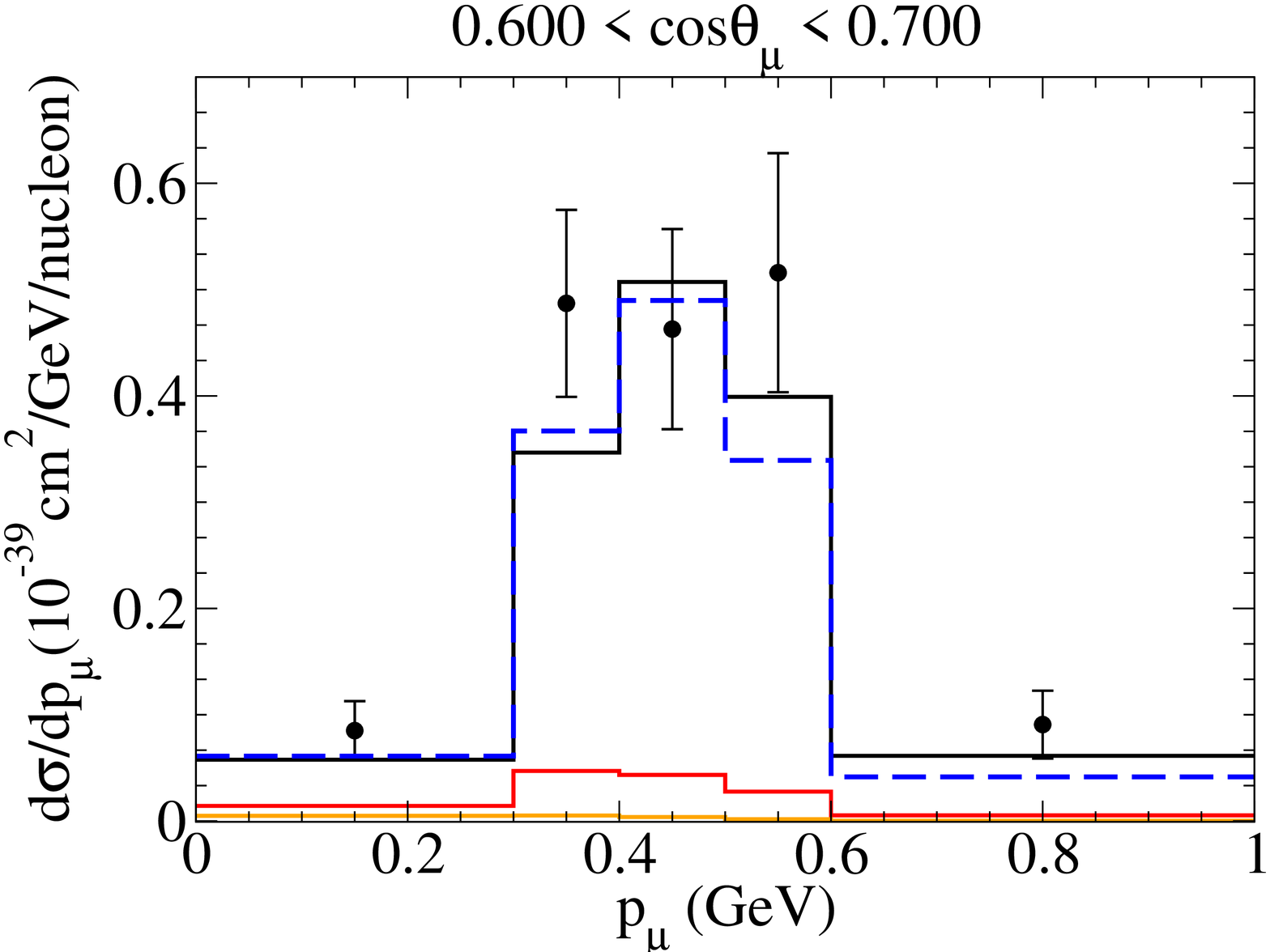}
		\includegraphics[width=0.349\linewidth, angle=0]{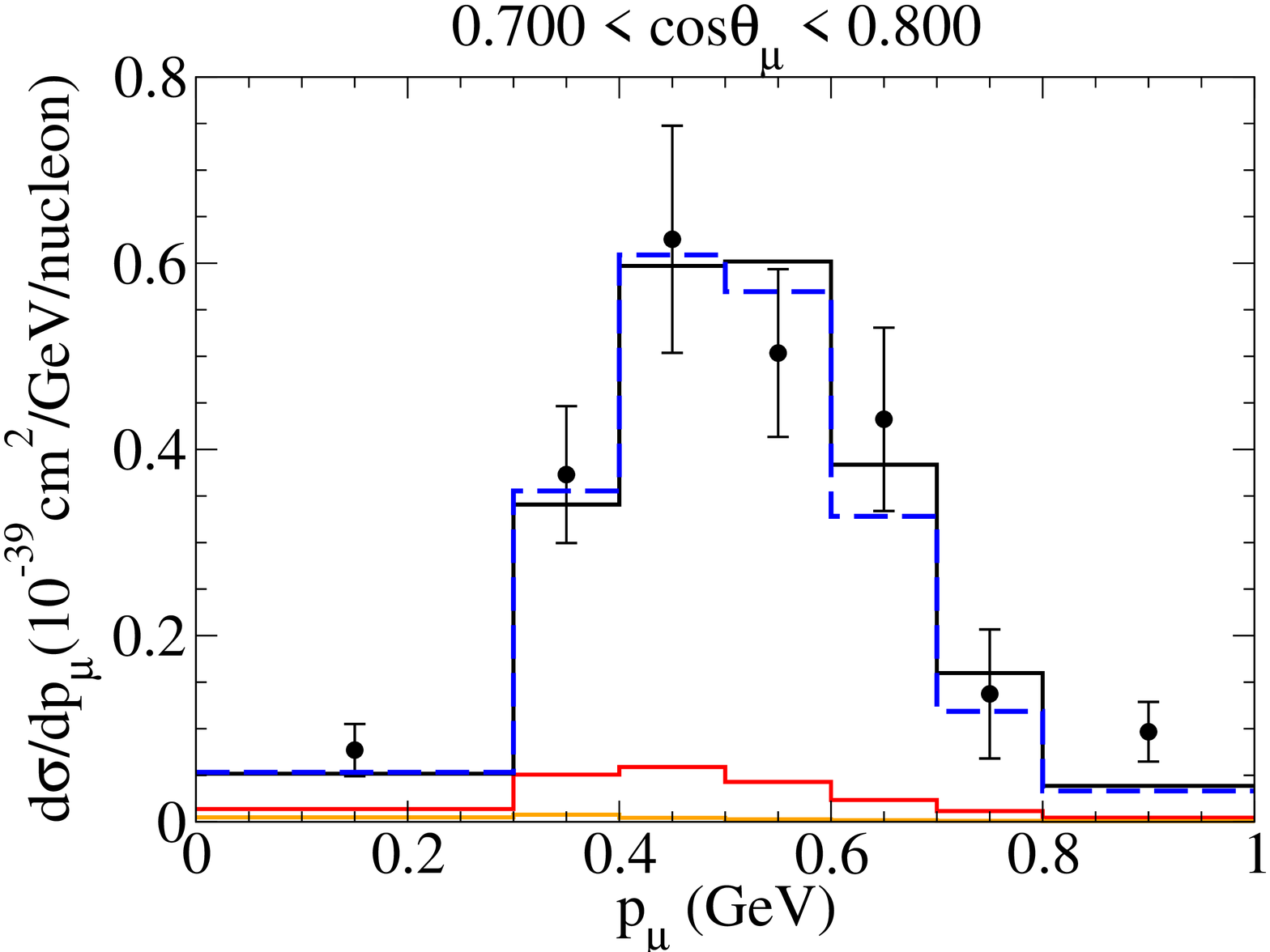}\hspace*{-0.295cm}
		\includegraphics[width=0.349\linewidth, angle=0]{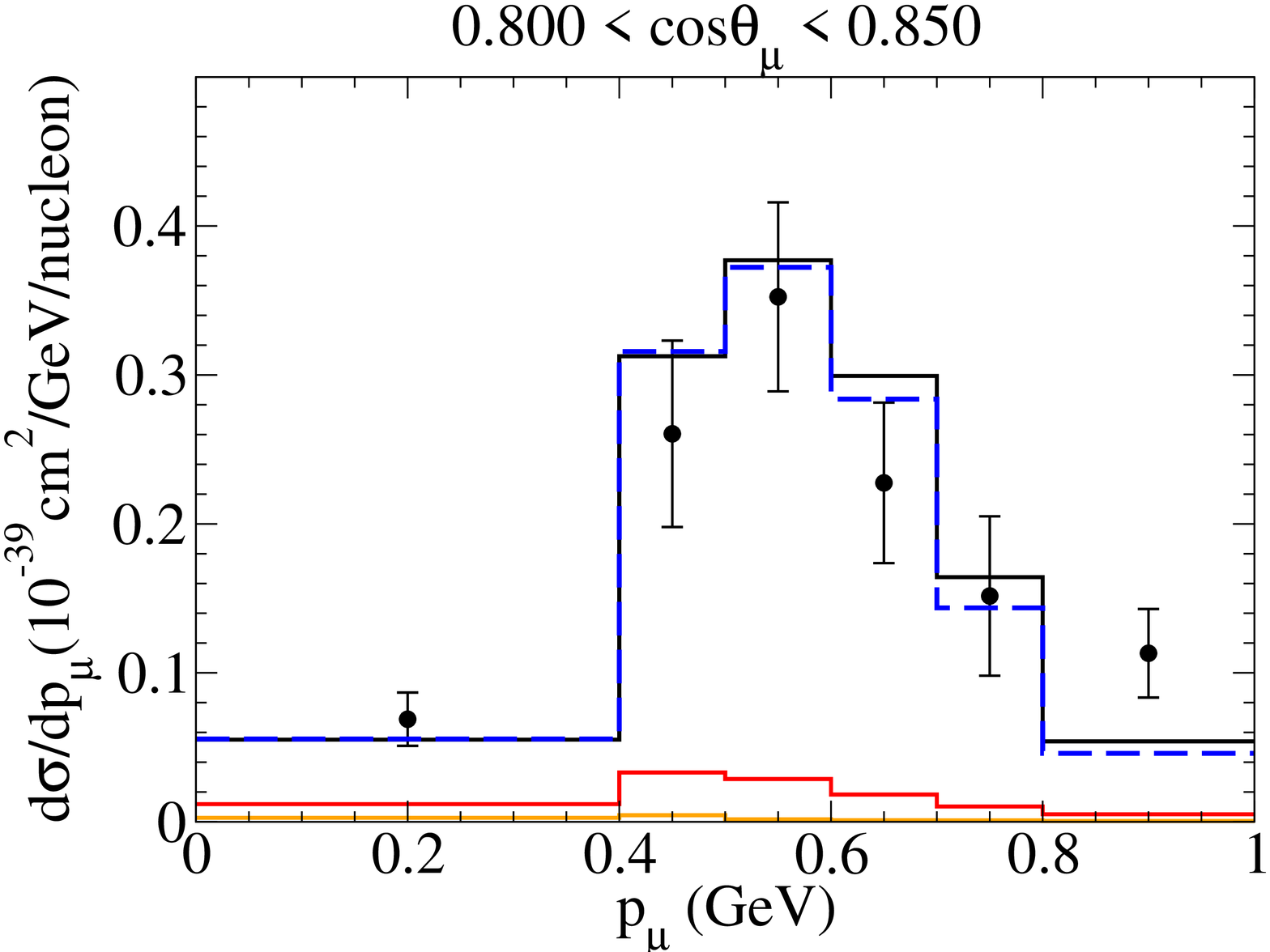}\hspace*{-0.295cm}
		\includegraphics[width=0.349\linewidth, angle=0]{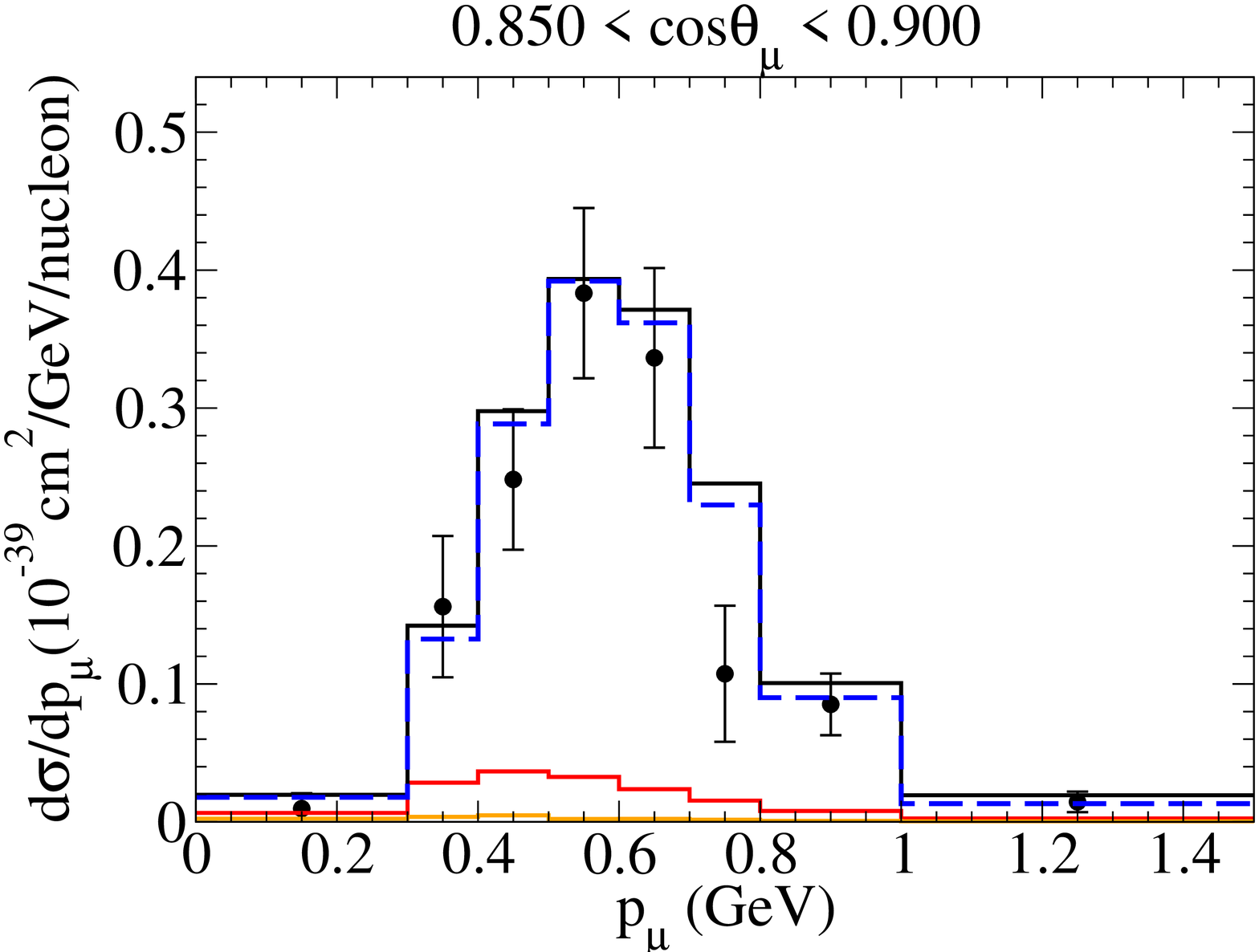}
		\includegraphics[width=0.349\linewidth, angle=0]{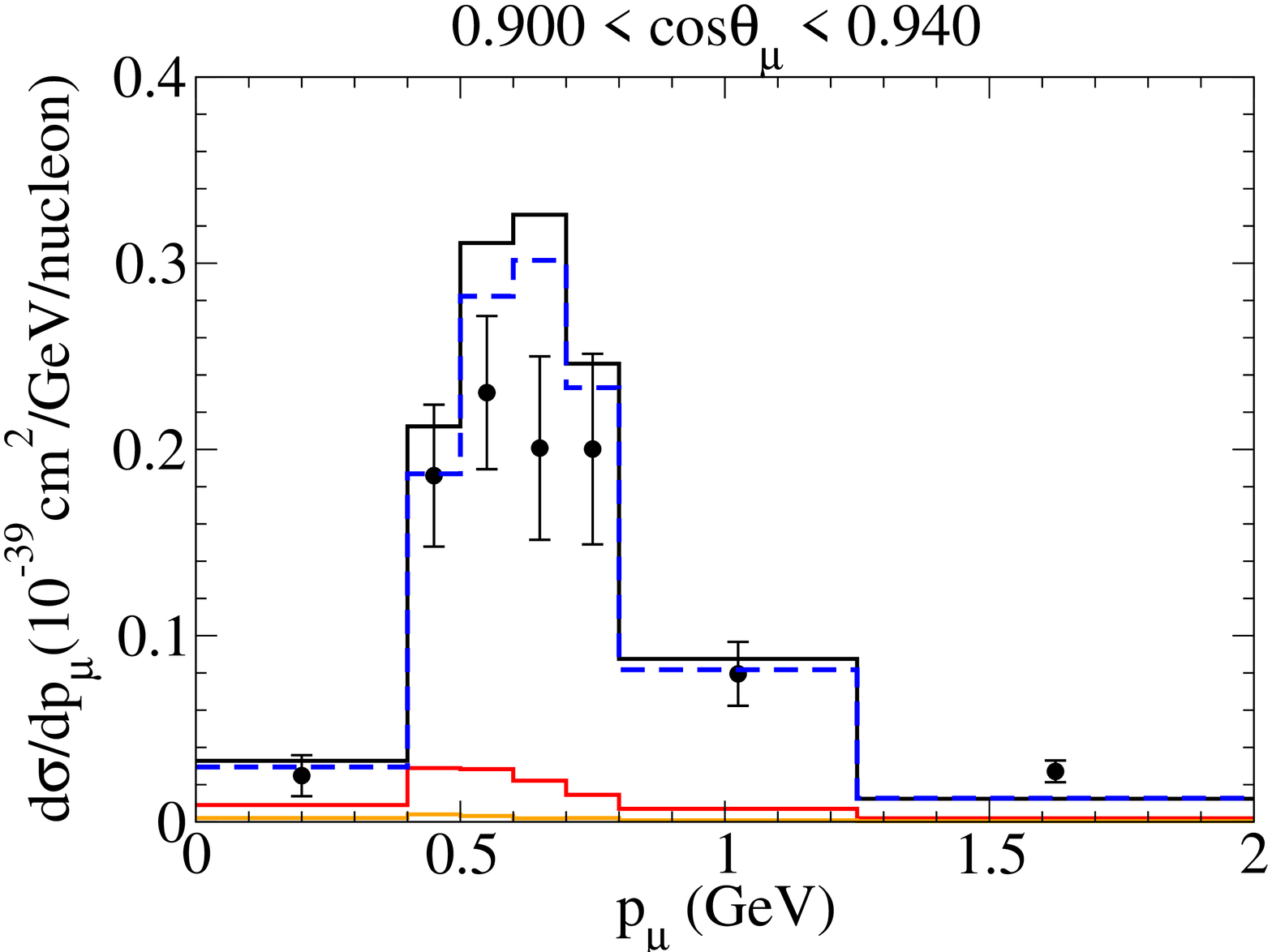}\hspace*{-0.295cm}
    	\includegraphics[width=0.349\linewidth, angle=0]{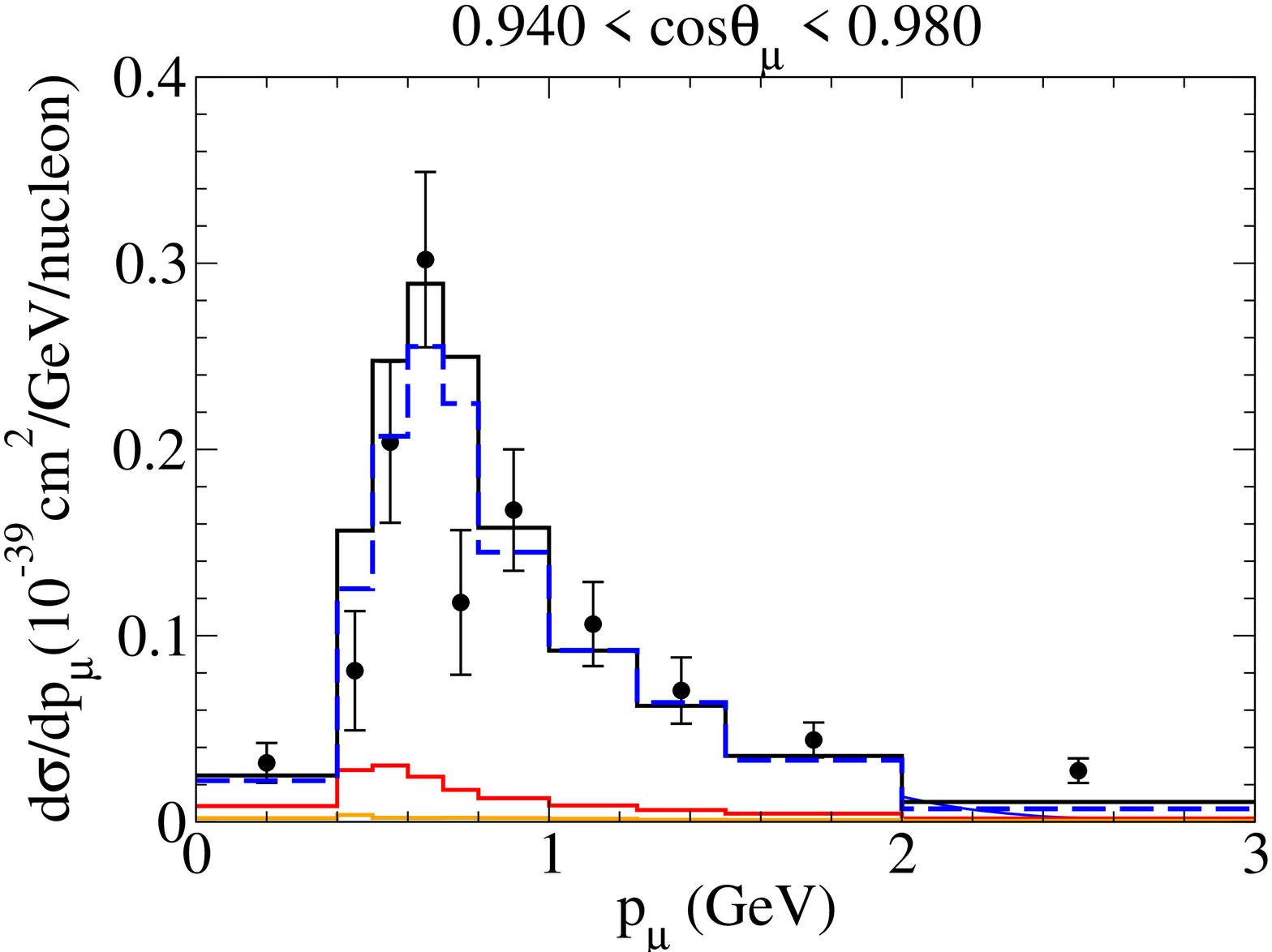}\hspace*{-0.295cm}
    	\includegraphics[width=0.349\linewidth, angle=0]{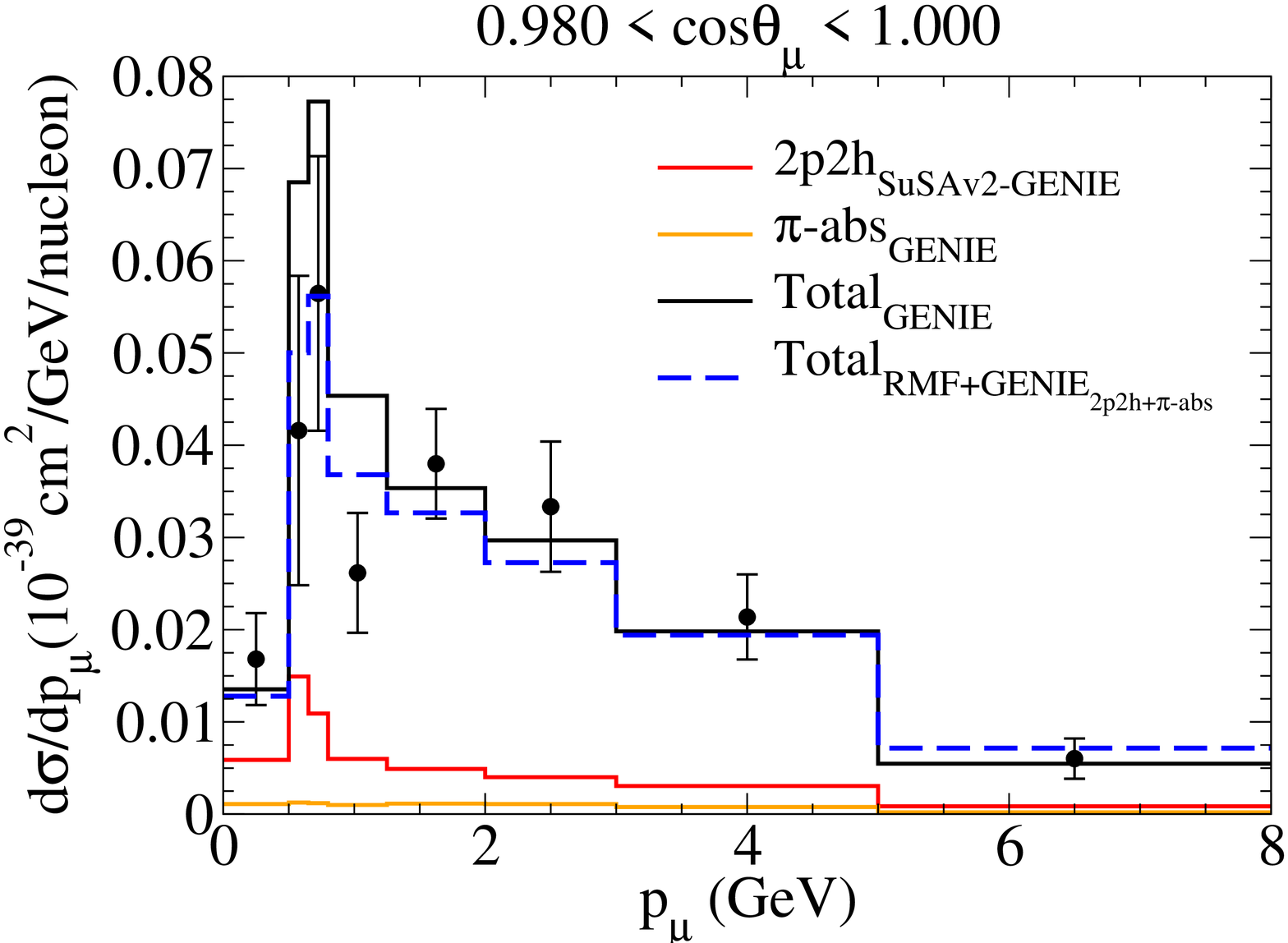}
	\end{center}
	\caption{Comparisons of data and model predictions for differential CC0$\pi$ muon-neutrino cross sections on $^{12}$C in the T2K neutrino beam as a function of the muon kinematics when there are no protons with momenta above 500 MeV. Two 1p1h predictions are shown (one from RMF, the other from SuSAv2 implemented in GENIE), in addition to the SuSAv2 2p2h and pion absorption contributions from GENIE. The total contributions when using each of the two 1p1h models is also shown. Goodness of fit are calculated to be $\chi^2_{RMF}=171.87$ (59 bins) and $\chi^2_{SuSA}=168.92$ (60 bins), where the latter includes a single extra bin from -1.0 to -0.3 $\cos{\theta}$ (not shown). The data points are taken from~\cite{Abe:2018pwo}.}
	\label{fig:ssincT2KComp}
\end{figure}
The previous T2K and MINERvA results are only related to the final lepton kinematics, the so-called inclusive measurements. At this point, it is worth mentioning that other models~\cite{Nieves:2011pp,Nieves:2011yp,Martini:2010ex,Martini12PRD,Gallmeister16,Meucci,Meucci15,Rocco16,Rocco:2018mwt,Golan12,Lovato:2015qka,Pandey:2016ee,Butkevich:2017mnc,Ivanov:2013saa,PhysRevD.99.093001,Martini:2011wp,Nieves:2013nubar,Martini:2013sha,Mosel:2014lja,Meucci:2014bva,MartiniCP,Mosel,Dolan:2018sbb} have been also developed to address these CC inclusive neutrino interactions and, although similar agreement with data can be obtained, they are based on different assumptions about the nuclear properties and dynamics. Combined analyses of these models with more exclusive neutrino measurements where hadron kinematics and other nuclear effects can be analyzed in more detail would help to improve model selection for data analysis (see also the discussions in Sect.~\ref{sec:semi_electrons}). The advantage of  SuSAv2-MEC (and RMF) is that of being a fully relativistic model that has shown an overall good agreement with electron and neutrino scattering data and that can be extended without further approximations to the full-energy range of interest for present and future neutrino experiments. 
For this reason, the SuSAv2-MEC model (1p1h and 2p2h) has been recently implemented in the GENIE neutrino event generator~\cite{Andreopoulos:2009rq} with the aim of improving the characterisation of the nuclear effects in neutrino cross section measurements and work is now in progress to implement this model together with the ED-RMF one in the NEUT event generator~\cite{Hayato:2009zz} for its application on the T2K oscillation analysis.

Based on these works, in Fig.~\ref{fig:ssincT2KComp} we study the T2K CC0$\pi$ measurement of interactions with protons less than 500 MeV~\cite{Abe:2018pwo} in comparison with the SuSAv2-1p1h, SuSAv2-2p2h and pion-absorption predictions from GENIE (see~\cite{Dol20} for details). At proton momentum below 500 MeV/c a clear dominance of the SuSAv2-1p1h channel is observed. Note that SuSAv2-GENIE's also shows an overestimation of data at very forward angles, as observed in Fig.~\ref{fig:T2K_d2snew} which is mainly due to low energy transfer scaling violations which are absent in the SuSAv2-model but present in the RMF theory. In order to analyze the low-energy nuclear effects and these scaling violations, we also show the comparison with the RMF model for the 1p1h channel together with the SuSAv2-2p2h and pion-absorption results from GENIE, noticing a remarkable improvement in the data comparison at very forward angles. 
Here, one of the drawbacks of the SuSAv2 model is that it only predicts lepton kinematics so its implementation in MC event generators has to rely on a factorization approach and on the information available in these generators to determine hadron kinematics (see~\cite{Dol20} for details). This limitation will be addressed in the forthcoming implementation of the ED-RMF model in event generators which will provide full information about lepton and hadron kinematics in the final state together with a consistent description of the nuclear dynamics.

Due to aforementioned dominance of the 1p1h channel on CC0$\pi$ interactions with low momentum protons observed in Fig.~\ref{fig:ssincT2KComp}, it remains difficult to draw clear conclusions about the goodness of the 2p2h description. 
\begin{figure}[htbp]
\begin{center}
\includegraphics[width=0.49\linewidth]{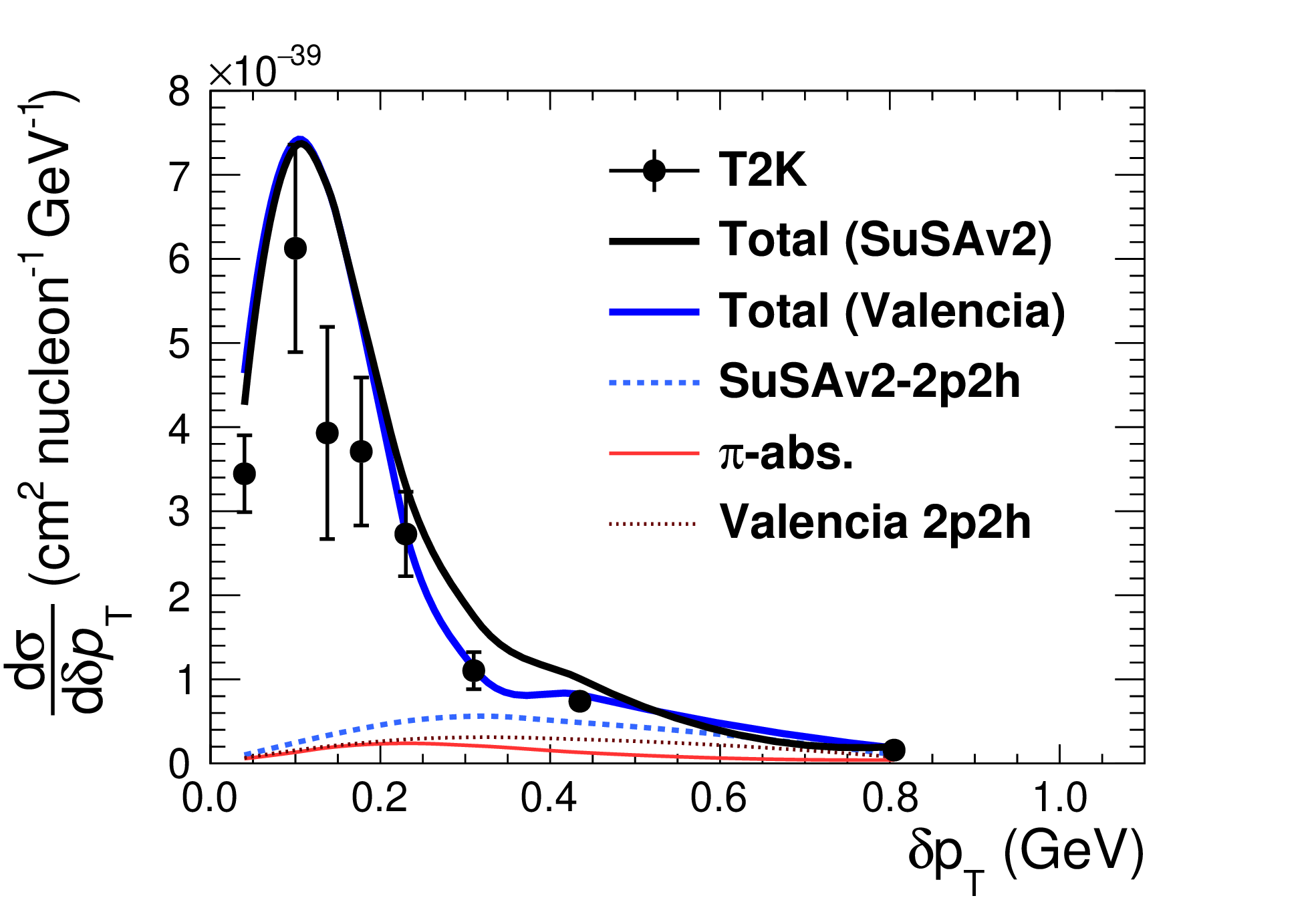}\hspace*{-0.15cm}
\includegraphics[width=0.509\linewidth]{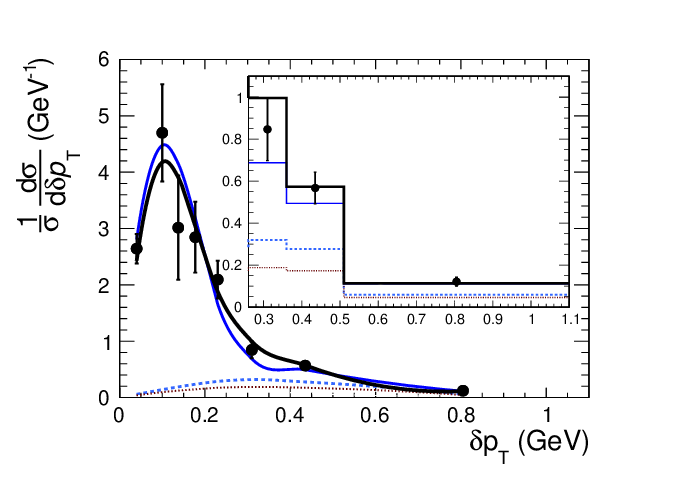}
\end{center}
\caption{The regularised T2K measurement of CC0$\pi$ muon-neutrino cross sections on $^{12}$C at T2K kinematics as a function of the Single Transverse Variable~\cite{Lu:2015tcr} $\delta p_T$ compared to predictions from the GENIE-implemented SuSAv2 and Valencia 1p1h+2p2h models, each of which is added to GENIE's pion absorption prediction. The total contributions when using SuSAv2 and Valencia models is also displayed. A shape only comparison is also shown (right panel). Goodness of fit are calculated as follows. For $\delta p_T$: $\chi^{2}_{SuSA}=20.5$, $\chi^{2}_{Valencia}=27.1$. The data points are taken from~\cite{Abe:2018pwo}.}
\label{fig:stvT2KComp}
\end{figure}
Nevertheless, this can be explored further using more semi-inclusive measurements with measurements of proton and muon kinematics as shown in Fig.~\ref{fig:stvT2KComp}, where the SuSAv2-MEC model in GENIE is compared with the so-called single transverse variables (STV), and in particular with the transverse momentum imbalance, $\delta p_T$, defined in terms of the momentum imbalances between the outgoing muon and highest momentum proton in the plane transverse to the incoming neutrino (see~\cite{Dol20,Lu:2015tcr} for details). These transverse kinematic imbalances allow one to study initial-state nuclear dynamics but also to better isolate the 2p2h channel. In $\delta p_T$ the 1p1h channel is not expected to contribute significantly beyond the initial state nucleon momentum ($\sim$230~MeV/c for carbon in a Fermi gas approach), thus implying that the high-$\delta p_T$ values will be dominated by 2p2h and other contributions, as observed in Fig.~\ref{fig:stvT2KComp}. The overestimation at high-$\delta p_T$ in the left panel may indicate that the 2p2h contribution is too strong. However, as discussed in~\cite{Dolan:2018zye}, this overall over-prediction could potentially be improved by stronger nucleon FSI, which may improve the data agreement in the tail. On the other hand, this may also be explained by the approximations taken to produce semi-inclusive predictions from inclusive models, as described in~\cite{Dol20}. It is expected that these drawbacks can be addressed in further works via the implementation of RMF models for the 1p1h channel and a full semi-inclusive 2p2h model. Apart from these limitations, it is worth noticing the almost perfect description of the shape of $\delta p_T$ in the right panel, improving the agreement reached by other descriptions also implemented in generators. 

To conclude this section, we show the comparison of the SuSAv2-MEC model with measurements that also consider non-QE contributions, mainly pion production. This reaction channel is also of relevance for present and future neutrino oscillation experiments. Neutral current $\pi^0$ production constitutes an important background in the electron neutrino and antineutrino appearance analyses but also neutrino-induced pions emitted can mimic QE-like events if they are not properly detected or if they are absorbed due to FSI effects. Thus, the analysis and detection of these pion production events in coincidence with the final lepton and other hadrons is of paramount importance for the neutrino energy reconstruction.
\begin{figure}[htbp]
\centering  
\includegraphics[width=.345\textwidth,angle=270]{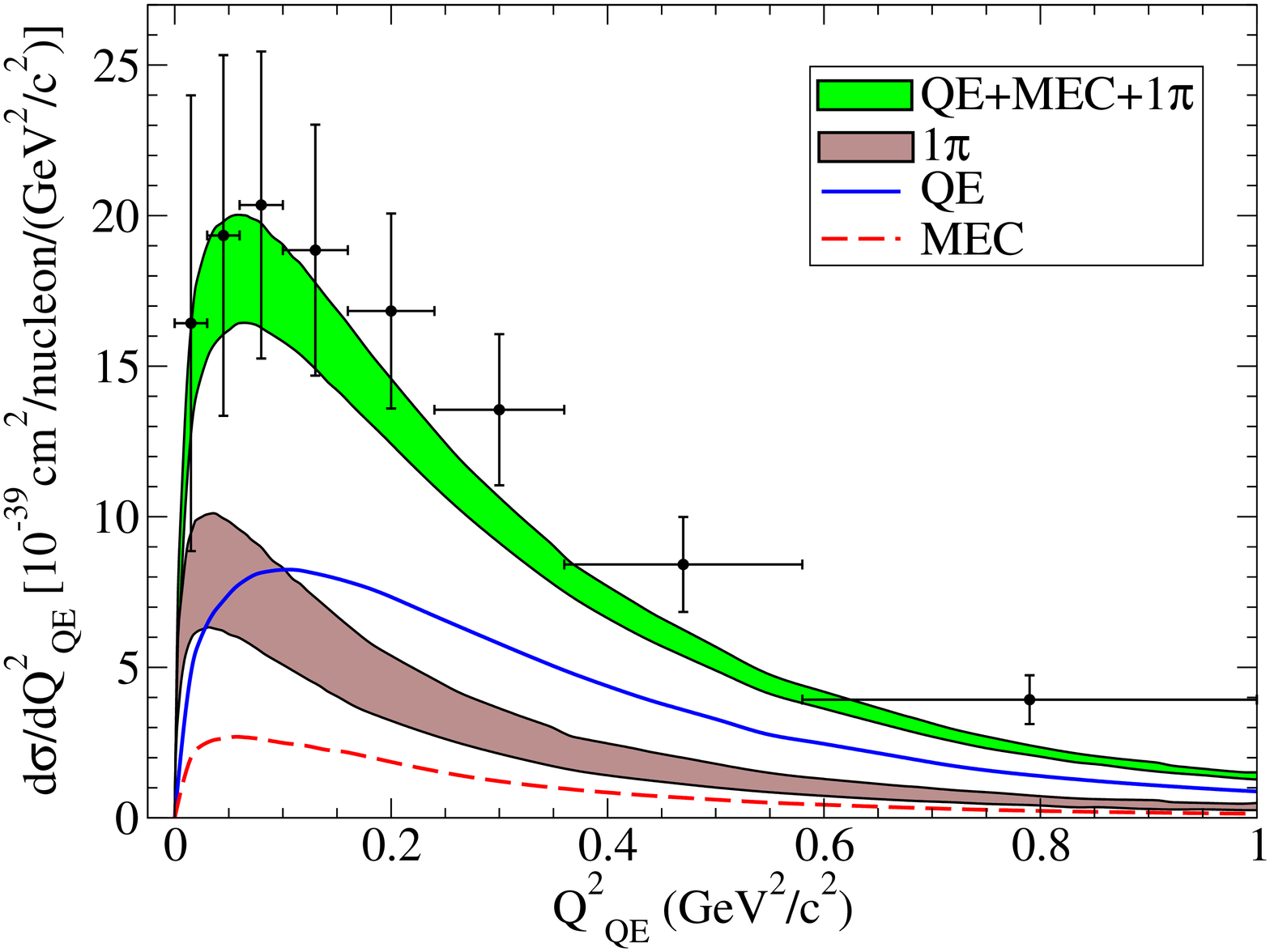}
\hspace{0.25cm}
\includegraphics[width=.345\textwidth,angle=270]{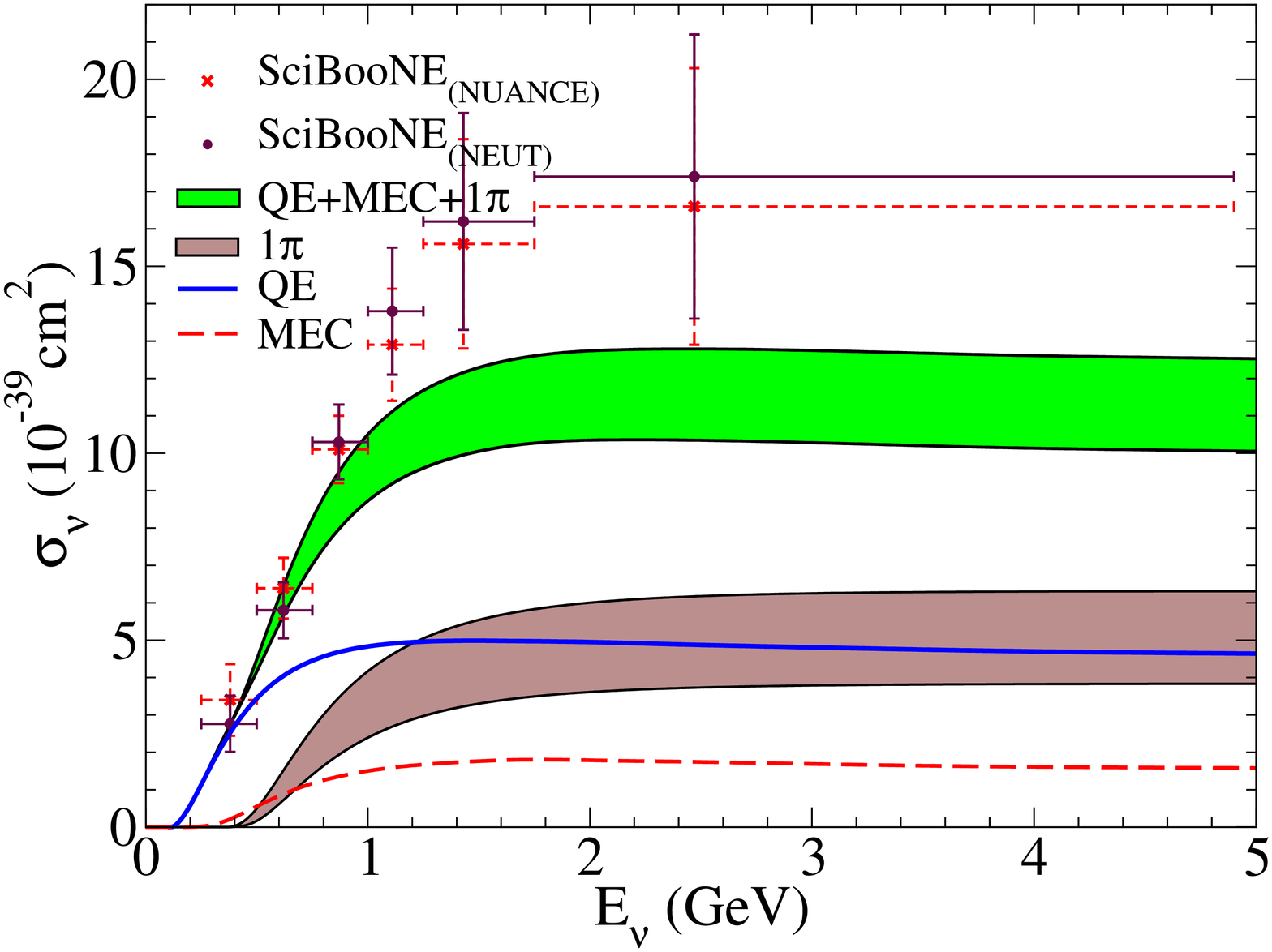}
\caption{ (Left panel) We show the CC-inclusive T2K flux-folded
  $\nu_e$-$^{12}$C $Q^2_{QE}$ differential cross section per
  nucleon. (Right panel) The CC $\nu_\mu$ total cross section on C$_8$H$_8$ is
  presented. Experimental data are from T2K~\cite{T2Kinclusive14}
  and SciBooNE~\cite{SciBooNE11}. Theoretical predictions for QE, non-QE
  (1$\pi$) and the 2p2h MEC are shown
  separately. 
  Plots  from Ref.~\cite{Amaro20}. }
\label{fig:T2KQE-SciB}
\end{figure}

In order to analyze the neutrino-induced pion production channel we make use of the phenomenological SuSA-$\Delta$ approach which extends the superscaling arguments observed in the QE regime to the $\Delta$ resonance region (see~\cite{Ivanov16} for details). In Fig.~\ref{fig:T2KQE-SciB} (left panel) we show the $\nu_e$-$^{12}$C inclusive differential cross section averaged with the T2K flux versus the reconstructed four-momentum transfer, $Q^2_{QE}$, and in Fig.~\ref{fig:T2KQE-SciB} (right panel) the $\nu_\mu$ total cross section on C$_8$H$_8$ target. Three contributions are shown in these plots, namely, the SuSAv2 QE and SuSAv2 2p2h-MEC channels together with the SuSA-$\Delta$ approach.
Although an overall good agreement with data is observed in both panels, some underestimations are present at large kinematics, {\it i.e.,} high $E_\nu$ and $Q^2_{QE}$, which reveal the need for including higher resonances and deep inelastic scattering in the description. To address this, the SuSAv2-inelastic, which has been successfully applied for the analysis of the full inelastic spectrum in electron scattering (see Sect.~\ref{sec:electron} for details), will be soon extended to the analysis of the neutrino sector.

\section{Conclusions}
\label{sec:concl}
In this paper we have summarized the basic ingredients that go into scaling analyses of inclusive electron scattering and charge-changing neutrino reactions with nuclei. Importantly, good agreement in the former case is viewed as a pre-requisite to being able to predict the latter --- clearly if one fails to account for inclusive electron scattering then it is unreasonable to expect that inclusive CC$\nu$ reactions will be adequately modeled. 

In particular, we have emphasized the use of scaling of the first kind (independence of $q$ at high energies) and of scaling of the second kind (independence of nuclear species). When both kinds of scaling are invoked we refer to it as Super-Scaling and accordingly our present focus has been placed on the so-called Super-Scaling Approach (SuSA) in its original form and in extensions of that original form. Such extensions have been introduced to account for the modest level of scaling violations seen in direct comparisons with inclusive electron scattering data.

In addition we have introduced the way 2p-2h MEC effects have been incorporated, showing that, while they are usually corrections to the cross sections being represented via the scaling functions obtained either phenomenologically or through use of specific models, they typically are required to get a successful picture of the response.

We have briefly summarized these basic ideas and then proceeded to show examples of the excellent agreement found with inclusive electron scattering cross sections measured for several light to medium-weight nuclei.

To place the discussions in context we have also provided a section outlining how semi-inclusive scattering and inclusive scattering are related, and specifically how the former probes particular regions of missing-energy and -momentum (characterized by so-called trajectories), while the latter involves integrations over the $E_m - p_m$ plane of the semi-inclusive response. On the one hand, inclusive scattering, being a total hadronic cross section, is less dependent on the underlying details of nuclear structure. In fact, given that the kinematics of the single-nucleon knockout are handled relativistically and that there are sum rules that determine the integral of the inclusive response, then a single parameter can be used to fix the width of the response. For this reason many models yield the rough behaviour of the inclusive cross section, even the relativisitic Fermi gas model. On the other hand, the semi-inclusive cross section depends critically on how the strength is distributed as a function of $E_m$ and $p_m$. Accordingly, models that fare reasonably well for inclusive scattering cannot be relied upon to properly represent the underlying nuclear structure needed for semi-inclusive reactions.

Finally, having discussed the foundations of Super-Scaling for electron scattering we show some selected results for CC$\nu$ reactions. The agreements with existing data are found to be excellent, giving us confidence that at least for inclusive neutrino reactions the problem appears to be well in hand. Note, however, from the statements made in the previous paragraph, that this is not to be taken as proof that semi-inclusive CC$\nu$ modeling should be expected to be robust. Indeed, different models yield quite different results for cross sections in which both a charged lepton and a nucleon are detected in coincidence (a trend in modern experiments), even when the corresponding inclusive cross sections do not differ significantly. The modeling of semi-inclusive reactions is part of our present projects.

\section{Acknowledgments}
This work was supported in part by the Project BARM-RILO-20 of University of Turin and from INFN, National Project NUCSYS (M.B.B.); by the Madrid Government and Complutense University under, project PR65/19-22430 (R.G.-J.); 
by the Spanish Ministerio de Ciencia, Innovaci\'on y Universidades and ERDF (European Regional Development Fund) under contracts FIS2017-88410-P, and by the Junta de Andalucia (grants No. FQM160 and SOMM17/6105/UGR) (J.A.C.); by the University of Tokyo ICRR's Inter-University Research Program FY2020\&FY2021 and by the European Union's Horizon 2020 research and innovation programme under the Marie Sk{\l}odowska-Curie grant agreement No. 839481 (G.D.M.); by the Spanish Ministry of Science through grant FIS2017-85053-C2-1-P, and by Junta de Andalucia (grant No. FQM-225) (J.E.A. and I.R.-S); by the Office of Nuclear Physics of the US Department of Energy under Grant Contract DE-FG02-94ER40818 (T.W.D.).

\end{document}